\documentclass[preprintnumbers,amsmath,amssymb,nofootinbib,twocolumn]{revtex4-1}

\usepackage{dcolumn}
\usepackage{bm}
\usepackage{graphicx}

\usepackage{color}
\usepackage{natbib}
\def\ie{\hbox{\it i.e.\,}{}}  

\def\eg{\hbox{\it e.g.\,}{}}  

\def\etal{\hbox{\it et al.}}

\def\GeV{\,{\rm GeV}}

\def \lta {\mathrel{\vcenter
     {\hbox{$<$}\nointerlineskip\hbox{$\sim$}}}}
\def \gta {\mathrel{\vcenter
     {\hbox{$>$}\nointerlineskip\hbox{$\sim$}}}}

\def\gap{\;\rlap{\lower 2.5pt
 \hbox{$\sim$}}\raise 1.5pt\hbox{$>$}\;}
\def\lap{\;\rlap{\lower 2.5pt
   \hbox{$\sim$}}\raise 1.5pt\hbox{$<$}\;}
\def\gsim{\;\rlap{\lower 2.5pt
 \hbox{$\sim$}}\raise 1.5pt\hbox{$>$}\;}
\def\lsim{\;\rlap{\lower 2.5pt
 \hbox{$\sim$}}\raise 1.5pt\hbox{$<$}\;}
\def\msun{{\rm\,M_\odot}}

\def\cm{{\rm\,cm}}
\def\sec{{\rm\,s}}

\def\sr{{\rm\,sr}}

\def\cm{{\rm\,cm}}

\def\kpc{{\rm\,kpc}}

\def\GeV{{\rm\,GeV}}
\def\TeV{{\rm\,TeV}}
\def\sec{{\rm\,s}}
\def\sr{{\rm\,sr}}

\def\spose#1{\hbox to 0pt{#1\hss}}
\def\lta{\mathrel{\spose{\lower 3pt\hbox{$\mathchar''218$}}
     \raise 2.0pt\hbox{$\mathchar''13C$}}}
\def\gta{\mathrel{\spose{\lower 3pt\hbox{$\mathchar''218$}}
     \raise 2.0pt\hbox{$\mathchar''13E$}}}
\newcommand{\beq}{\begin{equation}}
\newcommand{\eeq}{\end{equation}}
\newcommand{\be}{\begin{equation}}
\newcommand{\ee}{\end{equation}}
\newcommand{\nn}{\nonumber}
\newcommand{\ls}{\mathrel{\raise1.16pt\hbox{$<$}\kern-7.0pt 
\lower3.06pt\hbox{{$\scriptstyle \sim$}}}}         
\newcommand{\gs}{\mathrel{\raise1.16pt\hbox{$>$}\kern-7.0pt 
\lower3.06pt\hbox{{$\scriptstyle \sim$}}}}         

\long\def\comment#1{}

\def\msun{M_{\odot}}
\def\fun#1#2{\lower3.6pt\vbox{\baselineskip0pt\lineskip.9pt
  \ialign{$\mathsurround=0pt#1\hfil##\hfil$\crcr#2\crcr\sim\crcr}}}
\def\lap{\mathrel{\mathpalette\fun <}}
\def\gap{\mathrel{\mathpalette\fun >}}
\newcommand{\ba}{\begin{eqnarray}}
\newcommand{\ea}{\end{eqnarray}}

\newcommand{\ben}{\begin{eqnarray}}
\newcommand{\een}{\end{eqnarray}}
\newcommand{\bi}{\begin{itemize}}
\newcommand{\ei}{\end{itemize}}
\newcommand{\citeeq}[1]{Eq.~(\ref{#1})}
\newcommand{\citeeqs}[1]{Eqs.~(#1)}
\newcommand{\citeeqp}[1]{Eq.~\ref{#1}}
\newcommand{\citesec}[1]{Sect.~\ref{#1}}
\newcommand{\citesecs}[1]{Sects.~\ref{#1}}
\newcommand{\citetab}[1]{Tab.~\ref{#1}}
\newcommand{\citefig}[1]{Fig.~\ref{#1}}
\newcommand{\citefigs}[1]{Figs.~{#1}}

\newcommand{\vlii}{\mbox{\it Via Lactea II}}
\newcommand{\aquarius}{\mbox{\it Aquarius}}
\newcommand{\mchi}{\mbox{$m_{\chi}$}}
\newcommand{\sigv}{\mbox{$\langle \sigma v\rangle$}}

\newcommand{\gprop}{\mbox{${\cal G}$}}

\newcommand{\gtilde}{\mbox{$\widetilde{\cal G}$}}

\newcommand{\Msun}{\mbox{$M_{\odot}$}}
\newcommand{\Rsun}{\mbox{$R_{\odot}$}}
\newcommand{\xsun}{\mbox{$\vec{x}_{\odot}$}}
\newcommand{\rhosun}{\mbox{$\rho_{\odot}$}}

\newcommand{\Rvir}{\mbox{$R_{\rm vir}$}}
\newcommand{\MMW}{\mbox{$M_{\rm MW}$}}
\newcommand{\ofsub}[1]{\mbox{${#1}_{\rm sub}$}}
\newcommand{\ofsubtot}[1]{\mbox{${#1}_{\rm sub}^{\rm tot}$}}
\newcommand{\rhosm}{\mbox{$\rho_{\rm sm}$}}
\newcommand{\rhosub}{\mbox{$\rho_{\rm sub}$}}
\newcommand{\rhotot}{\mbox{$\rho_{\rm tot}$}}
\newcommand{\gsm}{\mbox{$g_{\rm sm}$}}
\newcommand{\gsub}{\mbox{$g_{\rm sub}$}}
\newcommand{\msub}{\mbox{$M_{\rm sub}$}}
\newcommand{\Mmax}{\mbox{$M_{\rm max}$}}
\newcommand{\Mmin}{\mbox{$M_{\rm min}$}}
\newcommand{\msubtot}{\mbox{$M_{\rm sub}^{\rm tot}$}}
\newcommand{\nsub}{\mbox{$N_{\rm sub}$}}
\newcommand{\fsub}{\mbox{$f_{\rm sub}^{\rm tot}$}}
\newcommand{\fref}{\mbox{$f_{\rm ref}$}}
\newcommand{\nsubtot}{\mbox{$N_{\rm sub}^{\rm tot}$}}
\newcommand{\boost}{\mbox{${\cal B}$}}

\newcommand{\prob}{\mbox{${\cal P}$}}
\newcommand{\mymeanlr}[1]{\mbox{$ \left\langle {#1} \right\rangle $}}
\newcommand{\mymean}[1]{\mbox{$ \langle {#1} \rangle $}}

\newcommand{\aap}{Astron. Astroph.}
\newcommand{\apjs}{Astrophys. J. Suppl. Series}
\newcommand{\apjl}{Astrophys. J. Lett.}
\newcommand{\mnras}{MNRAS}
\newcommand{\physrep}{Phys. Rept.}
\newcommand{\jcap}{J. Cosm. Astropart. Phys.}

\graphicspath{{./LastFigs/}}

\begin{document}

\title{Implications of High-Resolution Simulations on Indirect Dark Matter 
  Searches}

\author{Lidia Pieri$^{a}$, Julien Lavalle$^b$, Gianfranco Bertone$^a$, 
  Enzo Branchini $^c$}

\email{lidia.pieri@gmail.com,lavalle@to.infn.it}
\email{gf.bertone@gmail.com,branchin@fis.uniroma3.it}

\affiliation{$^a$ Institut d'Astrophysique de Paris, France. UMR7095-CNRS \\ 
  Universit\'e Pierre et Marie Curie, 98bis Boulevard Arago, 
  75014 Paris --- France}
\affiliation{$^b$ Dipartimento di Fisica Teorica, Universit\`a di Torino \& 
INFN, Via Giuria 1, I-10125 Torino --- Italia}
\affiliation{$^c$ Dipartimento di Fisica, Universit\`a di Roma 3, 
  Via della Vasca Navale 84, I-00146 Roma --- Italia}

\begin{abstract}
We study the prospects for detecting the annihilation products of
Dark Matter [DM]  in the framework of the two highest-resolution
numerical simulations currently available, \ie\ \vlii~and 
\aquarius. We propose a strategy to determine the shape and size of the 
region around the Galactic center that maximizes the probability of observing
a DM signal, and we show that although the predicted flux can differ 
by a factor of 10 for a given DM candidate in the two simulation setups, 
the search strategy remains actually unchanged, since it relies on the
angular profile of the annihilation flux, not on its normalization.
We present mock $\gamma$-ray maps that keep into account the diffuse
emission produced by unresolved halos in the Galaxy, and we show
that in an optimistic DM scenario a few individual clumps can be resolved 
above the background with the Fermi-LAT. Finally we calculate the 
energy-dependent boost factors for positrons and antiprotons, and show that 
they are always of $\cal O$(1), and therefore they cannot lead to the large 
enhancements of the antimatter fluxes required to explain the recent PAMELA, 
ATIC, Fermi and HESS data. Still, we show that the annihilation of 100 GeV 
WIMPs into charged lepton pairs may contribute significantly to the 
positron budget.
\end{abstract}

\maketitle

\preprint{DFTT 50/2009}

\section{Introduction}
\label{sec:intro}

Despite many observational and theoretical efforts, the nature of the 
Dark Matter, one of the main components of the universe, is still unknown. 

This has motivated the search for signals arising from the (weak) coupling
of the dark sector to ordinary matter and radiation, one of the most promising 
being self-annihilation. "Indirect" DM searches are based on the search for 
secondary particles (neutrinos, energetic electrons, antimatter and 
$\gamma$-rays), produced by the annihilation or decay of DM particles.

The spectacular increase in the positron ratio above 10 GeV measured by the 
PAMELA satellite \cite{2009Natur.458..607A} has given a boost to the 
phenomenological study of DM models and properties. The PAMELA excess can be 
interpreted in terms of standard astrophysical sources, see \eg\ 
Refs.~\cite{2009JCAP...01..025H,2008arXiv0812.4457P,2009arXiv0907.0373G,2009PhRvL.103e1104B,2009arXiv0905.0904P,2009PhRvL.103e1101Y,2010arXiv1002.1910D}, 
whereas its interpretation as DM annihilation signal requires unconventional DM 
particle models, see \eg\ \cite{2009NuPhB.813....1C,2009PhRvD..79a5014A}, and it
is rather severely constrained by the absence of an associated flux of IC 
photons, antiprotons and $\gamma$-rays \cite{2009JCAP...03..009B,2009PhRvD..79h1303B,2009arXiv0905.0372P,2009ApJ...699L..59B,2009arXiv0906.2251B,2009JCAP...07..020P,2010PhRvD..81l3522C}. 
Furthermore, all DM models with a high annihilation cross section, as needed
to reproduce the PAMELA data, would heat and ionize the baryons in the early 
universe; the constraints that can be set on these models from CMB data do not 
rely on uncertain assumptions on the DM distribution in virialized structures, 
and can therefore be regarded as robust and model-independent 
\cite{2009PhRvD..80023505,2009arXiv0906.1197S,2009arXiv0907.0719C}.

In this paper we discuss a self-consistent study of the antimatter flux 
arising from DM annihilations along with the associated $\gamma$-ray flux. A 
comprehensive outlook of the observable effect of a given DM candidates can 
indeed be useful in the case of a future claim for a DM signal.

Both messengers can indeed be produced from the hadronization (decay) of the DM 
annihilation final states if they are quarks (heavy bosons). 
In addition, the electrons or positrons injected from DM annihilation produce 
$\gamma$ rays from inverse Compton scattering with the interstellar radiation 
fields; although negligible in the case of hadronizing annihilation final 
states with respect to the $\pi^0$ decay yield, this $\gamma$-ray production 
could be sizable for leptophilic DM models. In fact, a $\gamma$-ray signal from 
DM annihilation would provide the 'cleanest' evidence for DM, since photons do 
not suffer deflection and energy losses in the local universe. Besides peculiar 
spectral features such as annihilation lines and final state radiation 
\cite{1998APh.....9..137B,2005JCAP...04..004B,2008JHEP...01..049B,2005PhRvL..94q1301B,2009arXiv0904.1442B}, an interesting smoking-gun for DM would be the 
detection of many  $\gamma$-ray sources with identical spectra and no 
counterpart at other wavelengths \cite{2000PhRvD..61b3514B,2004PhRvD..69d3512P,2004PhRvD..69l3501E,2004PhRvD..69d3501K,2005AIPC..745..434T,2006PhRvD..73f3510B,2005PhRvD..72j3517B, 2006PhRvD..74j3504H,2007PhRvD..75b3513C,2008IJMPD..17.1125F,2008MNRAS.384.1627P,2009A&A...496..351P,2008PhRvD..78j1301A}. 
In addition, a characteristic DM signature may also be found in the angular 
power spectrum of the diffuse $\gamma$-ray background 
\cite{2006PhRvD..73b3521A,2007PhRvD..75f3519A,2008PhRvD..77l3518C,2007JCAP...04..013C,2008JCAP...10..040S,2009PhRvL.102x1301S,2009arXiv0903.4685A,2009arXiv0903.2829D,2009arXiv0901.2921F}. 

We consider here four benchmark models, representative of the most commonly discussed DM candidates, and of the models that have been invoked to explain the cosmic 
leptons data discussed above.

The issue of the spatial distribution of DM can be tackled in different ways. 
Analytic methods based on the excursion set theory \cite{1993MNRAS.262..627L} 
provides a useful, though approximate, insight on the evolution of DM halos
\cite{2009arXiv0906.1730A}. N-body simulations are the best way to study the 
highly non-linear processes involved in the evolution of substructures. 
Unfortunately, they can only probe a limited range of halo masses and scales. 
The latest numerical simulations of a Milky Way (MW) - sized DM halos 
\cite{2007ApJ...657..262D,2008Natur.454..735D,2008Natur.456...73S,2008MNRAS.391.1685S}
are able to resolve $\sim 100,000$ substructures down to $\sim 10^{4.5} \msun$ 
at the present epoch. The evolution of micro-halos with size close to the free 
streaming mass can only be studied by simulating a small region at very high 
redshifts \cite{2005Natur.433..389D}.

As a consequence, modeling the properties of the galactic subhalos requires 
aggressive extrapolations which are usually performed by means of analytic, 
Monte Carlo or hybrid techniques and therefore are potentially affected by 
large theoretical uncertainties. In this work we rely on the results of the 
\aquarius~\cite{2008Natur.456...73S,2008MNRAS.391.1685S} and  
\vlii~\cite{2008Natur.454..735D} numerical simulations. For 
$\gamma$-rays, we then apply the hybrid method of  \cite{2008MNRAS.384.1627P} 
to compute the expected annihilation flux of  $\gamma$-ray photons produced 
within our Galaxy. For antimatter, we use the method developed in 
\cite{2007A&A...462..827L} to obtain the boost factor to cosmic-ray fluxes due 
to the presence of the same population of subhalos. In the photon flux 
prediction we also need to 
account for the extragalactic contribution and the diffuse Galactic foreground. 
In order to model such contributions we have scaled down the signal
measured by EGRET at E $>$ 3 GeV by 50\%. This reduction accounts for the fact 
that Fermi data do not confirm the so-called galactic excess measured by EGRET 
neither in the strips $10^{\circ}<|b|<20^{\circ}$ \cite{2009PhRvL.103y1101A} nor
at larger latitudes, as we have verified by comparing the Fermi maps made 
available by \cite{2009arXiv0910.4583D} with the EGRET data 
\cite{2005ApJ...621..291C}. The antimatter flux is 
obtained by solving the transport equation for high energy positrons and 
antiprotons produced by DM annihilation, and ignoring contributions from 
astrophysical sources. In this case, the background produced by spallation 
processes is taken from~\cite{2001ApJ...563..172D} 
and~\cite{2009A&A...501..821D}. 

The main aim of this work is to assess the reliability of this approach, \ie\ 
that modeling the expected DM annihilation flux extrapolating the results of 
state-of-the-art numerical simulation provides robust predictions that can be 
used to assess the possibility of detecting the annihilation signals, of both 
photons and antimatter, with current detectors.

The paper is organized as follows: In \citesec{sec:dm_mod} we describe our 
models for DM halos and their substructures which contribute to the cosmological
part of the DM annihilation signal. In \citesec{sec:bench} we introduce our 
particle physics benchmark models that determine  amplitude, shape and 
features of the annihilation spectrum. Our model predictions for the 
$\gamma$-ray and antimatter fluxes are described in \citesecs{sec:gamma} 
and \ref{sec:antimatter}, respectively. \citesec{sec:ics} shows the Inverse 
Compton Scattering computation for the particle physics benchmarks for which it 
is relevant. Finally, we discuss our results and 
conclude in Section \ref{sec:concl}.

In this work we adopt the WMAP-5yr \cite{2009ApJS..180..306D} flat, 
$\Lambda$CDM model ($\Omega_m=0.26$, $\Omega_{\Lambda}=0.74$, $\sigma_8$=0.79, 
$n_s$=0.96, H$_0=72\,{\rm km s^{-1} Mpc^{-1}}$).

\section{Modeling DM halos and subhalos}
\label{sec:dm_mod}

The N-body experiments \aquarius~
\cite{2008Natur.456...73S,2008MNRAS.391.1685S} and  \vlii~
\cite{2008Natur.454..735D} have simulated the DM halo of a MW-like 
galaxy in a flat, $\Lambda$CDM  model with cosmological parameters consistent, 
within the errors, with those that best-fit the WMAP-1yr and WMAP-3yr data,
respectively (\aquarius~used  $\Omega_m=0.25$, $\Omega_{\Lambda}=0.75$, 
$\sigma_8$=0.9, $n_s$=1, $H_0=73 \, {\rm km/s/Mpc}$, while \vlii~
used $\Omega_m=0.24$, $\Omega_{\Lambda}=0.76$, $\sigma_8$=0.88, $n_s$=0.97, 
H$_0=74 \,{\rm km/s/Mpc}$). Thanks to the unprecedented high resolution, these 
simulations were able to resolve substructures down to masses as small as  
$\sim 10^{4.5} \msun$ (\aquarius) and $\sim 10^5 \msun$  
(\vlii), to characterize their inner structure, to trace their 
spatial distribution within the main halo and to model the dependence of their 
shape parameter (the concentration) on the distance from the Galactic Center 
(GC). For the  \vlii~ we will consider subhalos selected by mass 
as in \cite{2008ApJ...686..262K} rather than by peak circular velocity as in 
\cite{2008Natur.454..735D}. We have checked that using either subhalo 
parameterizations  does not significantly affect our predictions for the 
detectability of the annihilation signal.

Following the results of numerical simulations, the DM distribution in the MW 
halo consists of two separate phases: a smoothly distributed component (main 
halo) and a clumpy component made of virialized substructures (subhalos). We 
therefore ignore the presence of caustics, streams and all other possible 
inhomogeneities that do not correspond to virialized structures.

\par The total density profile of the MW DM halo (smooth halo + subhalos) can 
be modeled by a Navarro, Frenk and White (NFW) profile in the case of \vlii:
\ben
\rho^{VLII}_{\rm tot}(R) =
\frac{\rho_s}{\frac{R}{r_s}\left(1+\frac{R}{r_s}\right)^2} ,
\label{eq:vl2smooth}
\een
and by a shallower Einasto profile with $\alpha=0.17$ for \aquarius:
\ben
\rho^{Aq}_{\rm tot}(R)=\rho_s \, 
\exp\left\{-\frac{2}{\alpha}
\left[\left(\frac{R}{r_s}\right)^{\alpha}-1\right]\right\}.
\label{eq:aqsmooth}
\een
where $R$ is the distance from the GC. The best fitting values of the scale 
density, $\rho_s$, and the scale radius, $r_s$, are listed in 
\citetab{tab:dm_setup}. The aforementioned density profiles are shown in 
\citefig{fig:rhodm} with solid lines.

\par Being $M_h$ and $\msub$ the main halo and subhalo masses, the joint 
spatial and mass subhalo distribution is given by 
\ben
\frac{d \rho_{\rm sh}(\msub,R)}{d\msub} = 
\rho_{\rm sub}(R) \, 
{\cal F}(\mu, \msub)\;.
\label{eq:rhosh}
\een
The previous equation provides the mass density in the form of subhalos per unit
subhalo mass. The normalized mass function ${\cal F}(\mu,\msub)$, which carries
units of inverse mass, is defined a bit farther in \citeeq{eq:mass_func}. \\

In the case of  \vlii, the global subhalo mass density profile $\rho_ {sub}(R)$ 
is best fitted by the so-called antibiased relation 
\cite{2008Natur.454..735D,2008ApJ...686..262K}:
\ben
\rho^{VLII}_{\rm sub}(R) = 
\frac{\rho^{VLII}_{\rm tot}(R)\,(R/R_a)}{\left( 1 + \frac{R}{R_a}\right)} \;,
\label{eq:vl2}
\een
Given the NFW overall profile $\rho^{VLII}_{\rm tot}(R)$, we see that the subhalo
distribution is cored below a scale radius $R_a$, while it asymptotically tracks
the smooth profile beyond. The procedure to obtain this antibiased profile is 
detailed in App.~\ref{app:antibiased}, where it is shown that $R_a$ is actually 
fixed by the mass fraction in the form of subhalos.

For \aquarius, an Einasto shape is also found for the spatial distribution of 
subhalos \cite{2008Natur.456...73S,2008MNRAS.391.1685S}, which leads to the
following global subhalo mass density profile:
\ben
\rho_{\rm sub}^{Aq}(R)  =  \rho_a  \, 
\exp \left\{-\frac{2}{\alpha} 
\left[\left(\frac{R}{R_{a}}\right)^{\alpha}-1\right]\right\} \;, 
\label{eq:aqclump}
\een
with $\alpha=0.678$, and where $\rho_a \equiv  k_V \, \msubtot = k_V\,\fsub\,
\MMW$ is fixed from the total subhalo mass (or the mass fraction, equivalently) 
and the parameter $k_V$, which normalizes the exponential term to unity within 
the Galactic volume.

The normalized subhalo mass function used in both subhalo distributions reads
\ben
{\cal F}(\mu,\msub) \equiv {\cal F}_0 \,
\left[\frac{\msub}{\Msun}\right]^{-\mu}\;,
\label{eq:mass_func}
\een
where ${\cal F}_0$, which carries units of inverse mass, allows the 
normalization of the mass integral of ${\cal F}$ to unity in the surveyed mass 
range. We will use $\mu = 2$ in the \vlii~configuration, and $\mu=1.9$ in the 
\aquarius~configuration.

Note that to get the subhalo number density from the mass density, one can use 
the trivial following relation:
\ben
\frac{dN_{\rm sh}(\msub,R)}{d\msub \, dV} = \frac{1}{\langle \msub \rangle }
\frac{d\rho_{\rm sh}(\msub,R)}{d\msub} \;,
\een
where $\langle M_{\rm sub} \rangle \equiv \int dm \, m \,{\cal F}(\mu,m) = 
\msubtot / \nsub $ is the average subhalo mass. This relation is valid for
any configuration.

In the following, we will consider that $\Mmin = 10^{-6}\Msun$ and 
$\Mmax = 10^{-2}M_h$. The logarithmic mass slope $\mu$ is steeper in 
the \vlii~configuration than in the \aquarius~configuration, which strongly 
increases the relative weight of the lightest subhalos to the total mass (and
therefore to the total annihilation rate) in the former case. All the 
parameters used for the above subhalo distributions are 
listed in \citetab{tab:dm_setup}. They are set to match the results of the 
corresponding N-body simulations in the resolved subhalo mass ranges. 
In the case of \vlii, we impose that 10 \% of the MW mass, $M_{h}$, consists of 
virialized structures with masses in the range $[10^{-5} M_{h}, 10^{-2} M_{h}]$. 
In the case of \aquarius, we require that 13.2 \% of $M_{h}$ is concentrated in 
subhalos with masses in the range $[1.8 \times 10^{-8} M_{h}, 10^{-2} M_{h}]$. 
The total mass fraction in the form of subhalos \fsub~is then such that:
\ben
\fsub \, M_h = \msubtot \equiv 
4\pi\int_{0}^{\Rvir} dr \, r^2 \int_{\Mmin}^{\Mmax} dm \, 
\frac{d\rho_{\rm sh} (m,r)}{dm}.\nn\\
\een
Finally, we can now define the smooth dark matter component for both 
configurations from the difference between the total and subhalo components:
\ben
\rho_{\rm sm}(R) = \rho_{\rm tot}(R) - \rho_{\rm sh}(R)\;.
\label{eq:def_smooth}
\een
\

\begin{figure}[t!]
 \centering
 \includegraphics[width=\columnwidth]{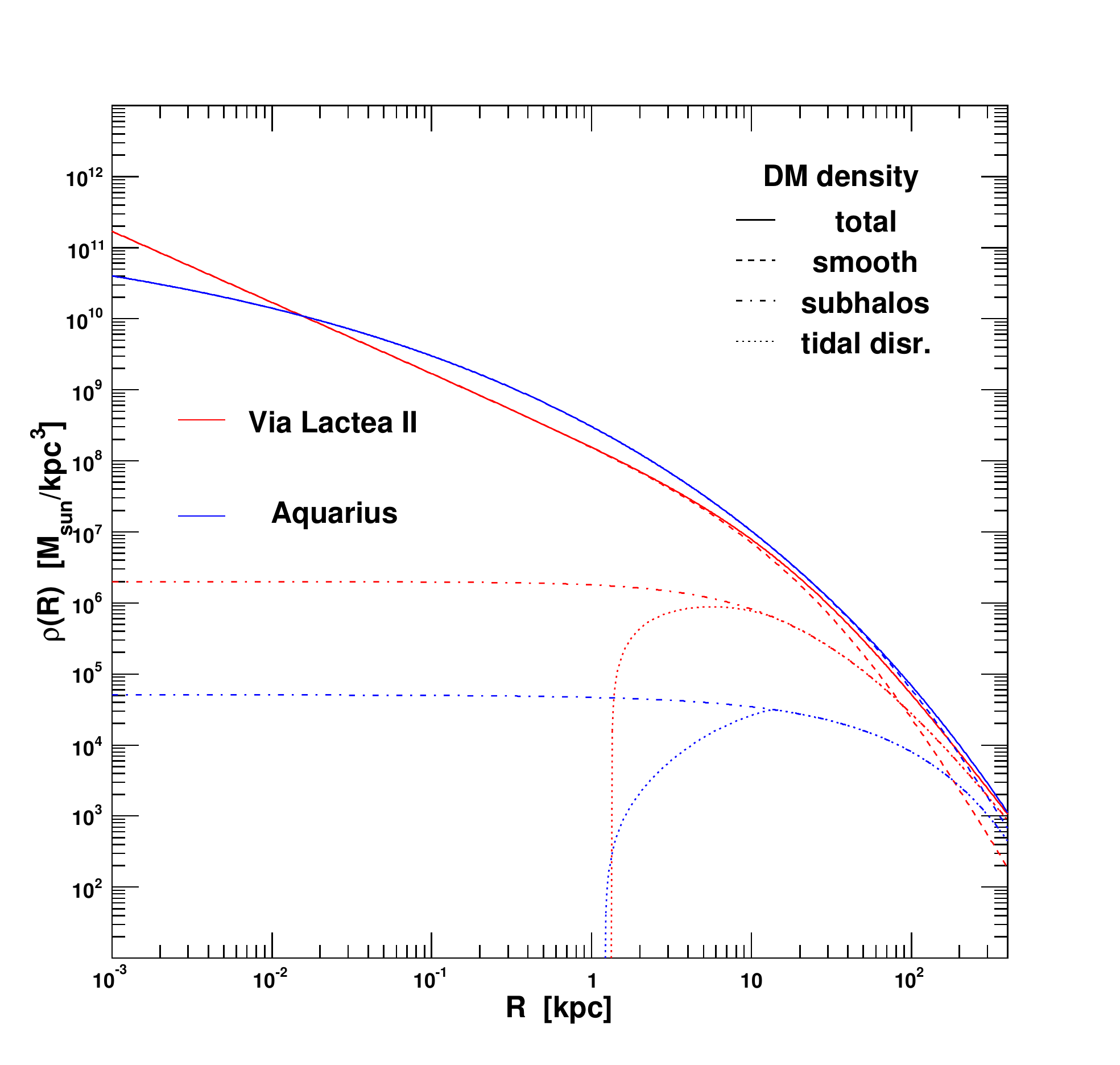}
\caption{Mass density profiles of the MW halo components for the 
  \vlii~and \aquarius~cases. For each setup, the solid line 
  represents the sum of all components, while the dashed line is the smooth 
  halo component and the dotted-dashed line accounts for the subhalo 
  component. The dotted line exhibits the subhalo component when the tidal
  disruption according to the Roche criterion is implemented.}
\label{fig:rhodm}
\end{figure}

We note that the MW mass in both simulations agrees, within the errors, with 
the recent observational estimates of \cite{2008MNRAS.384.1459L} based on the 
so-called Timing Argument \cite{1959ApJ...130..705K}.

\begin{figure}[t]
\begin{center}
\includegraphics[width=\columnwidth, clip]{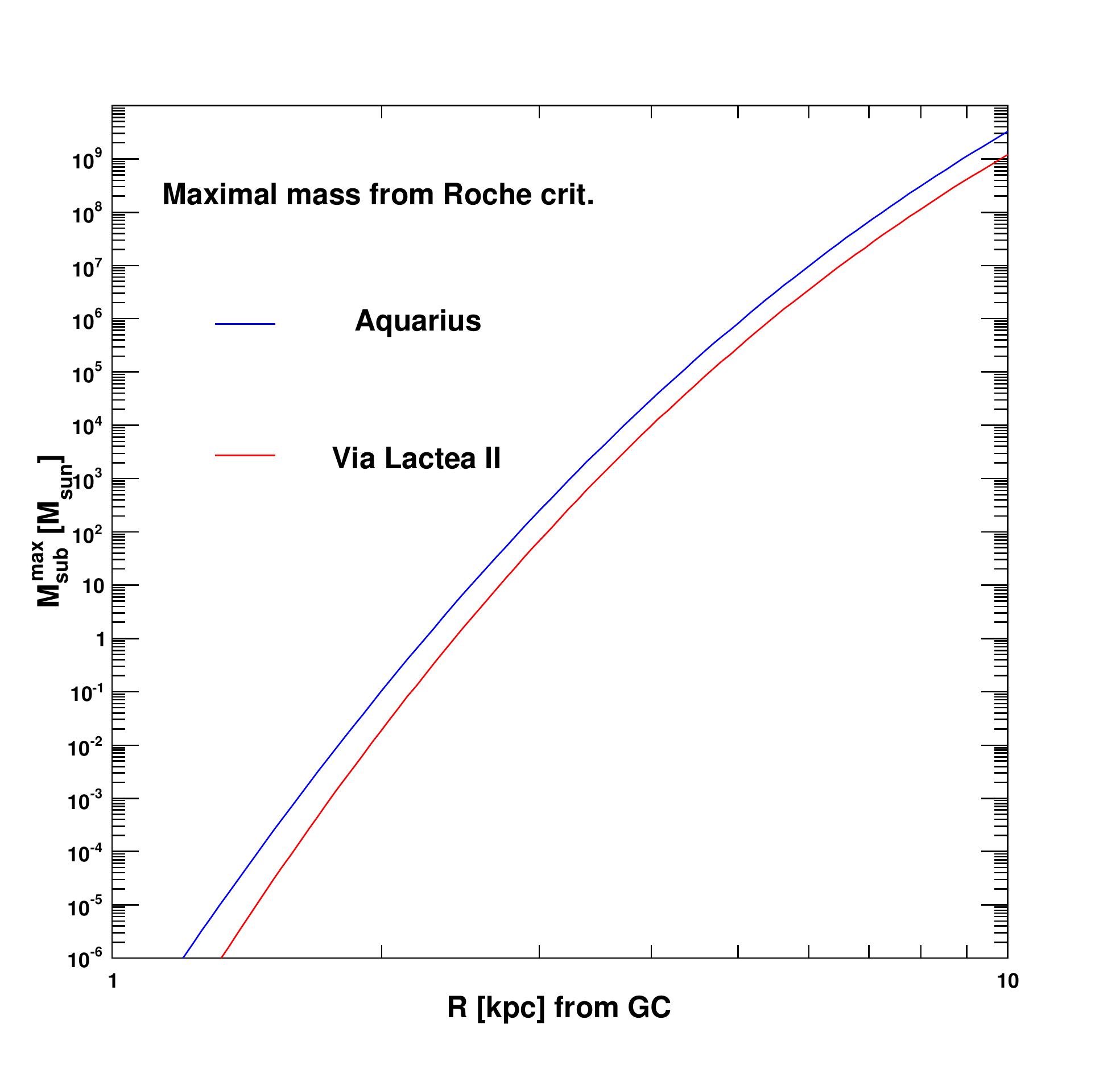}
\caption{Maximum subhalo mass that can be found at distance $R$ from the GC, 
  according to the Roche criterion used in this paper.}
\label{fig:roche}
\end{center}
\end{figure}

A word of caution is required for the subhalo distribution near the GC 
(\eg~\cite{2008PhRvD..77h3519B}). Since the subhalo number density at 
galactocentric distances of 8 kpc or less is poorly constrained by numerical 
simulations, we calculate this function by extrapolating the behavior at larger
distances. Tidal effects may disrupt subhalos in the central regions of the 
Galaxy, which severely depletes the subhalo population. To account for this 
effect we adopt the Roche criterion~\cite{2003ApJ...584..541H}: a subhalo is 
destroyed when its scale radius $r_s$ is larger than the tidal radius, \ie\ the 
radius at which the tidal forces of the host potential equal the self-gravity of
the subhalo:
\ben
r_{\rm tid}(R)= \left (\frac{\msub}{3 M_{h}} \right)^{1/3} R 
\label{eq:roche}
\een
Clearly, the amplitude of the effect depends on the subhalo mass and its 
distance from the GC. In \citefig{fig:roche} we plot the largest mass 
$M_{\rm sub}^{\rm max}$ of a subhalo that survives tidal disruption as a function
of $R$ --- the effect on the overall subhalo mass density profile is illustrated
by the dotted curves in \citefig{fig:rhodm}. Subhalos are shown to be almost 
completely disrupted within  $R\sim 2$ kpc. As shown by 
\cite{2004PhRvD..69d3512P}, the effect on the total $\gamma$-ray flux is 
negligible, since the main contribution at small distances 
from the GC is given by the smooth halo, while the subhalo contribution is 
subdominant. This turns out to be the case also for the antimatter flux 
as we will show, and as already pointed out by~\cite{2008A&A...479..427L}. We 
will discuss this point in greater detail in Section \ref{subsubsec:boosts}.\\

\begin{table}
\centering
\begin{tabular}{l|cc}
\hline
\hline
      & \vlii & \aquarius \\
\hline
$\Rvir \, [{\rm kpc}]$ & 402 & 433 \\
$M_{h} \, [\Msun]$ & $1.93 \times 10^{12}$ & $2.5 \times 10^{12}$ \\
$r_s \, [{\rm kpc}]$ & 21 & 20 \\
$\rho_s \, [10^6 \Msun\, \kpc^{-3}]$ & 8.1 & 2.8 \\
${\cal F}_0 \,[\msun^{-1}]$ & $10^{-6}$ & $3.6\times 10^{-6} $\\
$\rho_a \, [\Msun \, \kpc^{-3}]$ & - & $2840.3$ \\
$R_a [{\rm kpc}]$ & 85.5 & 199 \\
$\mymean{\rhosun} \, [{\rm GeV/cm^3}]$ & 0.42 & 0.57 \\
$\nsub $ & $2.8 \times 10^{16}$ & $1.1\times 10^{15}$ \\
$\msubtot (< \Rvir) \, [\Msun]$ & $1.05\times 10^{12}$ & $4.2\times10^{11}$\\
$\fsub (< \Rvir)$ & 0.53 & 0.17 \\
\hline
\end{tabular}
\caption{ Characteristics values for the smooth 
and clumpy components of the 
DM distribution modeled after the \vlii~and \aquarius~results. $\Rvir=$ virial 
radius, \ie\ the radius within which the numerical simulations define the halo. 
$M_{h}=$ MW mass. $r_s =$ scale radius of the overall DM distribution. 
$\rho_s=$ scale density of the overall DM distribution. 
${\cal F}_0=$ normalization to unity for the normalized subhalo mass function. 
$\rho_a =$ normalization of the subhalo mass density profile for the 
\aquarius~configuration. $R_a =$ scale radius of the subhalo distribution. 
$\mymean{\rhosun} = $ averaged local DM density (at 8 kpc from the GC).
$\nsub=$ total number of subhalos. $\msubtot=$ total mass in subhalos. 
$\fsub=$ clumpiness fraction, defined as $\msubtot/M_h$. Subhalo abundances 
are computed assuming the Roche criterion.}
\label{tab:dm_setup}
\end{table}

A very informative quantity that describes the inner shapes of subhalos is 
their concentration parameter, defined as the ratio between  $r_{200}$ (the 
radius which encloses an average density equal to 200 times the critical 
density of the universe) and the scale radius: $c_{200}\equiv r_{200}/r_s$. In 
general, this quantity is not constant but depends on the subhalo mass and on 
the distance from the GC. Following the numerical results of 
\cite{2008Natur.456...73S,2008MNRAS.391.1685S,2008Natur.454..735D,2008ApJ...686..262K} we can parameterize these dependences as follows:
\ben
\quad c_{200}(\msub,R)  &=& \left(\frac{R}{\Rvir}\right)^{- \alpha_R}  
\times \label{eq:cmr} \\ 
&&\left[C_1 \left[\frac{\msub}{\Msun}\right]^{- \alpha_{1}} + 
C_2 \left[\frac{\msub}{\Msun }\right]^{- \alpha_{2}}\right]. \nn
\een
The best fitting parameters for the \vlii~and \aquarius~simulations are listed 
in \citetab{tab:cvir_params}. In \citefig{fig:cm} we have plotted the mass 
dependence 
of the concentration parameter at the virial radius ($R = \Rvir$) that can be 
thought of as the concentration parameters of subhalos located at the edge of 
the simulated MW-like halo, \ie\ of fields halos. We have also plotted the 
concentration parameters computed at the Sun position ($R = 8 \kpc$), which 
provides additional information on the potential antimatter yield by featuring 
the local subhalo properties. From the plot we notice that subhalos in the 
\aquarius~experiment are more concentrated than in \vlii~at all masses; a 
discrepancy that reflects the larger power spectrum normalization ($\sigma_8$) 
assumed in the \aquarius~experiment.

\begin{table}
\centering
\begin{tabular}{l|cc}
\hline
\hline
 & \vlii & \aquarius \\
\hline
$\alpha_R$ & 0.286 & 0.237 \\
$C_1$ & 119.75 & 232.15 \\
$C_2$ & -85.16 & -181.74 \\
$\alpha_{1}$ & 0.012 & 0.0146 \\
$\alpha_{2}$ & 0.0026 & 0.008 \\
\hline
\end{tabular}
\caption{Parameters used for the fit to the concentration parameter of 
  subhalos.}
\label{tab:cvir_params}
\end{table}

\begin{figure}[t!]
 \centering
 \includegraphics[width=\columnwidth, clip]{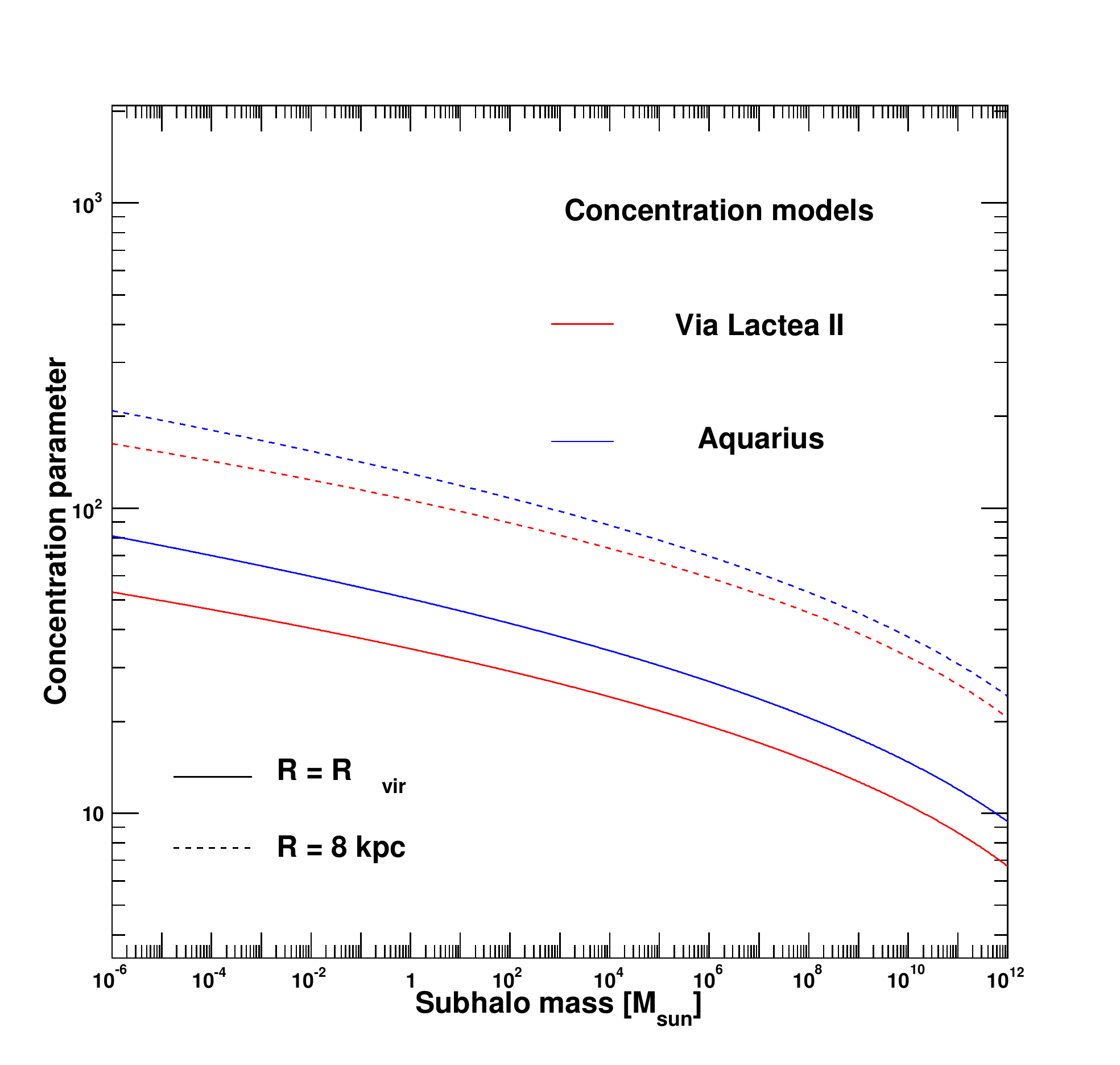}
\caption{Concentration parameter as a function of the halo mass as found in 
  \vlii~and \aquarius, computed at a 
  the virial radius $\Rvir$ and at the Earth-GC distance $R = 8$ kpc.}
\label{fig:cm}
\end{figure}

\section{Particle Physics Benchmarks}
\label{sec:bench}

In order to study the dependence of the results on particle physics parameters,
we show the results relative to four different benchmark models. 

{\bf Benchmark A} is representative of a class of models in the supersymmetric 
(SUSY) parameter space, that annihilate predominantly to $b \bar b$. In order 
to maximize the annihilation flux, we chose a light neutralino mass 
$m_\chi =40$ GeV. 

{\bf Benchmark B} is also representative of a class of SUSY models. The DM 
particle mass is in this case $m_\chi =100$ GeV, thus allowing annihilation to 
$W^+ W^-$, that is assumed to constitute the dominant annihilation channel. 

{\bf Benchmark C} provides a "minimal" solution to the rising positron ratio 
measured by PAMELA, without attempting to address higher energy (ATIC and 
Fermi) data. The mass is in this case $m_\chi =100$ GeV, thus barely above the 
PAMELA energy range, and the leading annihilation channel $e^+ e^-$.

{\bf Benchmark D}, finally, represents a class of candidates that attempt to 
explain the cosmic lepton data up to TeV energies. We have 
adopted, as \eg\ in Ref. \cite{2008PhRvD..78j3520B}, $m_\chi =2000$ GeV, and 
annihilation to $\tau^+ \tau^-$.

We have used in all cases a thermal annihilation cross section 
$\sigv  = 3 \times 10^{-26} \cm^{3} \sec^{-1}$. The parameters of the four 
benchmark models are summarized in \citetab{tab3}.

\begin{table}
\centering
\begin{tabular}{ccc}
\hline
\hline
  model & $m_\chi$ [GeV] & final state \\
\hline
A & 40 &  $b \bar b$ \\
B & 100 & $W^+ W^-$ \\
C & 100 & $e^+ e^-$ \\
D & 2000 & $\tau^+ \tau^-$ \\
\hline
\end{tabular}
\caption{Benchmark particle physics models. The annihilation cross-section is 
$\sigv  = 3 \times 10^{-26} \cm^{3} \sec^{-1}$.}
\label{tab3}
\end{table}

\section{Gamma-rays}
\label{sec:gamma}
The expected $\gamma$-ray flux from DM annihilation, $\Phi_\gamma$, can be 
factorized in two terms that depend on the  properties of the DM particle, 
$\Phi_{PP}$, and on their spatial distribution along the line-of-sight,  
$\Phi_{los}$: 
\ben
\Phi_\gamma(m_\chi,E_\gamma, M, r, \Delta \Omega) =
\Phi_{PP}(m_\chi,E_\gamma) \times \Phi_{los} (M, r, \Delta\Omega)\nn\\
\label{flusso}
\een
in units of inverse area and inverse time.
Here $m_\chi$ is the DM particle mass, $M$ the DM halo mass, r the 
position inside the halo and $\Delta \Omega$ represents the 
angular resolution of the instrument (in the case of Fermi, for energies above 
$\sim 1 \GeV$, one has $\Delta \Omega \sim 10^{-5} \sr$).

The term $\Phi_{PP}$ describes the number of photons yielded in a single 
annihilation, and can be written as:
\ben
\Phi_{PP}(M,E_\gamma) =
 \frac{1}{4 \pi} \frac{\sigv}{2 m_\chi^2} 
\int_{E_\gamma}^{m_\chi} \sum_{f} \frac{d N^f_\gamma}{d E_\gamma} B_f dE_\gamma ,
\label{flussopp}
\een
Here $f$ is the final state, $B_f$ is the branching ratio, and $\sigv$ 
denotes the thermal annihilation cross section which reproduces the observed 
cosmological abundance. 
$d N^f_\gamma / dE_\gamma$ is the differential annihilation photon spectrum 
that we take from \cite{2004PhRvD..70j3529F}.

The term $\Phi_{los}$ represents the number of annihilation events along the 
line-of-sight. It is obtained by integrating the square of the DM mass density:
\ben
\nn \Phi_{los} (M, r, \Delta\Omega) = \int \int_{\Delta \Omega} 
d \theta d \phi \int_{\rm los}  d\lambda \times \\ 
\left [ \frac{\rho_{DM}^2 (M, c , r(\lambda, \theta, \phi))} 
  {\lambda^{2}} J(x,y,z|\lambda,\theta ,\phi) \right] \,
\label{flussocosmo}
\een
Here $J$ is the Jacobian determinant and $c$ the 
concentration parameter. In the case of the smooth halo of the MW, $M = M_{h}$ 
and $c$ is fixed by the output of N-body simulations, while for the subhalos 
$M = \msub$ and $c$ is a function of mass and position: $c= c_{200}(\msub, R)$, 
as defined in \citeeq{eq:cmr}. The integration has been performed over a solid 
angle of $10^{-5} \sr$.

The $\gamma$-ray annihilation flux receives contributions from three different 
sources that we model separately: 
\begin{itemize} 
\item the DM smoothly distributed in the MW halo
\item the DM within Galactic subhalos
\item the DM in extragalactic halos and their substructure.
\end{itemize}

To compute the first contribution we simply consider the mass density profile
in \citeeq{eq:vl2smooth} or \citeeq{eq:aqsmooth} in the integral of
\citeeq{flussocosmo}.\\

To compute the second contribution one would need to consider all $\sim 10^{16}$
substructures down to the cutoff mass of  $10^{-6} M_{\odot}$, which is 
unfeasible. One possibility would be to integrate the subhalo distribution 
(\citeeqp{eq:vl2} or \citeeqp{eq:aqclump}). This represents the mean 
annihilation flux while one of the scopes of this work is to assess the 
possibility of detecting isolated DM subhalos. To circumvent the problem we 
follow the hybrid approach of \cite{2008MNRAS.384.1627P}, \ie\ we first compute 
the mean flux and then we use Monte Carlo techniques to account for the closest 
and brightest subhalos that one may hope to detect as isolated sources.

More specifically, the mean flux along the line-of-sight is obtained by 
integrating the following expression:
%
\begin{align}
&\Phi_{los}^{\rm sub}(M_{h}, R, \Delta \Omega) \propto \int_{M_ {sub}} 
d M_ {sub} \int_c d c \int \int_{\Delta \Omega} 
d \theta d \phi\times \nn\\
&\int_{\lambda}  d\lambda [ \rho_{sh}(M_{h},\msub,R) P(c)  \Phi_{los} J(x,y,z|\lambda,\theta ,\phi)]
\label{smoothphicosmo}
\end{align}
Here $\Phi_{los}$ represents the contribution from each subhalo computed by 
integrating \citeeq{flussocosmo}~in the range $[d - r_{tid}, d + r_{tid}]$, 
and $d$ is the distance to the object. 
This quantity is then convolved with the subhalo distribution 
(\citeeqp{eq:vl2} or \citeeqp{eq:aqclump}).

$P(c)$ represents the probability distribution function of the concentration 
parameter. Following \cite{2001MNRAS.321..559B}, we model it as a log-normal 
distribution with dispersion $\sigma_c$ = 0.14  and mean value $ \bar{c}$:
\ben
P(\bar{c},c) = \frac{1}{\sqrt{2 \pi} \sigma_c c} \, 
e^{- \left ( \frac{\ln(c)-\ln(\bar{c})} {\sqrt{2} \sigma_c} \right )^2}.
\label{pc}
\een

To model the fluctuations over the mean flux we compute the annihilation flux 
from the nearest and brightest subhalos in 10 independent Monte Carlo 
realizations. For this purpose we consider only those subhalos whose distance 
from the Sun is less than the maximum between: 1) the radius of the sphere 
centered on the sun within which lie about 500 halos, and 2) the distance at 
which the photon flux from a subhalo drops below the value of the average flux 
at the anticenter. The results may depend on the actual number of individual 
subhalos in the Monte Carlo simulations. To check the robustness of the results,
the authors of  \cite{2008MNRAS.384.1627P} have performed a number of 
convergence tests in which they demonstrated that increasing the number of 
individual halos does not change the estimate of their detectability. The 
scatter among the results is obtained from the different Monte Carlo 
realizations of the subhalo population. 

\begin{figure}
 \centering
 \includegraphics[width=\columnwidth, clip]{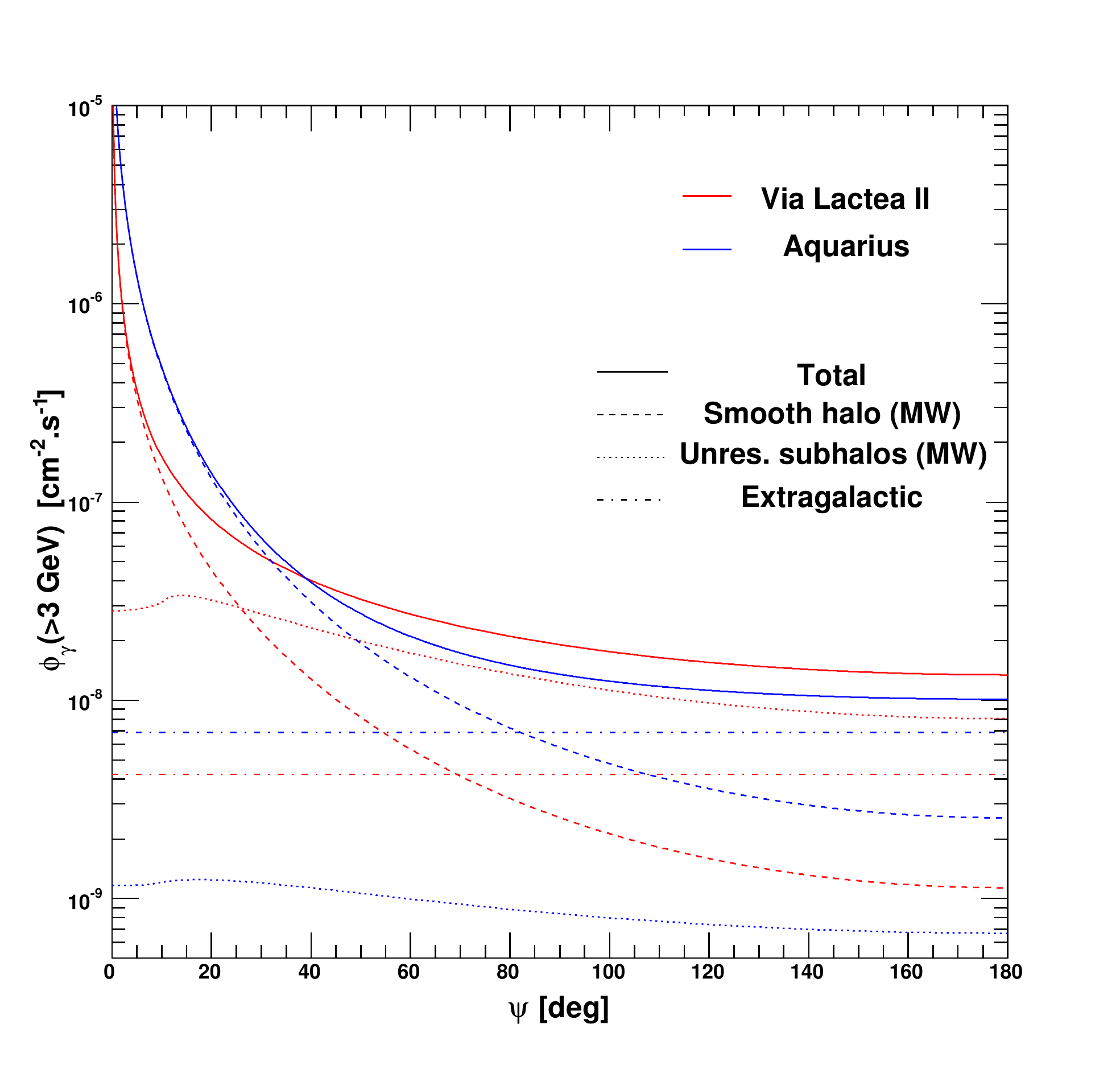}
\caption{Contribution to the $\gamma$-ray flux from DM annihilation, as a 
function of the angle of view from the GC, of the MW smooth (dashed curves), MW 
subhalo (dotted curves) and extragalactic (dotted-dashed curves) components for 
the \aquarius~(blue) and \vlii~(red) setup. The plain curves show the total
contribution for each configuration. Fluxes are measured 
over a solid angle of $10^{-5} \sr$. }
\label{fig:compareaq}
\end{figure}

Finally, to compute the extragalactic contribution to the annihilation flux we 
have used the formalism of \cite{2002PhRvD..66l3502U},  modified as in Sec. 
VI.C of \cite{2009arXiv0905.0372P} in order to account for the mass and radial 
dependence of the subhalo concentrations.

\begin{figure*}[t]
 \centering
 \includegraphics[width=\columnwidth]{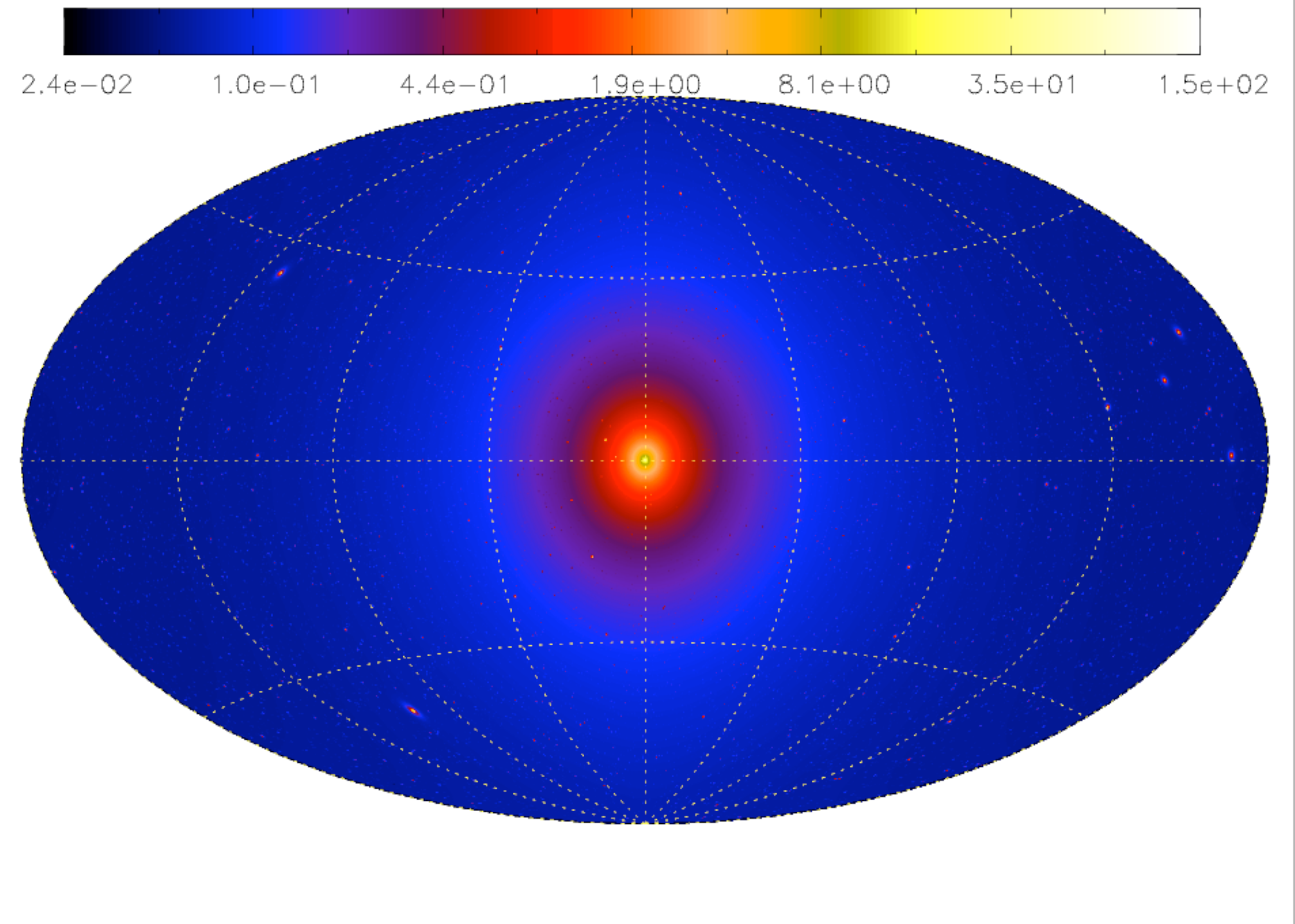}
 \includegraphics[width=\columnwidth]{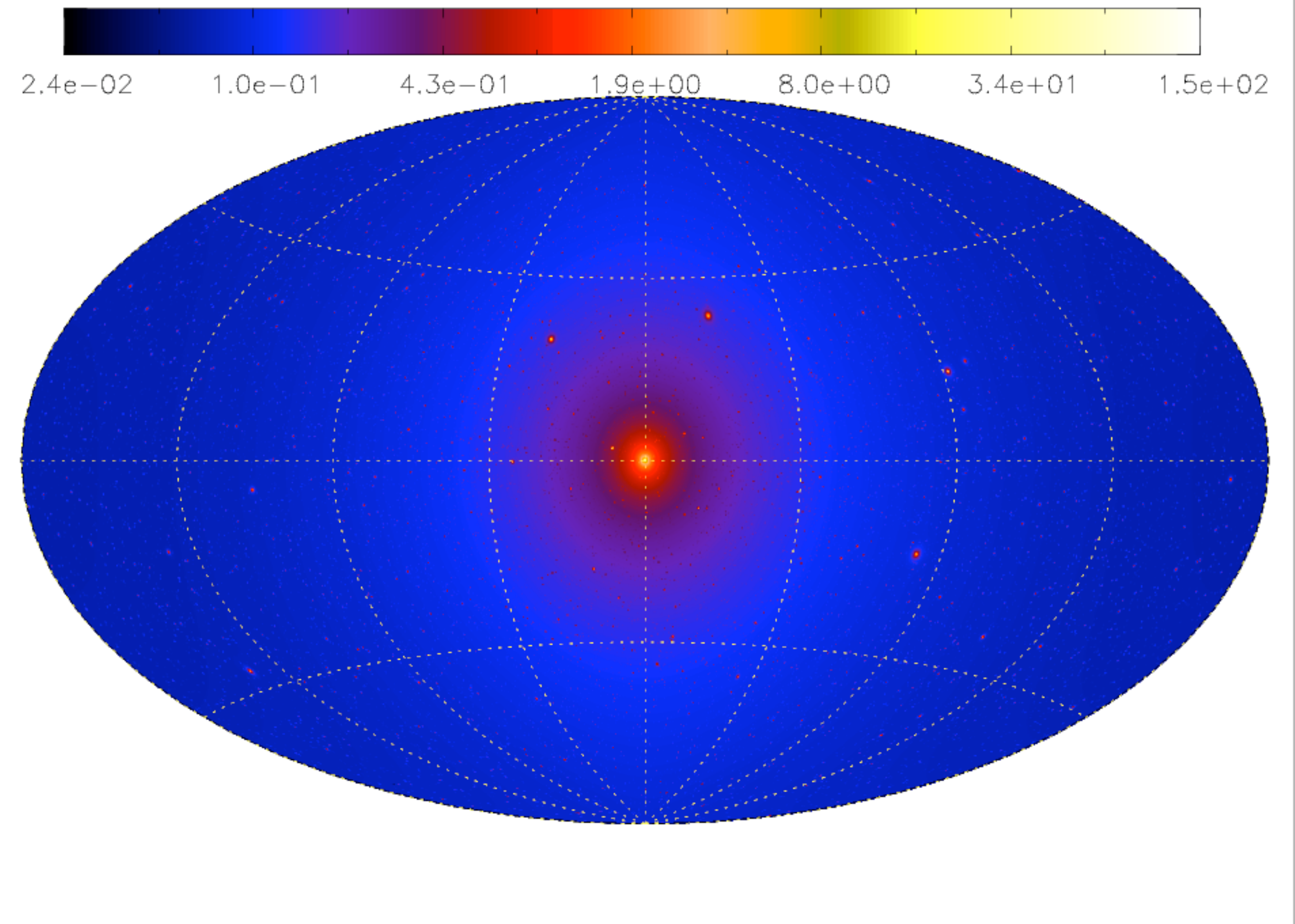}
 \caption{Full-sky map, in Galactic coordinates, of the number of photons 
   (above 3 GeV) produced by DM annihilation (benchmark A). The left (right) 
   panel shows the predicted flux in the \aquarius~(\vlii) setup.}
 \label{fig:sub}
\end{figure*}

\begin{figure*}
 \centering
 \includegraphics[width=\columnwidth]{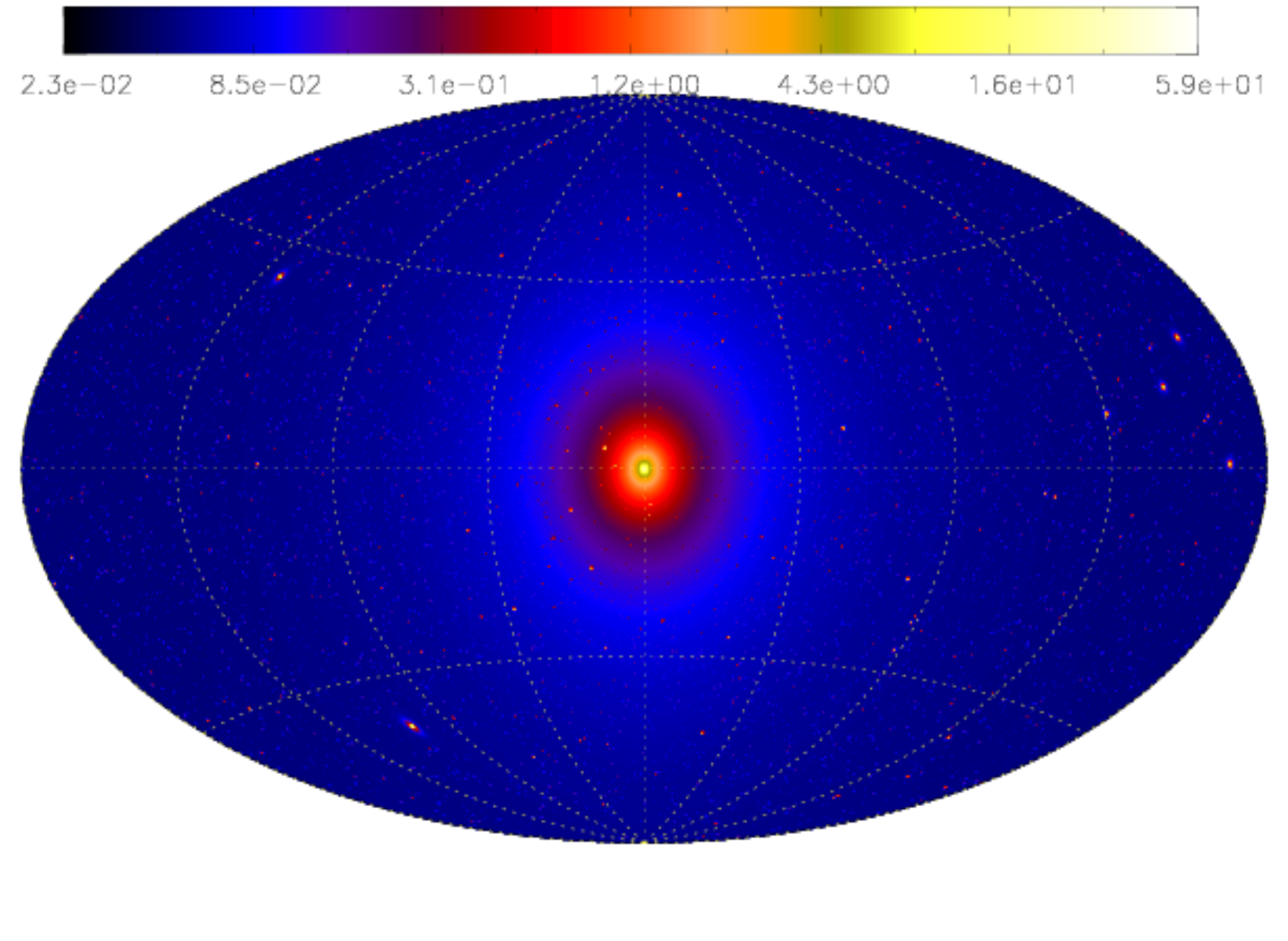}
 \includegraphics[width=\columnwidth]{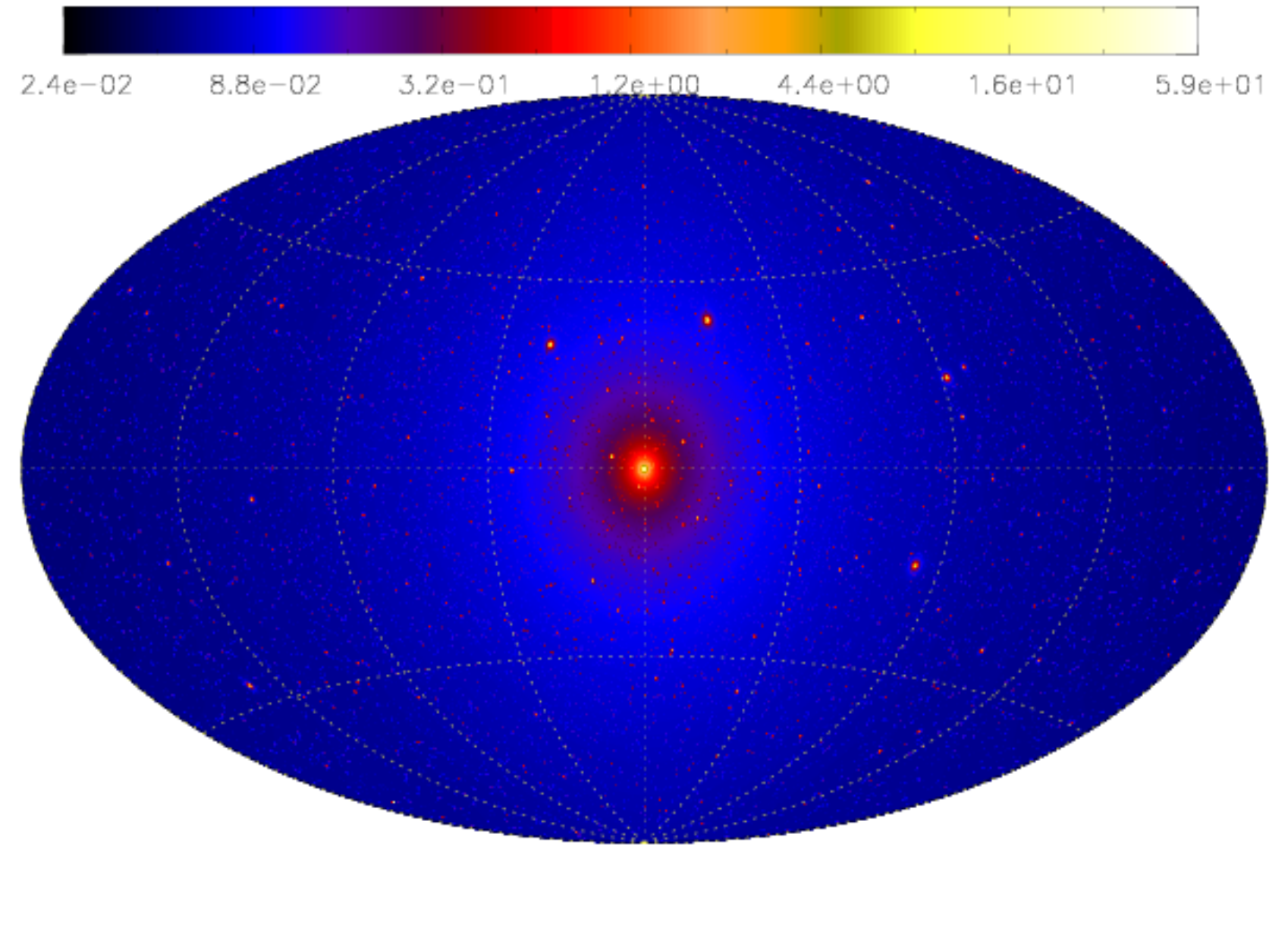}
 \caption{Same as \citefig{fig:sub}, but with the two simulation setups 
   rescaled to the same local density, same total mass and same fraction of 
   mass in substructures.}
 \label{fig:subRES}
\end{figure*}

\citefig{fig:compareaq} shows the predicted 
contribution of the smooth MW halo, MW subhalos and  extragalactic halos to the 
 $\gamma$-ray flux from DM annihilation integrated above 3 GeV, as a function 
of the angle of view $\psi$ from the GC in the case of \aquarius~and 
\vlii, respectively. The contribution from individual subhalos is computed by 
averaging over the 10 Monte Carlo realizations.

\begin{figure*}[t]
 \includegraphics[width=0.45\textwidth]{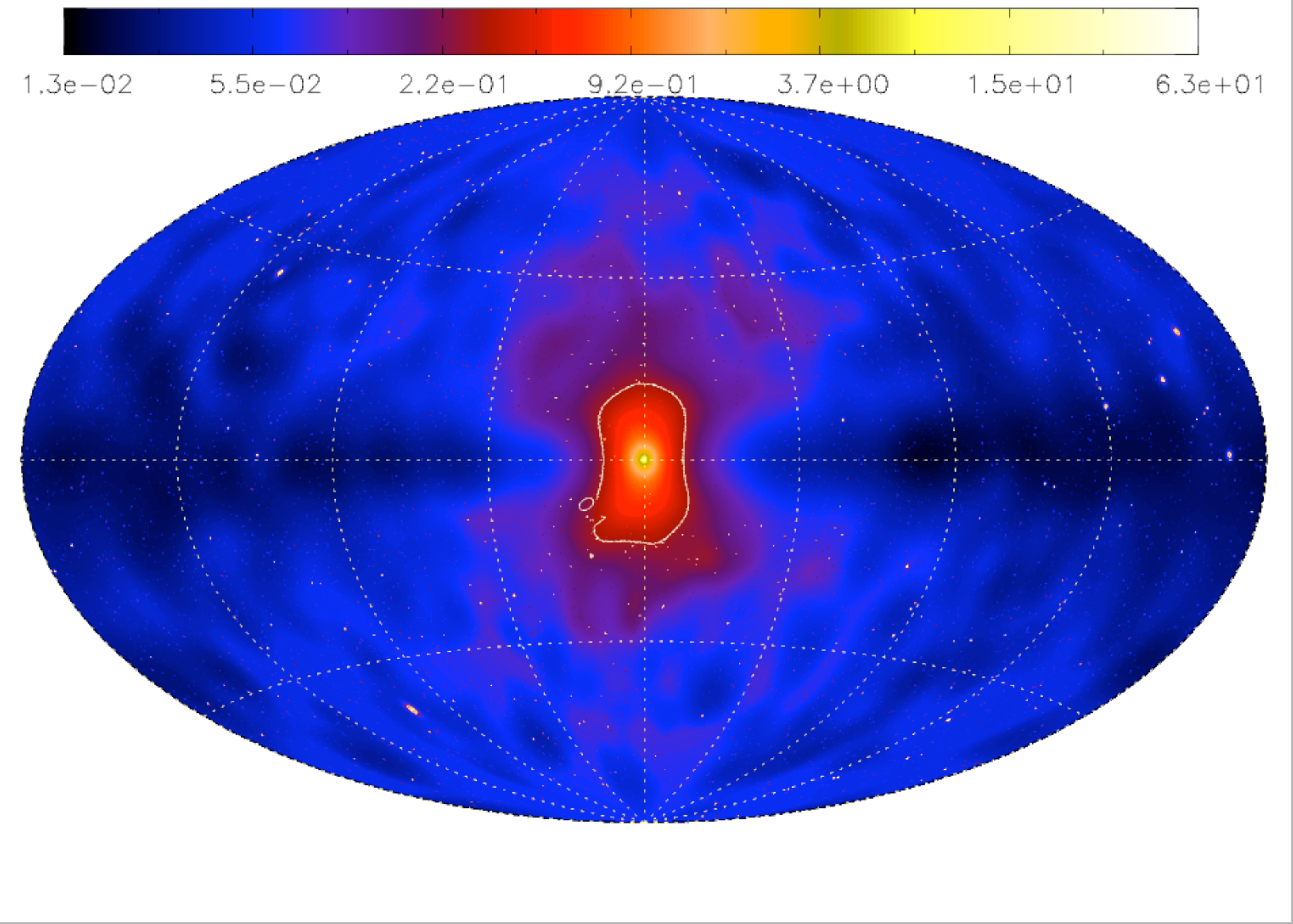}
 \includegraphics[width=0.45\textwidth]{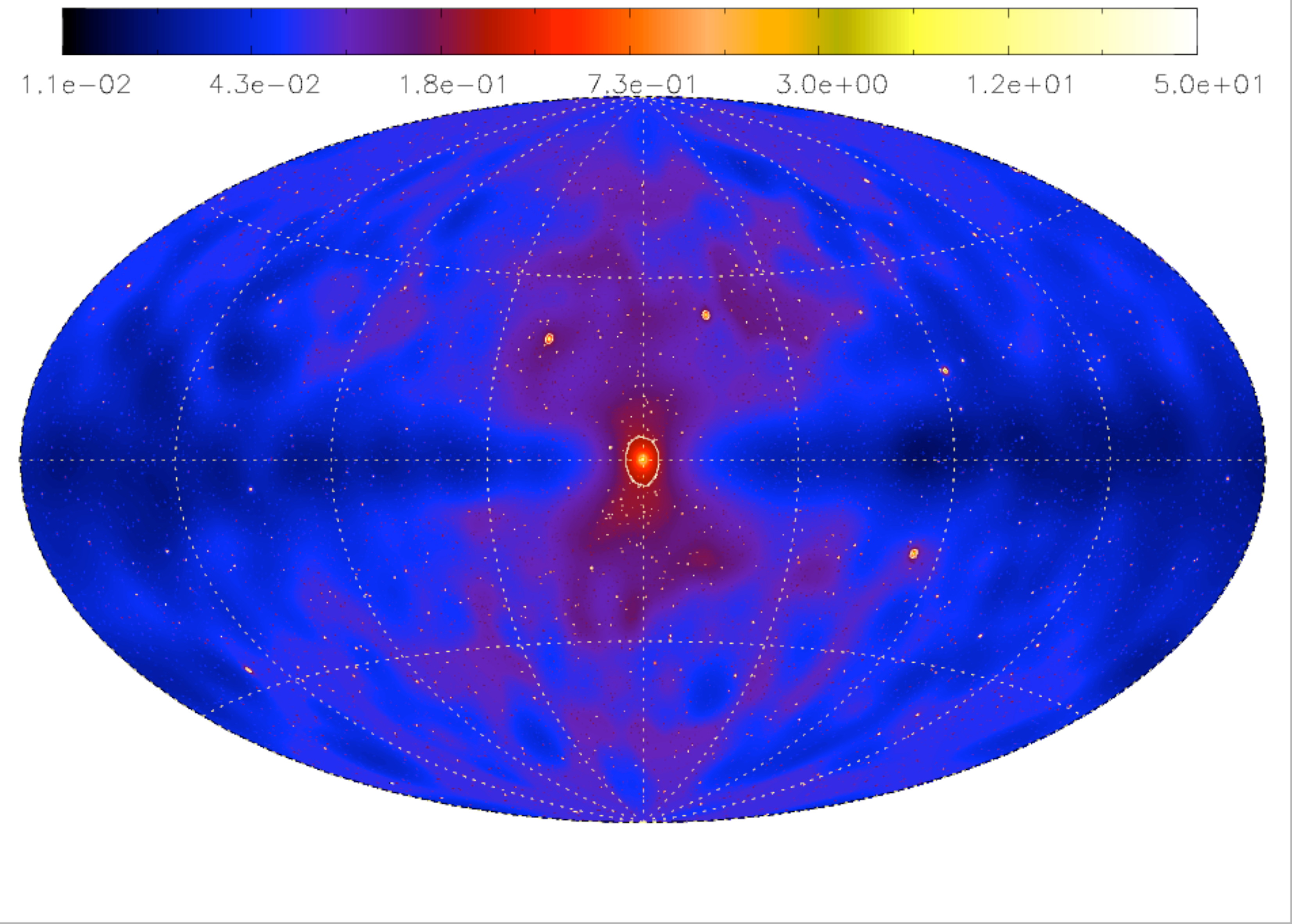}
 \caption{Sensitivity map, in galactic coordinates, for the \aquarius~and 
   \vlii~setups. The signal is as in \citefig{fig:sub}, while the 
   background is obtained through a suitable rescaling of the EGRET maps 
   (see text for further details).}
\label{fig:sens}
\end{figure*}

In the central part the annihilation signal is dominated by the MW smooth 
component in both  \aquarius~(at $\psi < 50^\circ$) and  \vlii~(at 
$\psi < 20^\circ$). Away from the GC the dominant contribution is provided by 
DM subhalos in the \vlii~case and by extragalactic background in the \aquarius~ 
case. The different behaviors simply reflect the different amount of Galactic 
substructures in the two simulations.

We note that our predictions satisfy the observational constraint represented 
by the diffuse Galactic signal. Indeed the {\it mean} diffuse Galactic flux 
above 3 GeV that should be measured by Fermi ($\sim 5.3 \times 10^{-7}$ ph 
cm$^{-2}$ s$^{-1}$ sr$^{-1}$) is safely above the expected annihilation signal 
in both simulation setups.

The flux in the innermost regions is higher for the \aquarius~simulation, 
as clear from from the full sky maps shown in \citefig{fig:sub}, where we 
show the total annihilation flux (MW smooth + galactic subhalos + extragalactic
halos and subhalos). The fact that the annihilation signal at the GC is higher 
in the \aquarius~case is mostly due that in this simulation the DM density in 
the solar neighborhood is larger than in \vlii. This fact propagates in a 
mismatch between the two fluxes proportional  to the density squared, \ie\ 
$[\rho_{\rm sm}^{Aq}(\odot)/\rho_{\rm sm}^{VL2}(\odot)]^2=[0.57/0.42]^2=1.84$.
An additional source of discrepancy is the fact that the total mass of the MW
in the \vlii~simulation is smaller than in \aquarius, as reported in 
\citetab{tab:dm_setup}. However, as shown in \citefig{fig:subRES}, the two 
predictions can be brought in agreement by requiring that (i) both \vlii~and 
\aquarius~have the same local density $\rhosun$ 
(we have taken the recent estimate $\rhosun = 0.385\;{\rm GeV/cm^3}$ 
from \cite{2010JCAP...08..004C,2009arXiv0906.5361S}), (ii) the same 
subhalo mass fraction ($\fsub=0.18$) is adopted and (iii) the same mass profile
is assumed.

\subsection{Experimental detectability}

In order to assess the detectability of the  $\gamma$-ray annihilation flux 
with the Fermi-LAT satellite, we have to specify what the signal, background or 
noise are.

If we are interested in finding a signal above the astrophysical backgrounds, 
the signal is contributed by the sum of all the aforementioned components of 
the annihilation flux (MW smooth mass distribution + galactic subhalos + 
extragalactic halos and subhalos). We focus on photons with energies larger
than 3 GeV and we assume an exposure time of 1 year, 
which corresponds to about 5 years of data taking with Fermi, and we 
assume an effective detection area of $10^4 \cm^2$. We don't consider here any 
dependence on the photon energy nor on the incidence angle. The background or 
noise is contributed by 
the diffuse Galactic foreground and the unresolved extragalactic background. 
As mentioned in \citesec{sec:intro}, to model such contributions we have 
rescaled the EGRET data at $E>3$ GeV by 50\%. We remind that this reduction 
reflects the fact that the Fermi data do not confirm the so-called galactic 
excess measured by EGRET. The expected sensitivity is simply given by  
$\sigma_{\rm DM}$ = $N^\gamma_{\rm DM} \over \sqrt{N^\gamma_{\rm background} + N^\gamma_{\rm DM} }$,
where $N^\gamma_{\rm DM}$ is the total number of annihilation photons and 
$N^\gamma_{\rm background}$ are the photons contributed by the astrophysical 
background. The left and right panels of \citefig{fig:sens} 
show the resulting sensitivity maps in Galactic coordinates for the 
\vlii~and the \aquarius~simulations, respectively. The sensitivity maps have a 
sharp peak near the GC, as expected. Iso-sensitivity contours are not symmetric 
but have a characteristic 8-shape around the GC due to the disk-like 
Astrophysical Galactic Foreground that contribute to the noise term. The 
sensitivity maps after the rescaling procedure adopted to produce the two maps 
in \citefig{fig:subRES} would look very similar to the left panel of 
\citefig{fig:sens}.

This procedure can be useful to estimate the regions that optimize the 
signal-to-noise in DM searches. In fact, while the Fermi data become available, 
one can just take the DM template presented here, and divide the actual Fermi 
data by such template. One can then calculate the signal-to-noise ratio, S/N, 
for the iso-flux contours, and determine the size and the shape of the region 
that maximizes the S/N.

Disentangling the annihilation signal from the astrophysical one near the GC 
might however be difficult due to the presence of a strong astrophysical
background. An alternative strategy is to look for individual subhalos, \ie 
isolated bright spots in the $\gamma$-ray sky. In this case the signal is 
given by the annihilation photons produced in the nearest and $\gamma$-ray 
brightest subhalos  in our  Monte Carlo realizations. The noise is contributed 
by all  remaining sources of $\gamma$-ray photons, including those from DM 
annihilation (MW smooth + average subhalo component + extragalactic). The 
sensitivity is in this case given by $\sigma_h$ = $N^\gamma_{\rm signal} \over 
\sqrt{N^\gamma_{\rm back+fore}}$ where $N^\gamma_{\rm signal}=
N^\gamma_{\rm subhalos,MC}$ are the annihilation photons produced within the 
resolved subhalos and $N^\gamma_{\rm back+fore}= N^\gamma_{\rm MW,smooth} + 
N^\gamma_{\rm subhalos,average} + N^\gamma_{\rm extragalactic} + 
N^\gamma_{\rm background}$ includes contributions from the smooth MW halo, 
unresolved subhalos, diffuse extragalactic background and Galactic foreground, 
respectively.\\

In \citetab{tab4} we list the number of  3 $\sigma$ and 5 $\sigma$ detections
expected with an exposure of 1 yr with  the Fermi-LAT. Both 
the \aquarius~and the \vlii~cases are considered. The number of detectable 
subhalos is small but significantly different from zero in benchmark models A 
and B. No individual subhalos are expected to be detected if the case of models 
C and D.\\

\citefig{figobservables} shows the number of the subhalos detectable at 
3 $\sigma$ for the benchmark model A, as a function of the subhalos 
mass. The symbols and error bars represents the mean and the scatter over the 
10 Monte Carlo realizations. All detectable subhalos have masses above 
$10^5 \msun$, in some cases comparable to the estimated masses of the local 
dwarf galaxies, suggesting that these DM-dominated objects may indeed be good 
targets for DM indirect detections. Furthermore, the results show that the 
only detectable halos are those already resolved in the \aquarius~and 
\vlii~simulations, \ie\ the results presented here are independent on the 
aggressive extrapolations required to model the properties of low-mass subhalos.

\begin{figure}
\centering
\includegraphics[width=\columnwidth]{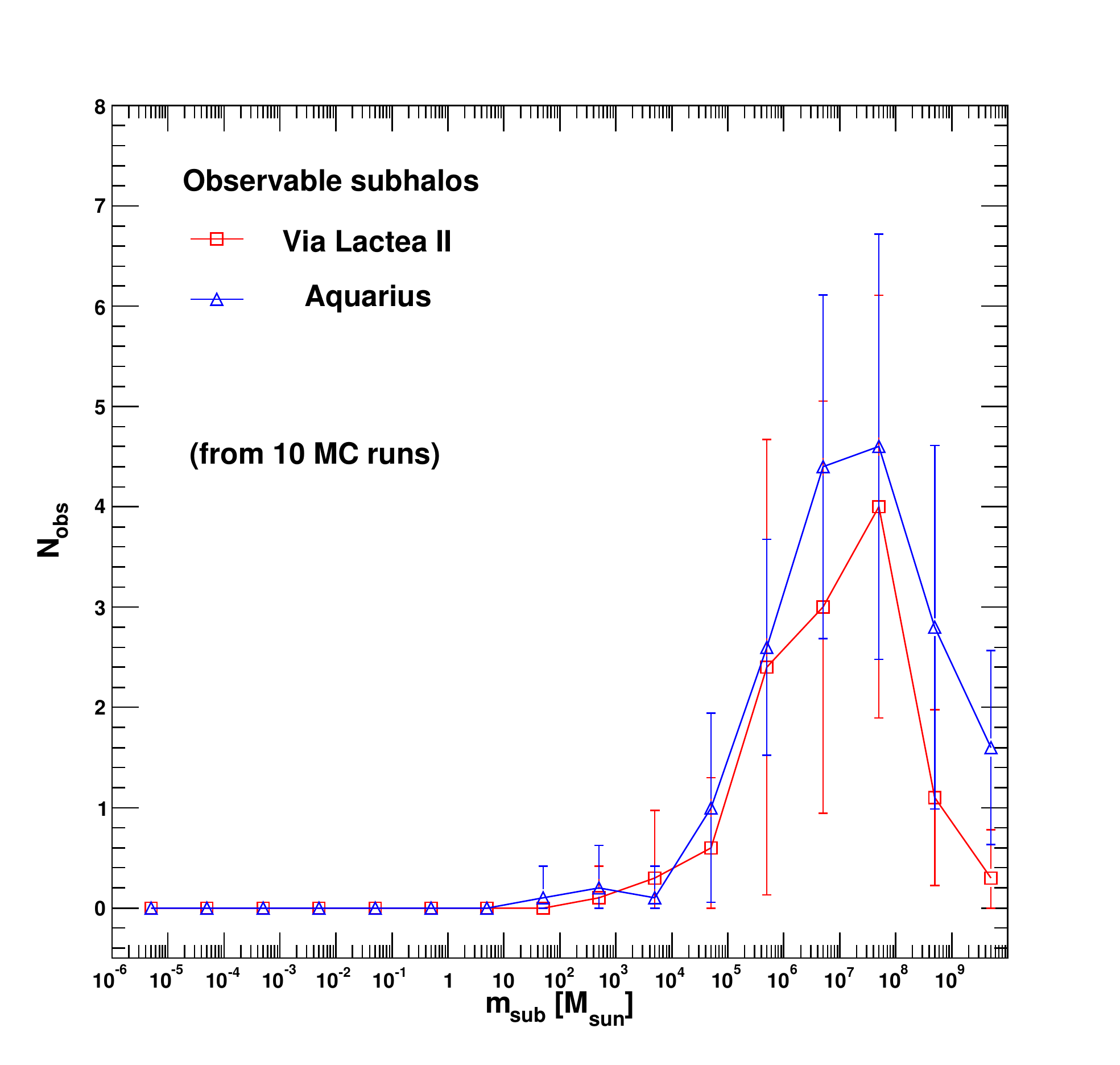}
\caption{Number of halos that can be detected at 3 $\sigma$ with an exposure 
of 1 year on-axis (corresponding to approximately 5 years of data taking) with 
the Fermi-LAT, in the case of the \vlii~and \aquarius~
simulations and the particle physics benchmark model A, as a function of the 
halo mass. Errors represent the scatter over 10 Monte Carlo realizations.}
\label{figobservables}
\end{figure}

\begin{table}
\centering
\begin{tabular}{ccccc}
\hline
\hline
model & {\it VLII} 3 $\sigma$ & {\it VLII} 5 $\sigma$  & 
{\it Aq} 3 $\sigma$ & {\it Aq} 5 $\sigma$ \\
   \hline
A & $9.2 \pm 2.6$ & $4.1 \pm 1.3$ & $13.5 \pm 2.5$ & $7.3 \pm 2.4$ \\
B & $3.1 \pm 1.1$ & $1.4 \pm 0.8$ & $6.2 \pm 2.1$ & $2.9 \pm 1.4$ \\
\hline
\end{tabular}
\caption{Number of subhalos detectable at 3 and 5 $\sigma$ with the Fermi-LAT 
satellite in 5 years of data taking, computed for the benchmark models A and 
B. Errors represent the scatter over 10 Monte Carlo realizations.}
\label{tab4}
\end{table}

\section{Antimatter}
\label{sec:antimatter}

Antimatter cosmic rays [CRs] have long been considered as potential tracers 
of DM annihilation because they are barely produced in standard astrophysical 
processes~\cite{1984PhRvL..53..624S}. Indeed, most of the standard contributions
are expected to be of secondary origin, \ie\ produced by nuclear interactions of
standard CRs (protons and light nuclei) with the interstellar gas (hydrogen and 
helium). This picture is essentially true for antiprotons, because their 
propagation scale is rather large compared to the spatial fluctuations of the 
interstellar medium and CR sources, but should be taken with caution for 
positrons, since their high energy component is strongly sensitive to time and 
spatial fluctuations of the local environment. Moreover, some astrophysical 
sources, like pulsars, are known to have the capability to produce positrons 
from pair creations in strong magnetic fields (see an early discussion on this 
in~\cite{1989ApJ...342..807B}). Because there are such sources in abundance 
within the kpc scale around the Earth, one can expect them to be significant 
contributors to the high energy positron budget~
\cite{2008arXiv0812.4457P,2010arXiv1002.1910D}. 
Nevertheless, many theoretical as well as observational uncertainties still 
affect the modeling of most of astrophysical sources~\cite{2010arXiv1002.1910D}.
It is consequently 
important to scrutinize the potential imprints that DM annihilation could 
provide in the antimatter budget, in addition to those in $\gamma$-rays. In 
this section, we will shortly review the transport of CRs in the Galaxy, and 
then focus on the positron and antiproton fluxes at Earth that DM annihilation 
may generate. To model the sources of the annihilation products we will again 
use the highest resolution N-body simulations discussed in the previous 
sections, including the effect of subhalos. For more details on the 
semi-analytical method used to follow the antimatter transport within the 
Galaxy that will be used in the following, we refer the 
reader to~\cite{2007A&A...462..827L,2008A&A...479..427L}.

\subsection{High energy positron and antiproton transport}
\label{subsec:cr_prop}

Antimatter CRs in the GeV-TeV energy range, like other charged cosmic 
rays, diffuse on the inhomogeneities of the Galactic magnetic field. Because 
those inhomogeneities are not fully confined in the Galactic disk, cosmic 
rays can pervade beyond this tiny region and diffuse away up to a few kpc 
upwards and downwards (see \eg~\cite{1994hea..book.....L} for a pedagogical 
insight). This turbulent volume actually defines the diffusion 
zone inside which CRs are confined, and beyond which they escape 
forever; it can be featured like a disk-like slab with radial and vertical 
extensions, $R_{\rm slab}$ and $L$, respectively. In the following, we will 
fix $R_{\rm slab}=20$ kpc, and $L = 4$ kpc~\cite{2001ApJ...555..585M}. CRs 
experience different processes during their journey, depending on their nature: 
in addition to spatial diffusion, (anti)nuclei will be mostly affected 
by convection and spallation in the interstellar medium localized in the disk, 
processes which are more efficient at low energy, while electrons and positrons 
will have their transport dominated by energy losses above a few GeV. 
Formally, given a source ${\cal Q}$, all CRs obey the same continuity 
equation~\cite{berezinsky_book_90},
\ben
\widehat{\cal D} \widehat{\cal J} = 
\partial_\mu {\cal J}^{\mu} + \partial_E {\cal J}^{E} +\Gamma {\cal N} = 
{\cal Q}(\vec{x},E,t)\;,
\een
where ${\cal N}=dn/dE$ is the CR density per unit energy, 
${\cal J}^{\mu}$ is the space-time current, ${\cal J}^{E}$ the energy current 
and $\Gamma$ stands for a destruction rate (spallation for nuclei).
The time current is merely the CR density ${\cal J}^t\equiv {\cal N}$, 
while the spatial current is reminiscent of the Fick law, and accounts for 
the spatial diffusion and convection ${\cal J}^{\vec{x}}\equiv 
\left\{K_x({\cal R})\vec{\nabla}-\vec{V}_c\right\} {\cal N}$. The spatial 
diffusion coefficient, describing the stochastic bouncing interactions with 
the magnetic inhomogeneities, is usually parameterized as $K_x({\cal R})=
K_0({\cal R}/1\,{\rm GV})^\delta $ where the dependence on the CR 
rigidity ${\cal R}\equiv p/Z$ is explicit. The energy current carries the 
energy loss and reacceleration terms ${\cal J}^E\equiv \left\{dE/dt - 
K_p(E)\partial_E \right\}{\cal N}$.

Although all CR species obey the same transport equation, some of the 
processes mentioned above will be negligible in the GeV-TeV energy range, 
depending on the species. Beside spatial diffusion, antiprotons will mostly 
suffer spallations and convection, but almost not energy losses. 
On the contrary, energy losses will dominate the positron transport, mainly 
due to the inverse Compton scattering with the CMB photons and the interstellar 
radiation fields, and to synchrotron losses with the Galactic magnetic field. 
In both cases, reacceleration is negligible above a few 
GeV~\cite{2001ApJ...555..585M,2009A&A...501..821D}. Neglecting the irrelevant 
terms and assuming steady state ($\partial_t {\cal N}=0$), the Green functions 
--- or propagators --- associated with the transport equation can be derived 
analytically for antiprotons and positrons. We refer the reader 
to~\cite{2008A&A...479..427L} for the detailed expressions of the propagators. 
In the following, ${\cal G}$ will denote the Green function, such that 
$\widehat{\cal D}{\cal G}(\vec{x},E \leftarrow \vec{x}_S,E_S) = \delta^3(\vec{x}-\vec{x}_S)\delta(E-E_S)$; $S$ indexes the source quantity. 
More details on CR propagation can be found in 
\eg~\cite{1964ocr..book.....G,berezinsky_book_90,1994hea..book.....L,2007ARNPS..57..285S,2002astro.ph.12111M}.

Most of the propagation parameters can be constrained from measurements of the 
ratios of secondary to parent primary species, and we will use the 
\emph{median} set derived in~\cite{2004PhRvD..69f3501D}: 
$K_0 = 1.12\times 10^{-2}{\rm kpc^2 Myr^{-1}}$, $\delta = 0.7$ and 
$V_c = 12\,{\rm km/s}$. The energy loss rate ascribed to positrons will be 
$dE/dt = -b(E) = - b_0E^2$, with $b_0 =(\tau_{\rm loss}\times 1\,{\rm GeV})^{-1}=
10^{-16}{\rm s^{-1}GeV^{-1}}$, which is a reasonable approximation accounting 
for the inverse Compton loss on CMB, starlight and dust radiation and for the 
synchrotron loss~\cite{1994hea..book.....L,1998PhRvD..59b3511B,2009A&A...501..821D}. However, we remind that there are theoretical uncertainties and 
degeneracies among those propagation parameters~\cite{2001ApJ...555..585M}, so 
that predictions for primaries and secondaries may vary by large factors (see 
\cite{2004PhRvD..69f3501D} for primary antiprotons 
and~\cite{2008A&A...479..427L,2008PhRvD..77f3527D} for primary positrons).

An important consequence of the differences in the propagation histories among 
species is that the corresponding characteristic propagation scales also 
differ. For antiprotons, the propagation scale is set by the spatial current, 
$\lambda_{\bar{p}} = K({\cal R})/V_c\sim 1\,(T/{\rm GeV})^{\delta}{\rm kpc}$, 
and increases with energy. At variance with antiprotons, positrons have their 
propagation scale set by transport in both space and momentum, 
$\lambda_{e^+} \propto \left\{\int dE \,K(E)/b(E)\right\}^{1/2}$ which roughly
scales like $\sim 3\,(E/{\rm GeV})^{(\delta-1)/2} {\rm kpc} $ for a loss of 
half the injected energy. Therefore, the positron range decreases with energy 
(see \eg\ Fig.~2 of \cite{2008PhRvD..78j3526L}). Consequently, low energy 
antiprotons and high energy positrons observed at Earth must originate from the 
very local environment. For instance, a positron injected in the Galaxy 
at 200 GeV and detected at 100 GeV has a probability to originate from regions 
farther than $\sim$1.5 kpc from the observer which is gaussianly suppressed.

Beside the primary signals due to DM annihilation that we will discuss below, 
one should also be aware of the backgrounds and their theoretical uncertainties.
Since we disregard here the conventional astrophysical sources of CRs, our 
positron and antiproton backgrounds are those secondaries resulting from 
spallation processes of cosmic protons and nuclei with the interstellar matter 
located in the disk. We refer the reader to~\cite{2001ApJ...563..172D} and
\cite{2009A&A...501..821D} for thorough discussions on those secondary 
components and related theoretical uncertainties. Throughout the paper, we will 
use the median secondary backgrounds derived in those references.

\subsection{Smooth and clumpy DM contributions: boost factors}
\label{subsec:cr_boosts}

The fact that the DM spatial distribution is not smooth but actually 
fluctuates due to the presence of subhalos leads to local fluctuations in the 
annihilation rate~\cite{1993ApJ...411..439S}. Formally, any flux estimated 
from a smoothly averaged DM profile should therefore be enhanced by a factor 
$\langle \rho_{\rm dm}^2 \rangle_{V_{\rm cr}} / 
\langle \rho_{\rm dm} \rangle_{V_{\rm cr}}^2$ to account for those fluctuations, 
the average being performed in a volume $V_{\rm cr}$ that depends on the 
CR propagation scale. Such an enhancement must be, therefore, quite 
different from what has been previously discussed for $\gamma$-rays, simply 
because the averaging volume for the latter is the resolution 
cone carried by the line of sight instead of a propagation volume.

The antimatter flux at the Earth originating from the annihilation of a single, 
smoothly distributed DM component is given by the following expression:
\ben
\phi_{\rm sm}(E) = \frac{\beta\, c}{4\pi}\,{\cal S} 
\int_{\rm slab} d^3\vec{x}_S\, \gtilde (\xsun \leftarrow \vec{x}_S,E) 
\left( \frac{\rho(r)}{\rhosun}\right)^2 \;,\nn\\
\een
where $\gtilde$ is the convolution of the Green function ${\cal G}(E \leftarrow 
E_s)$ with the injected spectrum $dN/dE_S$, 
${\cal S} \equiv (\sigv/2) (\rhosun/\mchi)^2 $ and 
$\rhosun \equiv \rho_{\rm tot}(\xsun)$ is the  DM density near the Sun inferred 
from the smooth profile $\rho_{\rm tot}$ in \citeeqs{\ref{eq:vl2smooth} and 
\ref{eq:aqsmooth}}.

Besides, we need to determine the average contribution due to the population 
of subhalos. The overall mass density profiles of subhalos given by  
\citeeqs{\ref{eq:vl2} and \ref{eq:aqclump}} can be used to obtain 
the corresponding normalized probability function as follows:
\ben
\frac{d\rho_{\rm sh}(\msub,R)}{d\msub} &=& \nsubtot \mymean{\msub} 
\frac{d{\prob}_V(R)}{dV}
\frac{d{\prob}_m(\msub,R)}{d\msub}\nn\\
 &=& \msubtot \frac{d{\prob}_V(R)}{dV} \frac{d{\prob}_m(\msub,R)}{d\msub}\;,
\een
where $\nsubtot$ is the total number of subhalos and $\mymean{\msub}$ their
mean mass. This expression is such that the product of $d{\prob}$s corresponds 
to a normalized probability function
\ben
\int d\prob_{\rm tot} &=& \int_{\rm MW} d^{3}\vec{x}\, \frac{d{\prob}_V(R)}{dV} 
\int_{m_{\rm min}}^{m_{\rm max}}dm\, \frac{d{\prob}_m(m,R)}{dm}\nn\\ 
&=& 1 \;.
\label{eq:normprob}
\een
Defining such a probability function allows us to treat each quantity related 
to a single subhalo as a stochastic variable~\cite{2007A&A...462..827L}. Notice 
that though the upper bound in the integral of the mass distribution is fixed 
to $m_{\rm max}$, the mass distribution itself is in fact a function of $R$. 
Such a dependence arises because of the tidal disruption of subhalos that is 
implemented in the present analysis according to the Roche criterion 
(see \citeeqp{eq:roche}) that we have discussed in \citesec{sec:dm_mod}.

Before inferring the overall subhalo flux, we need to define the luminosity 
of a single object. Given a subhalo of inner profile $\rhosub^{DM}$, the 
corresponding annihilation rate will be proportional to the \emph{annihilation 
volume}
\ben
\ofsub{\xi}(\msub,R) \equiv 4\pi \int_{\ofsub{V}} dr \, r^2 
\left[ \frac{\rhosub^{DM}(r,\msub,R)}{\rhosun}\right]^2 ,\nn\\
\label{eq:def_xi}
\een
which is the volume that would be necessary to obtain the whole subhalo 
luminosity from the local DM density $\rhosun=\rho_{\rm tot}(\odot)$ associated 
with the DM setup of interest. As detailed in previous sections, the 
concentration parameter $c(\msub,R)$, which characterizes the shape of the 
inner density profile, depends on the subhalo mass $\msub$ and position $R$ 
in the Galaxy.

\begin{figure}[t]
\begin{center}
\includegraphics[width=\columnwidth, clip]{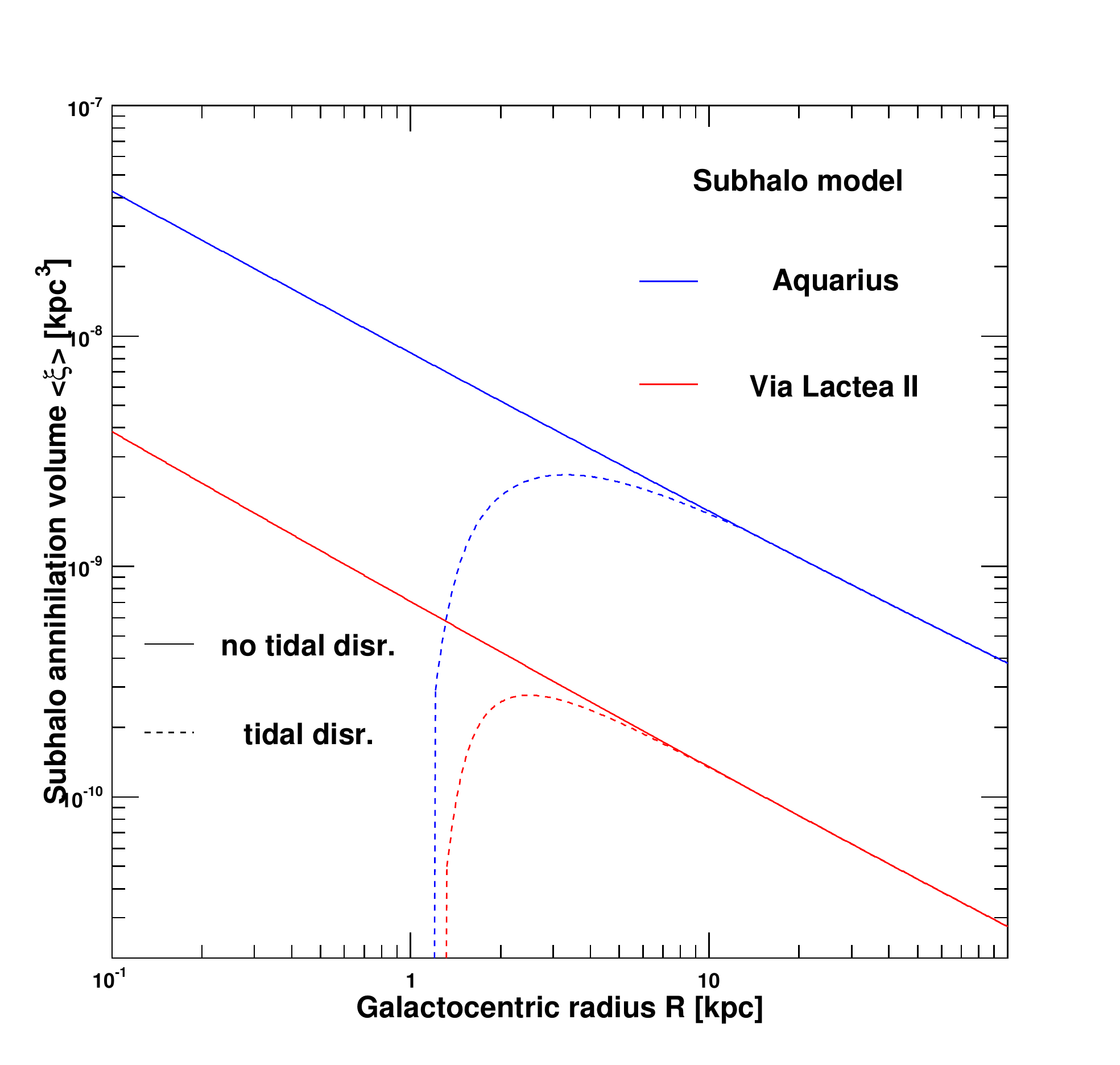}
\caption{Mean value of the subhalo annihilation volume \mymean{\xi}~defined in 
\citeeq{eq:def_xi} as a function of the galactocentric radius $R$. The 
solid/dashed curves show the effect of unplugging/plugging the subhalo tidal 
disruption for both the \vlii~(red) and \aquarius~(blue) configurations. For 
the sake of comparison, we have used $\rhosun = 0.3\,{\rm GeV/cm^3}$ in 
\citeeq{eq:def_xi} for both configurations.}
\label{fig:ximean_vs_r}
\end{center}
\end{figure}

\begin{widetext}
Therefore, the CR flux for a subhalo population reads
\ben
\ofsubtot{\phi}(E) &=& \frac{\beta \,c}{4\pi}\, {\cal S}\, \nsubtot
\int_{\rm slab} d^{3}\vec{x}_s \gtilde (\vec{x}_S,E) 
\frac{d{\prob}_V(R)}{dV} \int_{m_{\rm min}}^{m_{\rm max}}dm \, \ofsub{\xi}(m,R)
\frac{d{\prob}_m(m,R)}{dm}\nn\\
&=& \frac{\beta \,c}{4\pi}\, {\cal S}\, \nsubtot \,
\int_{\rm slab} d^{3}\vec{x}_s \, \gtilde (\vec{x}_S,E)\, 
\frac{d{\prob}_V(R)}{dV} \,\mymeanlr{\ofsub{\xi}}_{\msub} (R)\nn\\
&=& \frac{\beta \,c}{4\pi}\, {\cal S}\, \nsubtot \, 
\mymean{ \gtilde(\vec{x}_S,E) \, \mymeanlr{\ofsub{\xi}}_{\msub} (R)}_{\rm slab}
= \nsubtot \, \mymean{\ofsub{\phi}} \;.
\label{eq:phitot}
\een
$\mymean{x}_{m/V}$ means an average of the variable $x$ according to the 
mass/spatial distribution, respectively. By writing this equation, we made 
the implicit assumption that a subhalo can be treated as a 
stochastic point-like source. This assumption is valid while the typical 
propagation scale is larger than the scale radius of any subhalo, which is 
fully the case here: in the \aquarius~setup, a $10^{-6}/10^6\Msun$ subhalo 
has a scale radius of $\sim 10^{-6}/3\times 10^{-2}$ kpc, respectively. We 
obviously recover the same result as in~\cite{2008A&A...479..427L} that the 
total subhalo flux is given by the total number of subhalos times the mean 
flux from a single subhalo, consistently with the stochastic treatment. 
Nonetheless, the spatial average can no longer be separated from the mass 
average in the present study. Not only does this come from the full 
implementation of the tidal effects, but also from the spatial dependence of the
concentration parameter (the two effects are related). Consequently, the average
luminosity of a subhalo, $\mymeanlr{\ofsub{\xi}}_{\msub} $, does explicitly 
depend on its location in the Galaxy, even when the spatial disruption \`a la 
Roche is neglected. We illustrate the radial dependence of 
$\mymeanlr{\ofsub{\xi}}_{\msub} (R)$ in \citefig{fig:ximean_vs_r}, where we 
explicitly show the effect of the tidal disruption: the average luminosity of a 
subhalo is quickly turned off in the central part of the Galaxy when Roche 
criterion is applied.

Because the flux derived above is in fact a {\em mean} quantity, it can 
be associated with a statistical variance. Taking the single subhalo flux
as a stochastic quantities, \ie\ assuming subhalos are not spatially 
correlated, the relative fluctuation of the antimatter CR flux is given from
Poissonian statistics by~\cite{2007A&A...462..827L}:
\ben
\frac{\ofsubtot{\sigma}}{\ofsubtot{\phi}}(E) = \frac{1}{\sqrt{\nsubtot}} 
\frac{\ofsub{\sigma}}{\mymeanlr{\ofsub{\phi}}}(E) = 
\frac{1}{\sqrt{\nsubtot}} \left\{ 
\frac{\mymean{ \gtilde^2(\vec{x}_S,E) \, 
\mymeanlr{\ofsub{\xi^2}}_{\msub}  (R)}_{\rm slab}}
{\mymean{ \gtilde(\vec{x}_S,E) \, \mymeanlr{\ofsub{\xi}}_{\msub}  
(R)}^2_{\rm slab}}
-1 \right\}^{\frac{1}{2}}\;.
\label{eq:sigphitot}
\een
\end{widetext}
This quantity, associated with the whole population of subhalos, 
is linked to the fluctuation of the flux of a single object in a standard 
manner, with the factor $1/\sqrt{N}$. A large value would express the fact 
that a small number of subhalos may have a large impact on the predictions, 
while a small value ensures that the predictions are typified by contributions 
of a large number of objects. The relative fluctuation obviously increases with 
decreasing propagation horizon, \ie\ with increasing (decreasing) energy for 
antiprotons (positrons, respectively)~\cite{2008A&A...479..427L}. It is
important to stress that for positrons, the quantity 
$\mymean{ \gtilde^2}/\mymean{ \gtilde}^2$ does actually diverge like 
$1/\lambda^3$ in the limit of vanishingly small propagation scale (\ie\ 
$E\rightarrow E_S$). Although this affects an infinitely small part of the 
propagated spectrum in the case of an injected positron line, this divergence 
spreads over the whole propagated spectrum in the case of the injection of 
positrons at all energies below $m_\chi$. This divergence would have a physical
meaning if there was a non-zero probability that a subhalo were located exactly 
at the positron detector, which is obviously not the case. To avoid this 
calculation artefact, we conservatively assume that there is no DM source 
fluctuation within 50 pc around the earth, which prevents any computation
crash.

The overall antimatter CR flux is the sum of the subhalo component plus the 
smooth component, the latter being somewhat corrected for not carrying the 
whole DM mass anymore: $\phi_{\rm tot} = \phi_{\rm sm} + 
\ofsubtot{\phi}$, which is not expected to equal to $\bar{\phi}_{\rm tot}$, 
\ie\ the flux computed from the overall smooth halo featured by 
\citeeq{eq:vl2smooth} or (\ref{eq:aqsmooth}). The so-called boost factor 
is then merely defined from their ratio
\ben
\boost(E) &=& \frac{\phi_{\rm tot}}{\bar{\phi}_{\rm tot}} 
\approx 1 + \frac{\ofsubtot{\phi}(E)}{\phi_{\rm sm}(E)}\;,
\label{eq:def_boost}
\een
where we have used $\phi_{\rm sm}\approx \bar{\phi}_{\rm tot}$ in the latest 
approximation. We emphasize that the boost factor does depend on 
energy. The smooth component will dominate at large propagation scale, 
when the dense regions close to the Galactic center come into play. On the 
contrary, the contribution of subhalos may be significant at small propagation 
scales because the smooth contribution is set by the squared local DM density. 
There is a close parallel to make with the boost factor as computed for 
$\gamma$-rays, for which the relevant physical variable is not the energy, but  
the angle between the line-of-sight and the Galactic center direction $\psi$:
the boost is negligible at small angles because of the large contribution of 
the central part of the Galaxy. Finally, note that $\boost(E)$ also depends on 
the injection spectrum for positrons because of energy losses. This is not the 
case for antiprotons, though their associated boost factor still depends on the 
energy because of spatial diffusion.

The boost factor is associated with a statistical fluctuation that is 
straightforwardly connected to that of the total subhalo flux:
\ben
\sigma_{\boost}(E) = \frac{\ofsubtot{\sigma}(E)}{\bar{\phi}_{\rm tot}(E)}\;.
\een
We see that even if fluctuations of the subhalo flux were found to be 
large compared to the subhalo flux itself, the boost factor would have a 
sizable variance only if those fluctuations are greater than the smooth 
flux. This mostly characterizes the large propagation scale regime, where the 
smooth contribution can strongly dominate the signal and completely overcome 
the statistical variance expected from the subhalo flux.

It turns out possible to derive an analytical expression for the boost 
factor, or equivalently for the total subhalo flux, in the vanishingly small 
propagation scale limit~\cite{2008A&A...479..427L}. This asymptotic expression 
is very convenient not only to check numerical computations, but also because it
usually corresponds to the maximal \emph{mean} value of the boost factor --- 
which can of course fluctuate around its mean value. This analytical limit 
relies on the fact that at very short propagation scale, the Green function 
$\gprop(\xsun\leftarrow\vec{x})\xrightarrow{\propto} \delta(\vec{x}-\xsun)$, 
so that $\gtilde\xrightarrow{\propto} \delta(\vec{x}-\xsun) \frac{dN}{dE}$. 
We are therefore left with local quantities:
\ben
\boost_\odot &=& 
\left[\frac{\rho_{\rm sm}(\Rsun)}{\rho_{\rm tot}(\Rsun)}\right]^2 + 
\nsubtot\,\mymeanlr{\ofsub{\xi} (\Rsun)}  \, 
\frac{d\prob_V(\Rsun)}{dV}\nn \\
&\simeq& 1 + \nsubtot\,\mymeanlr{\ofsub{\xi} (\Rsun)}  \, 
\frac{d\prob_V(\Rsun)}{dV}\;.
\label{eq:local_boost}
\een
We emphasize that this expression is valid for \emph{any} CR species 
and for \emph{any} set of propagation parameters, in the regime of vanishingly 
small propagation scale. Nevertheless, we also remind that such a regime is 
generally associated with large statistical fluctuations of the boost factor, 
because the average number of subhalos in such a small volume can be of the 
order of unity or  less: the actual boost can be much larger if we sit on the 
top of a subhalo, or much lower if no bright object wanders  in the 
neighborhood.

\subsection{Benchmark results and discussion}
\label{subsec:crs_discussion}
In this section, we discuss the results we have obtained for the 
overall positron and antiproton fluxes and corresponding boost factors, 
using the benchmark WIMPs defined in \citesec{sec:bench} 
and the DM distributions associated with the \vlii~and \aquarius~configurations.

\subsubsection{Boost factors}
\label{subsubsec:boosts}

In \citefig{fig:boost_crs}, we show the results obtained for the boost factors
and the corresponding 1-$\sigma$ statistical bands, for both the positron and 
the antiproton signals. 

The left panel represents the boost factor as a function of the positron energy 
given different DM distributions (\vlii- or \aquarius-like) and assuming the 
injection of a 100 GeV positron line at source. The \vlii~subhalo configuration 
is shown to have much more impact than the \aquarius~one, though the averaged 
boost factor is still modest, reaching a value of $\sim 3$ asymptotically. 
The subhalo contribution has little effect in the \aquarius~setup, 
except from the related statistical fluctuations at high energy (colored
areas). These fluctuations size the probability that a nearby (or few) massive 
subhalo dominates the signal. The dashed curves illustrate the effect of 
applying the Roche disruption in the central region of the Galaxy. Since it is 
mostly efficient in the inner 2 kpc of the Galaxy (see \citefig{fig:rhodm}), the
effect is almost negligible in terms of averaged
boost factor, as expected. Nevertheless, the variance is strongly depleted 
(from the colored to the hatched areas), due to a significant decrease of the 
fluctuations related to $\xi$ in \citeeq{eq:sigphitot}. Indeed, the available 
mass range is locally squeezed down when the Roche criterion is applied, as 
shown in \citefig{fig:roche}; 
likewise, tidal disruption is most effective in the inner kpc of our 
Galaxy where the bulk of subhalos is made by small objects and therefore
preferentially affects those substructure that are more concentrated
(see \citefig{fig:ximean_vs_r}).

The middle panel shows the boost factors for the positron fluxes associated 
with all our benchmark WIMP models (see \citesec{sec:bench}), the injection 
spectra of which differ significantly. These boost factors are computed in the 
frame of the \vlii~setup only and are plotted as functions of 
$x\equiv E/m_{\chi}$ for a more convenient reading. It is remarkable that the 
mean values converge towards the same limit at high energy, as expected from 
the short range limit given in \citeeq{eq:local_boost}. The energy dependence of
the averaged boost factor is a bit steeper in the case of a positron line due to
that (i) the spatial origin of positrons at the Earth is more strongly 
connected to their energy in that case, so that the low energy part mostly 
comes from the smooth GC component (instead, injection occurs at all energies 
$<\mchi$ for a continuous spectrum, which alleviates this segregation as much 
as the spectrum gets softer); (ii)  a given value of $x$ corresponds to 
different energies and therefore to different propagation scale given
the WIMP mass (especially relevant for the $\tau^+\tau^-$ model). The previous 
arguments also explain the more pronounced differences arising in the 
variance band shapes. Still, an additional point can help better understanding 
the differences in the variance shapes, which is inherent
to continuous spectra. As mentioned below \citeeq{eq:sigphitot}, the subhalo
flux fluctuations are intrinsically divergent in the limit $E\rightarrow E_S$: 
although made finite thanks to our spatial cut-off, they are still large and
spread over the whole energy range in the case of continuous spectra (instead, 
they are confined around the monochromatic injection energy for the 
positron-line model).

Finally, the right panel of \citefig{fig:boost_crs} exhibits the boost factors 
for antiprotons in the \vlii~and \aquarius~configurations, which are 
independent of the injection spectra. We remark that the asymptotic value
at low energy is the same as what is obtained for high energy positrons 
(see left panel). This illustrates the validity of the asymptotic limit
arising in the short propagation range regime and given in 
\citeeq{eq:local_boost}. Fluctuations are shown to be large at low energy,
where convection and spallations are important; nevertheless, further accounting
for solar modulation effects will be shown to deplete these large fluctuations
significantly and shift them to much lower energies, irrelevant for 
measurements at the Earth.

\begin{figure*}[t]
\begin{center}
\includegraphics[width=0.67\columnwidth, clip]{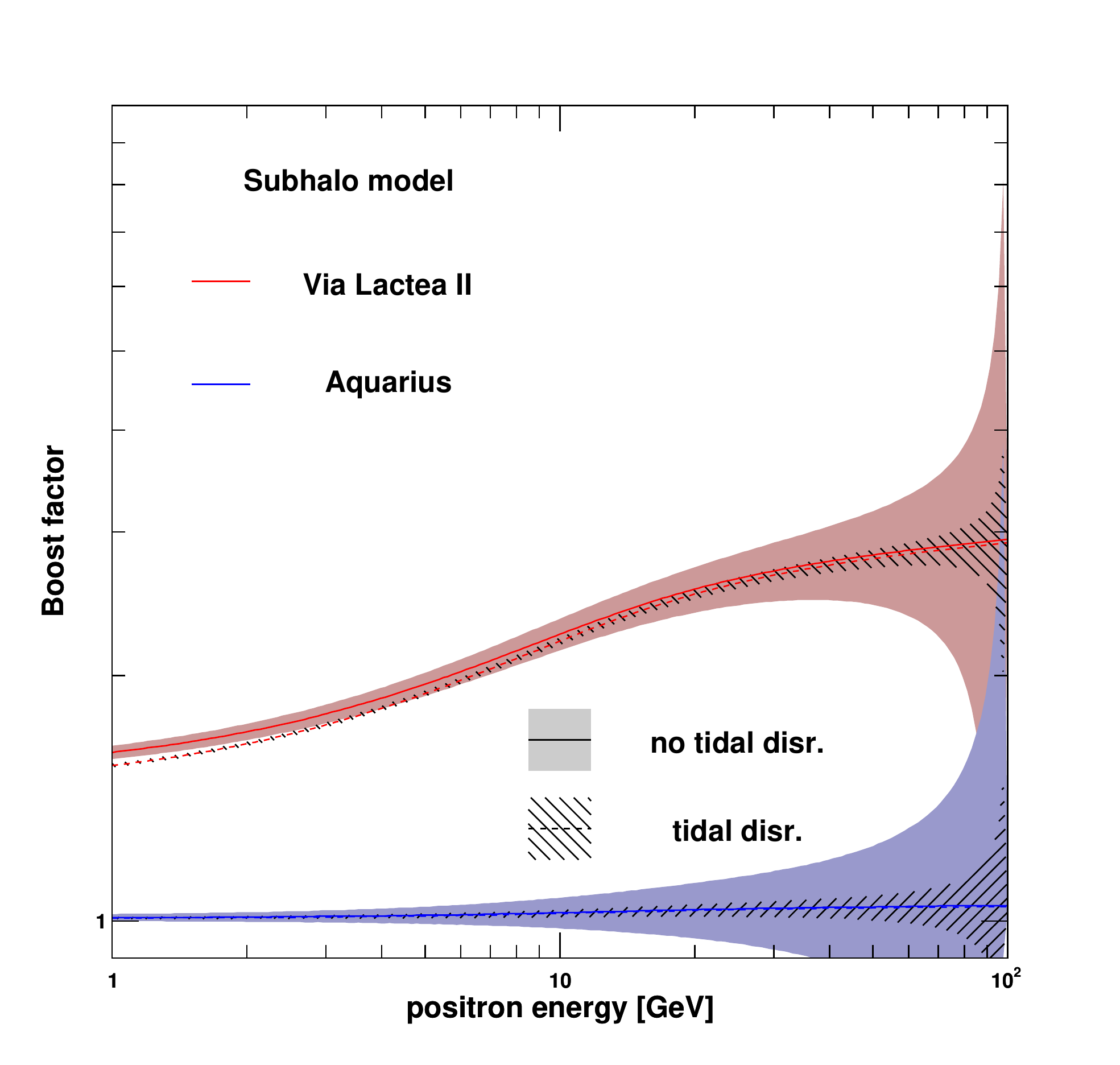}
\includegraphics[width=0.67\columnwidth, clip]{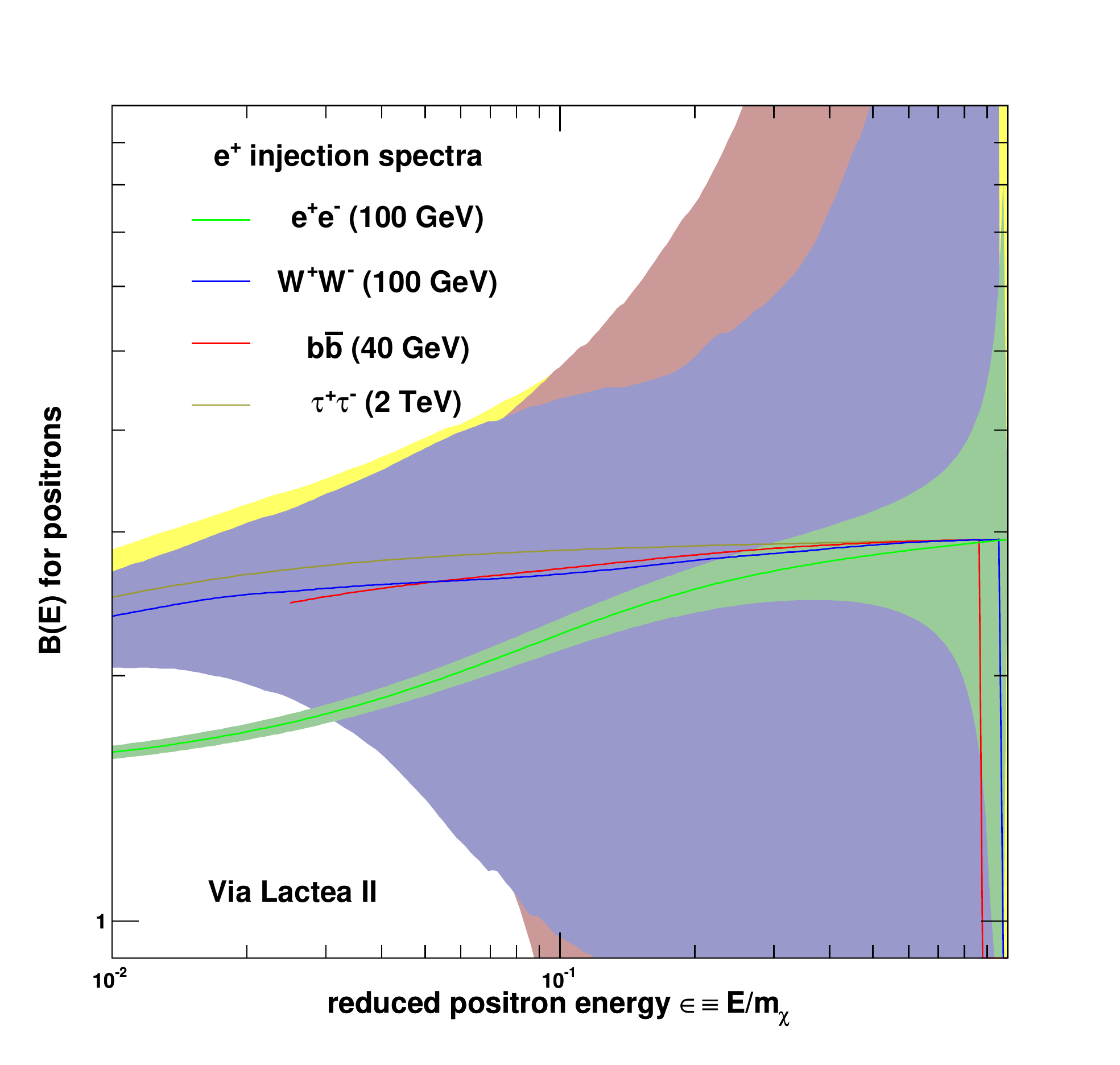}
\includegraphics[width=0.67\columnwidth, clip]{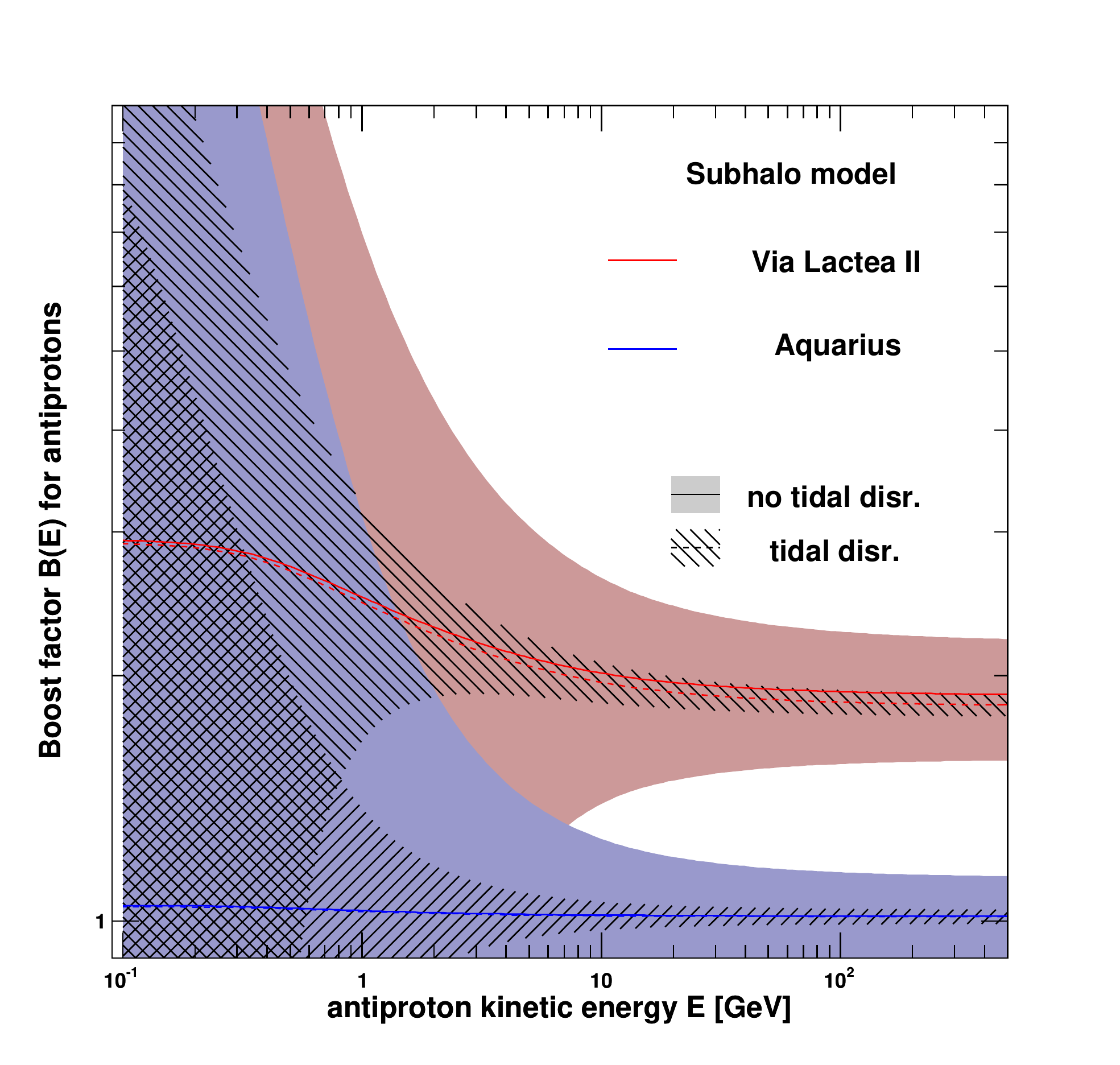}
\caption{\small Left: boost factor on positrons for different (sub)halo 
models, \vlii~(red) and \aquarius~(blue), assuming a 100 GeV positron 
line injected at source. Middle: boost for positrons for different injected 
spectra and for the \vlii~case. Right: boost for antiprotons, 
and for the same (sub)halo modes as in the left panel. All boosts are computed 
without solar modulation; colored bands account for a $1\sigma$ fluctuation, 
which shrink to the hatched areas once the Roche criterion is plugged.}
\label{fig:boost_crs}
\end{center}
\end{figure*}

\subsubsection{Predictions of the antiproton and positron fluxes}
\label{subsubsec:cr_fluxes}

\begin{figure*}[t]
\begin{center}
\includegraphics[width=0.67\columnwidth, clip]{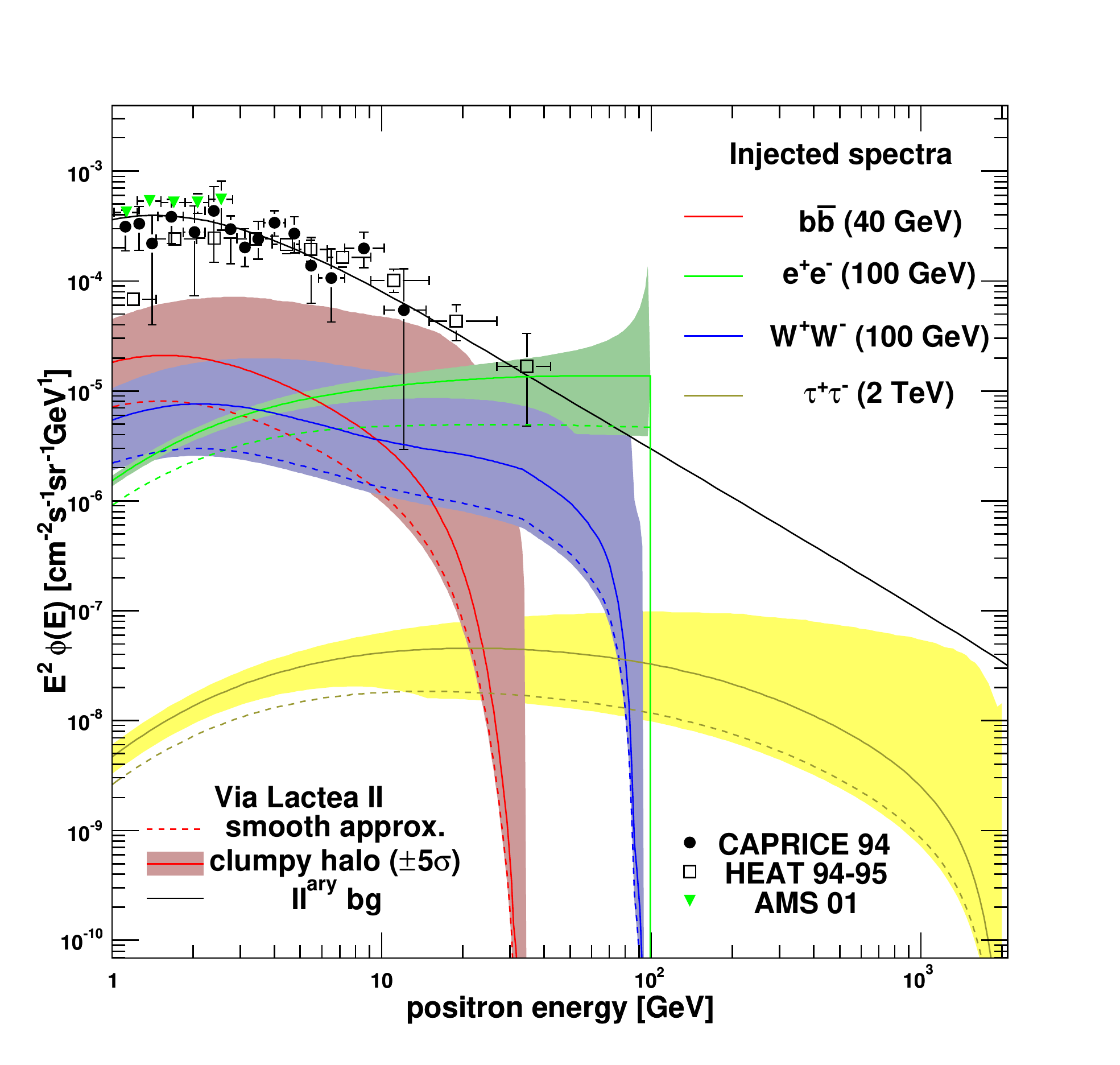}
\includegraphics[width=0.67\columnwidth, clip]{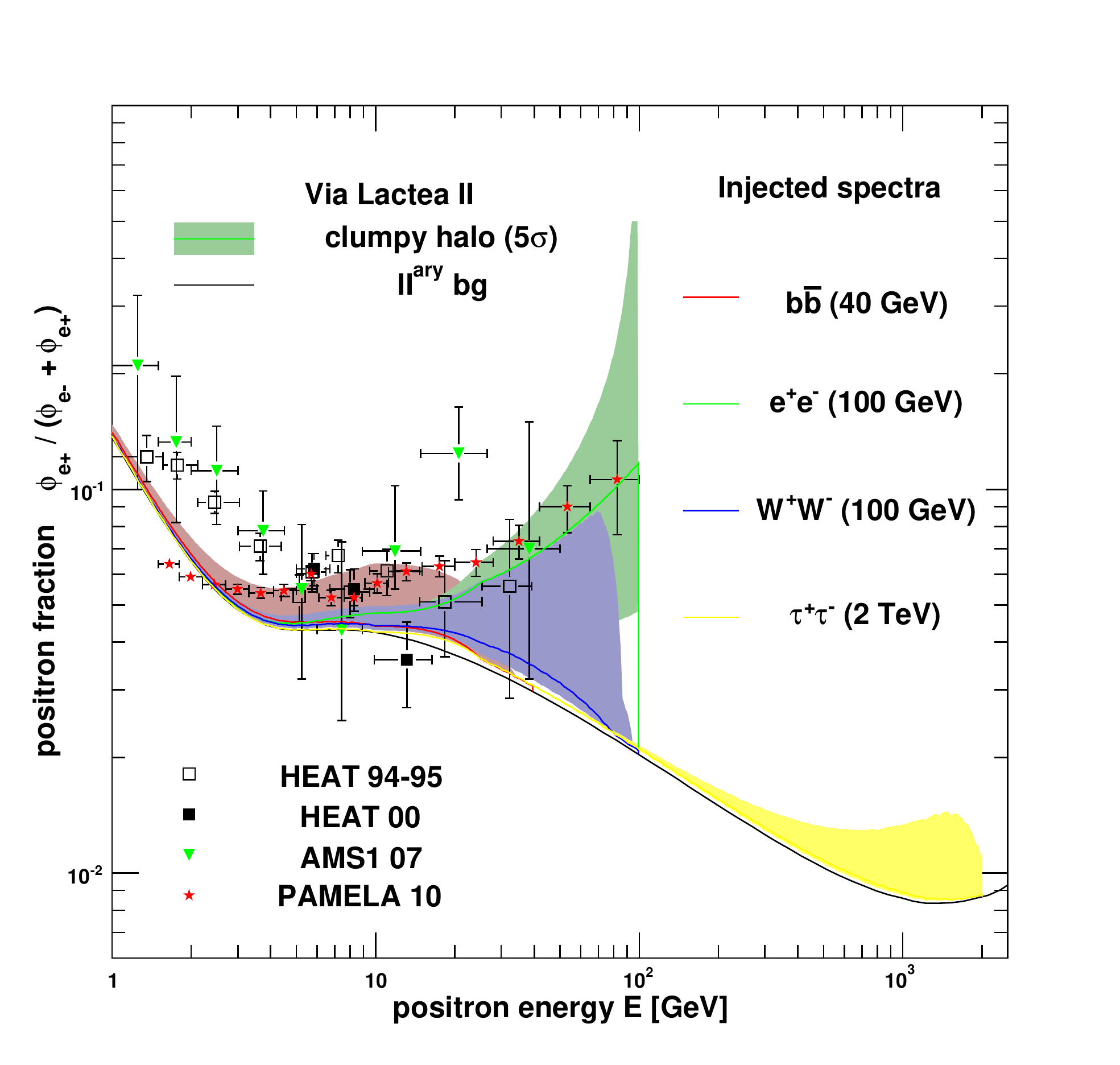}
\includegraphics[width=0.67\columnwidth, clip]{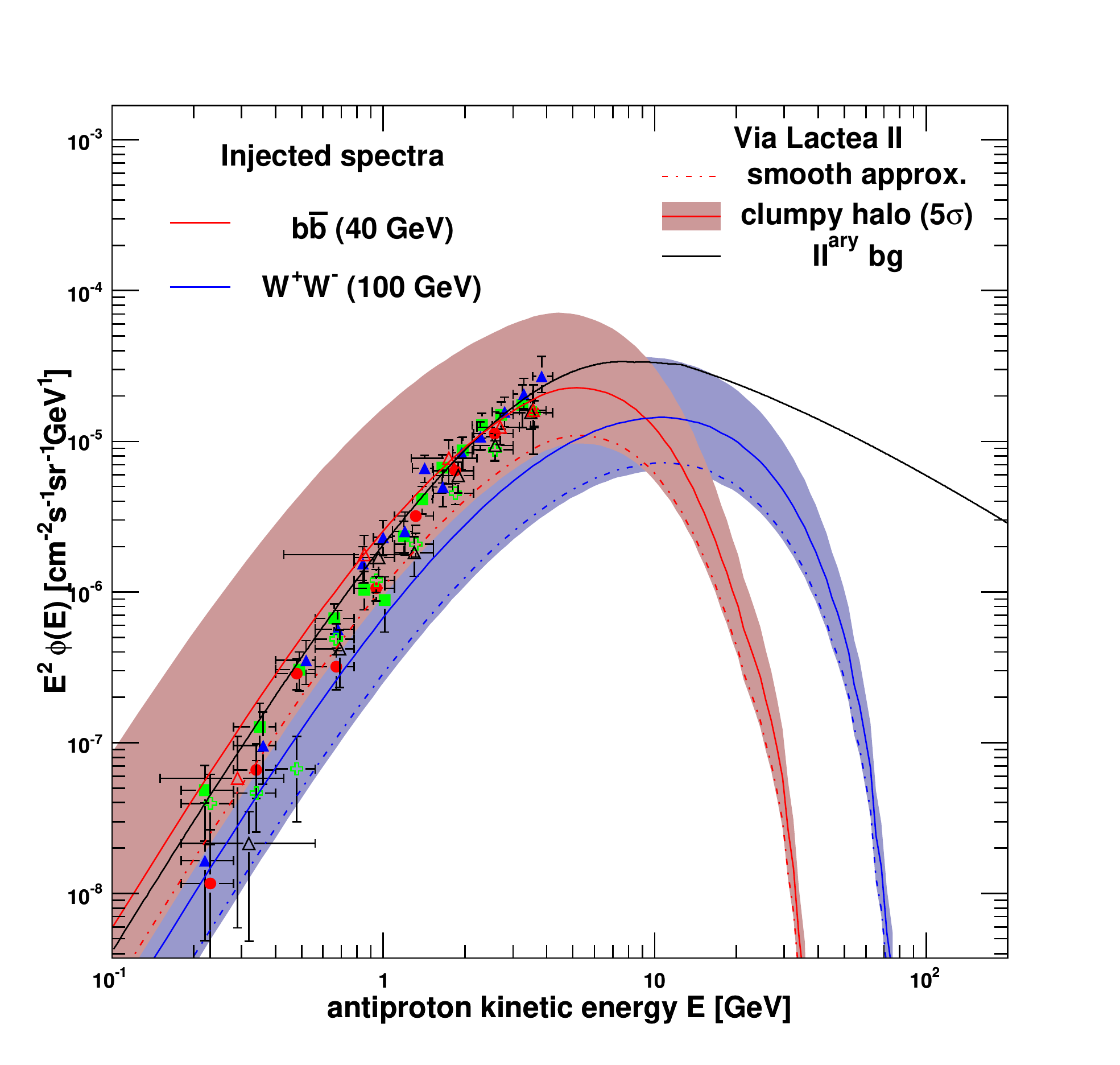}
\includegraphics[width=0.67\columnwidth, clip]{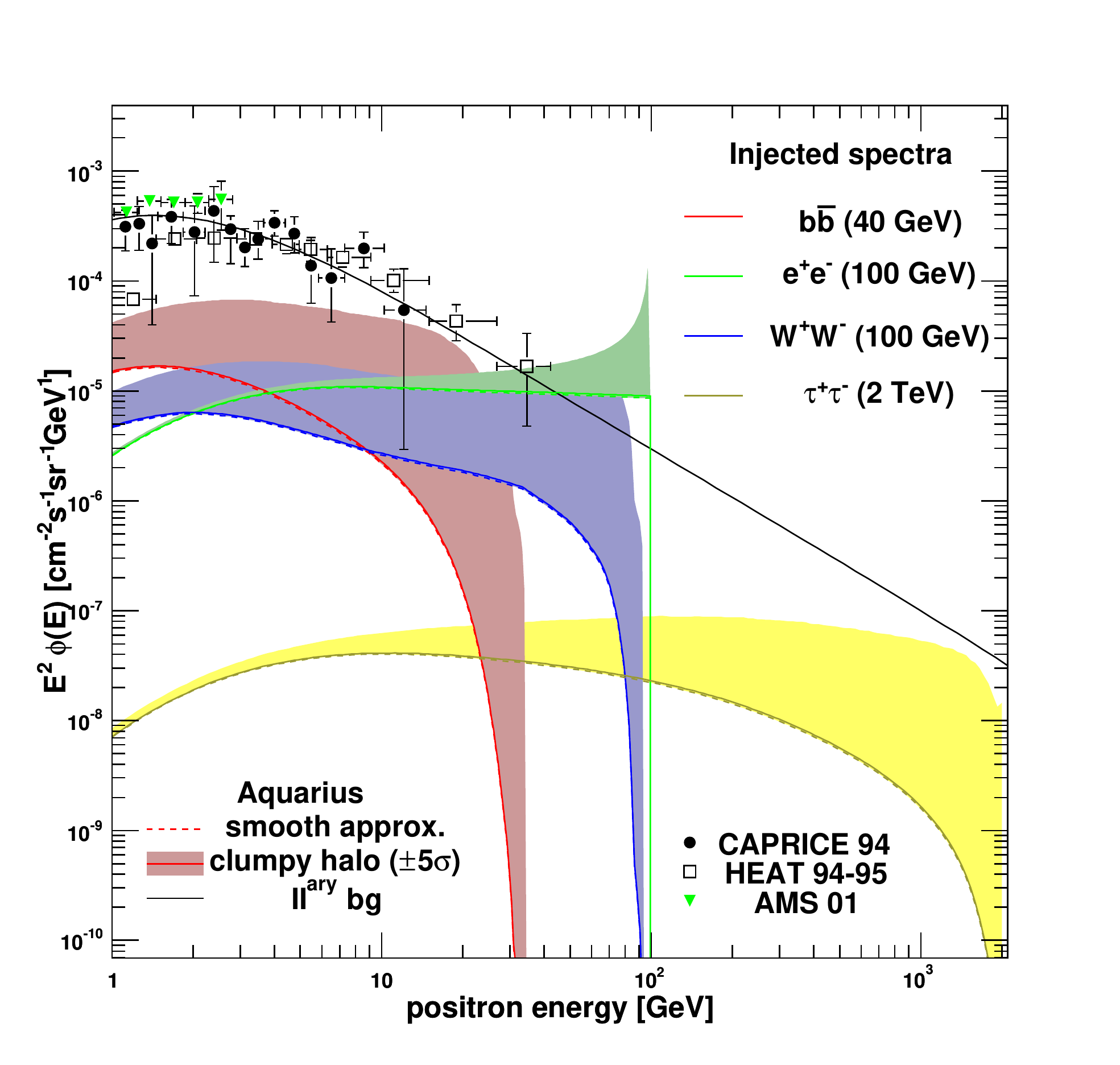}
\includegraphics[width=0.67\columnwidth, clip]{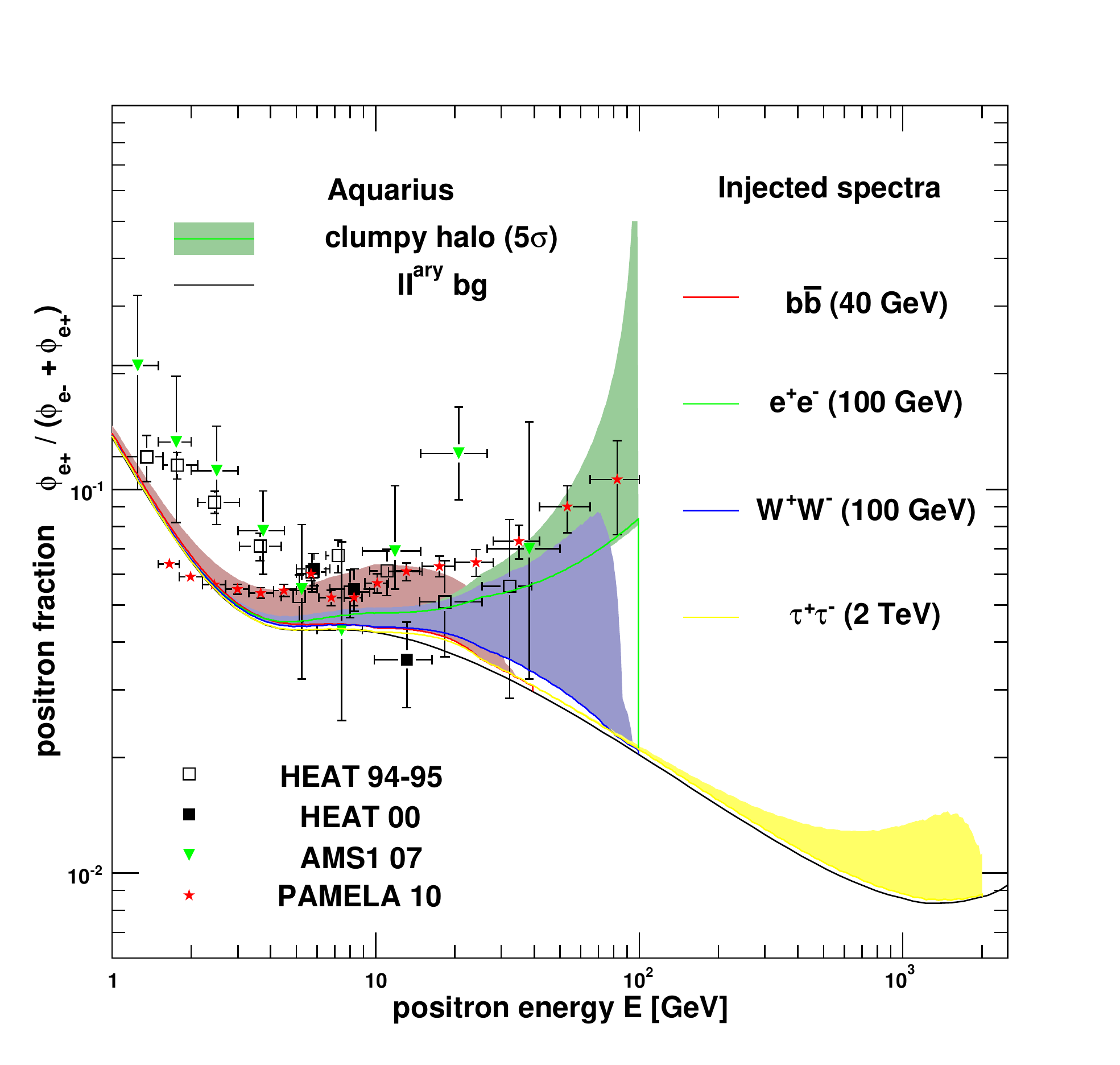}
\includegraphics[width=0.67\columnwidth, clip]{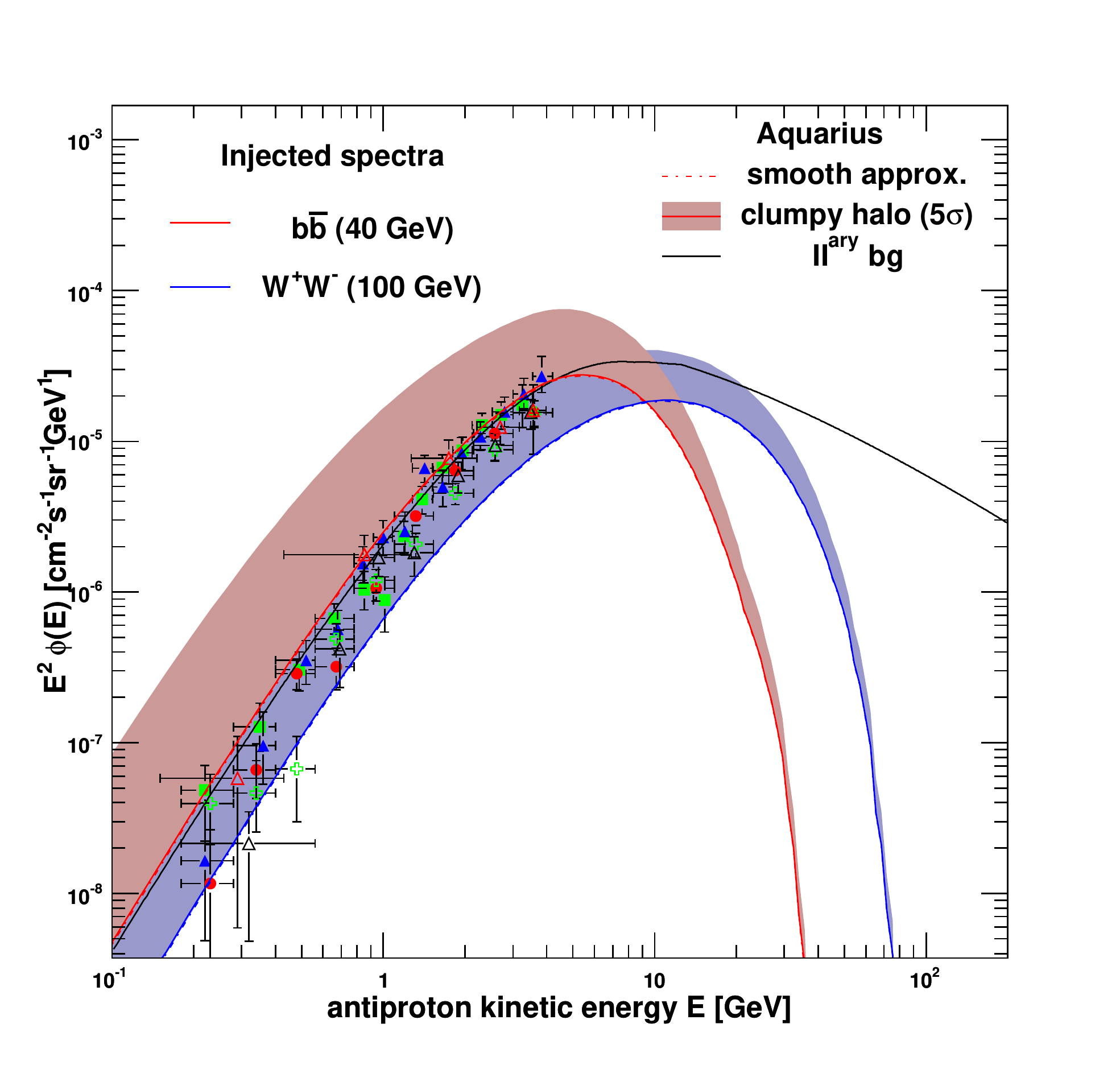}
\caption{\small Positron flux (left panel) and positron fraction (middle panel)
for our benchmark WIMP models. Right: antiproton fluxes for the WIMP models
annihilating into $b\bar{b}$ and $W^+W^-$. The top (bottom) row corresponds to 
the results obtained for the \vlii~(\aquarius) setup when the tidal disruption 
\`a la Roche is unplugged. The effects of tidal disruption are shown in 
\citefig{fig:cr_flux_tide}.}
\label{fig:cr_flux}
\end{center}
\end{figure*}

\begin{figure*}[t]
\begin{center}
\includegraphics[width=0.67\columnwidth, clip]{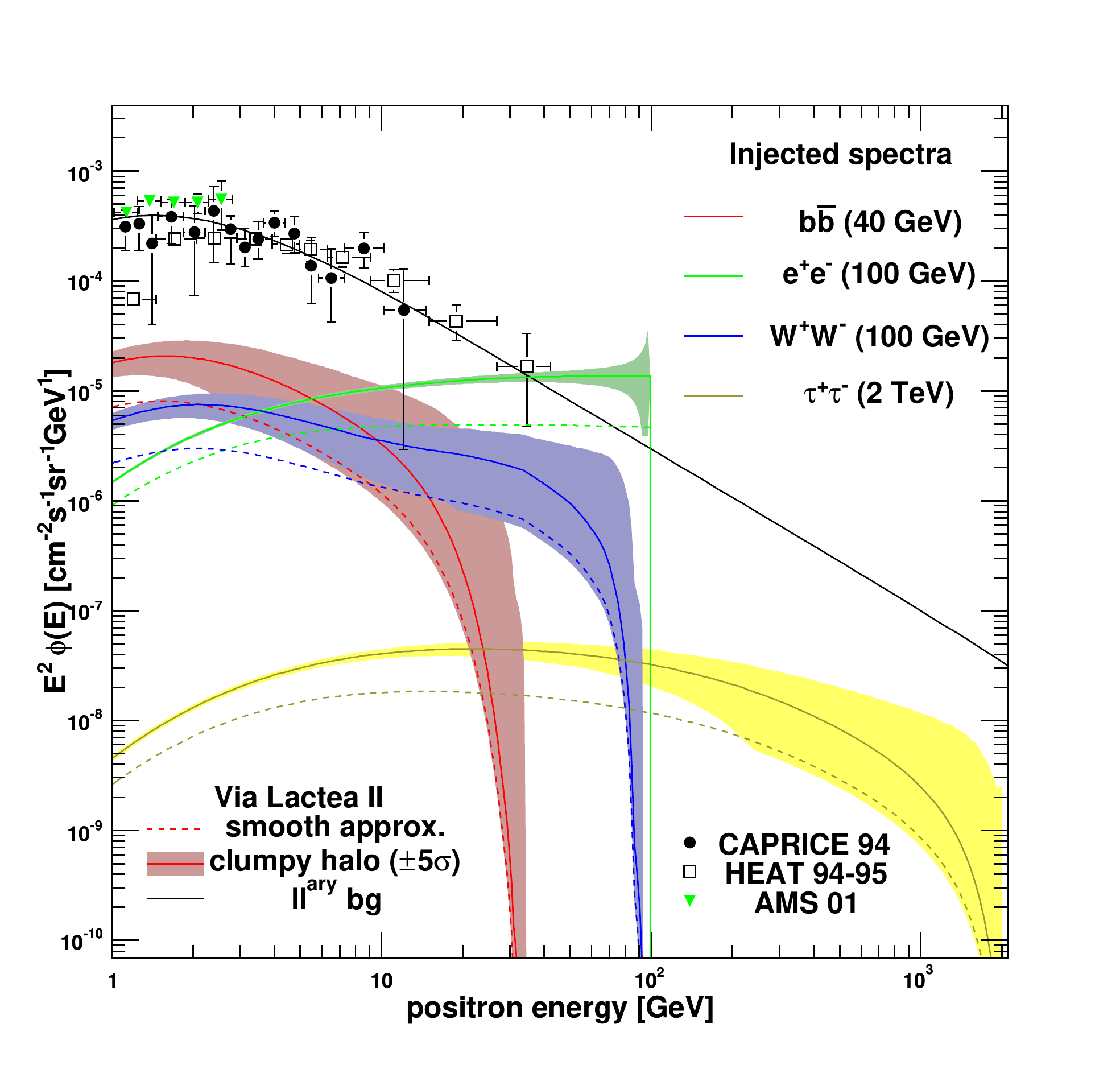}
\includegraphics[width=0.67\columnwidth, clip]{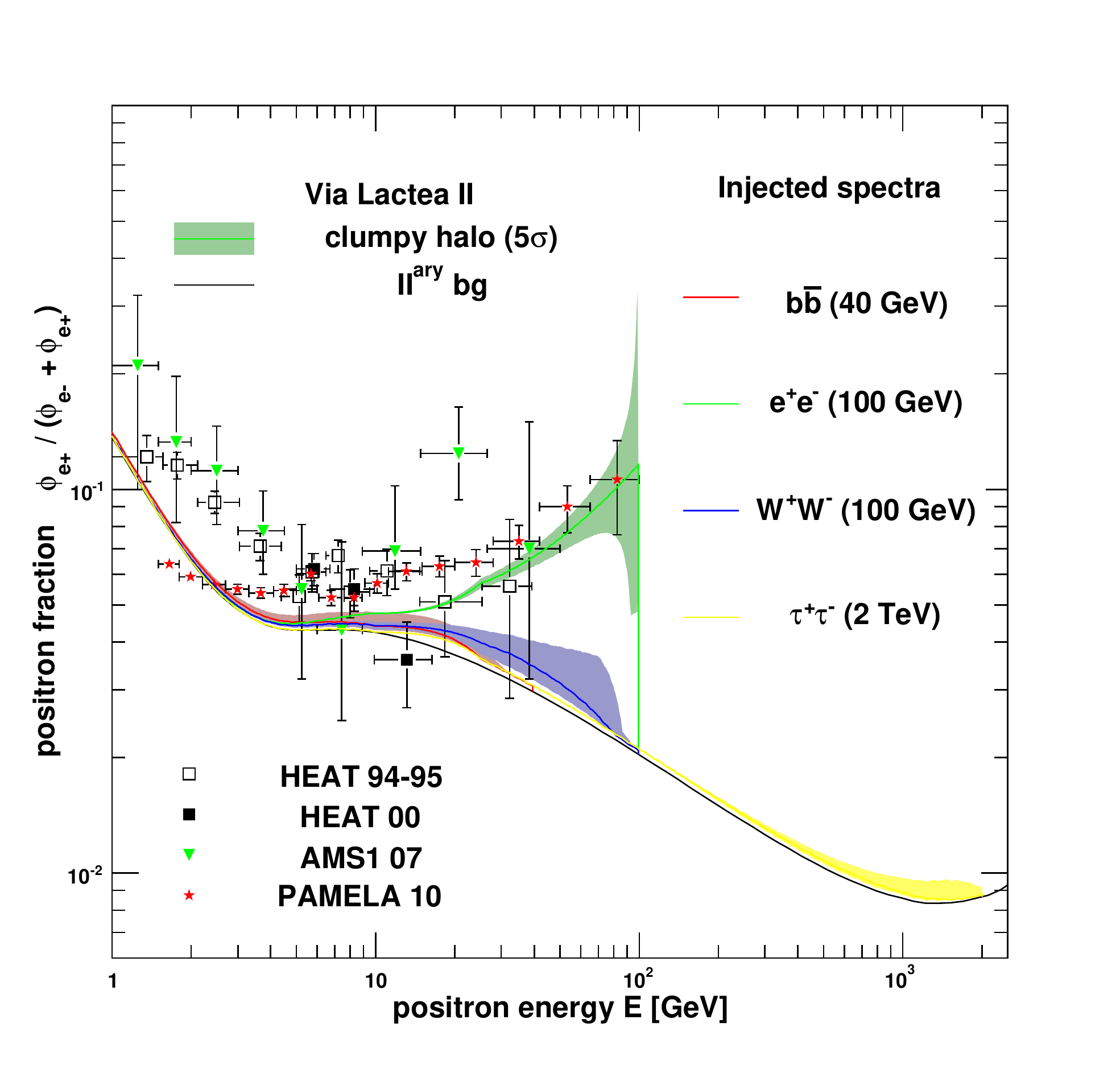}
\includegraphics[width=0.67\columnwidth, clip]{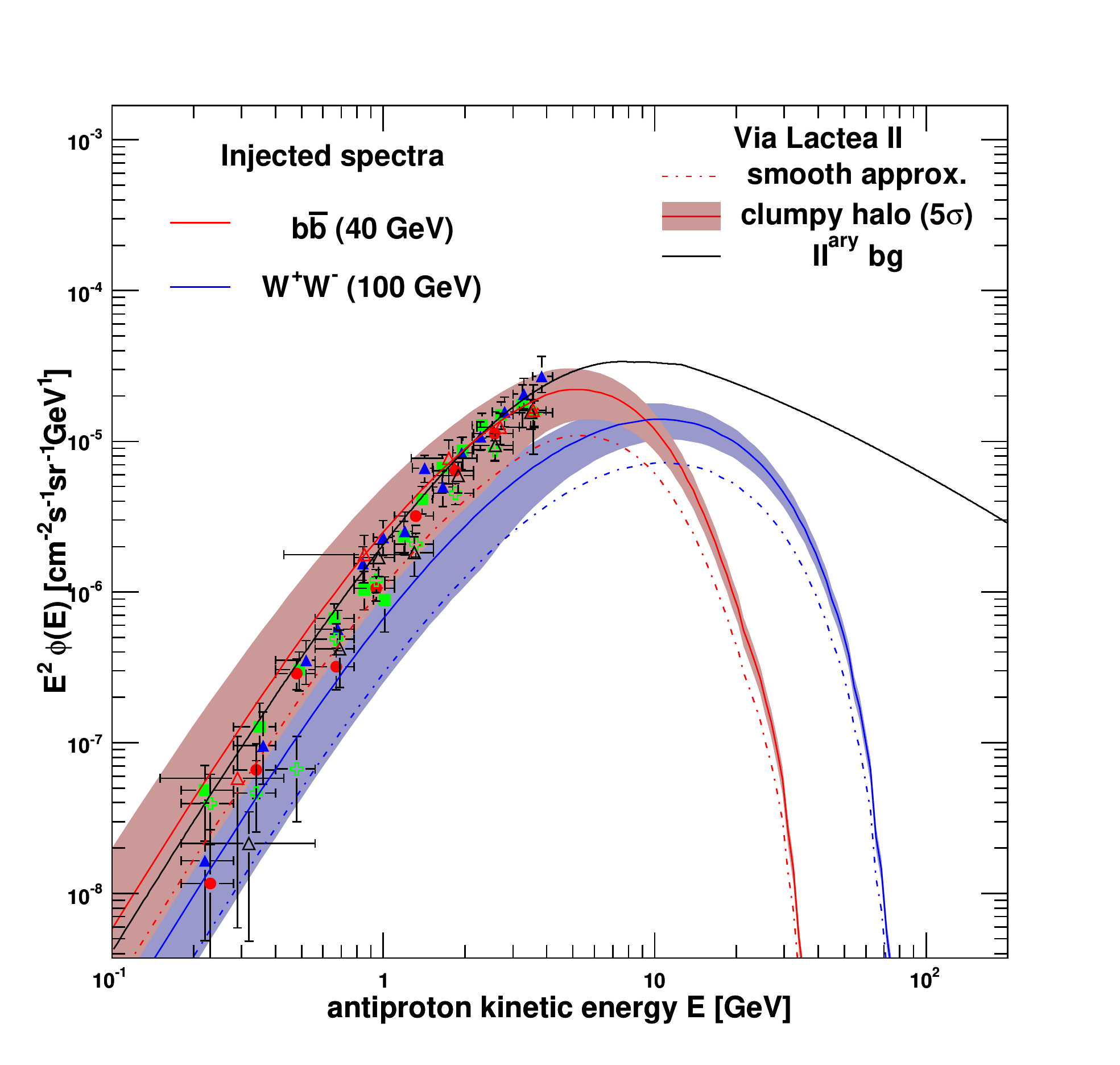}
\includegraphics[width=0.67\columnwidth, clip]{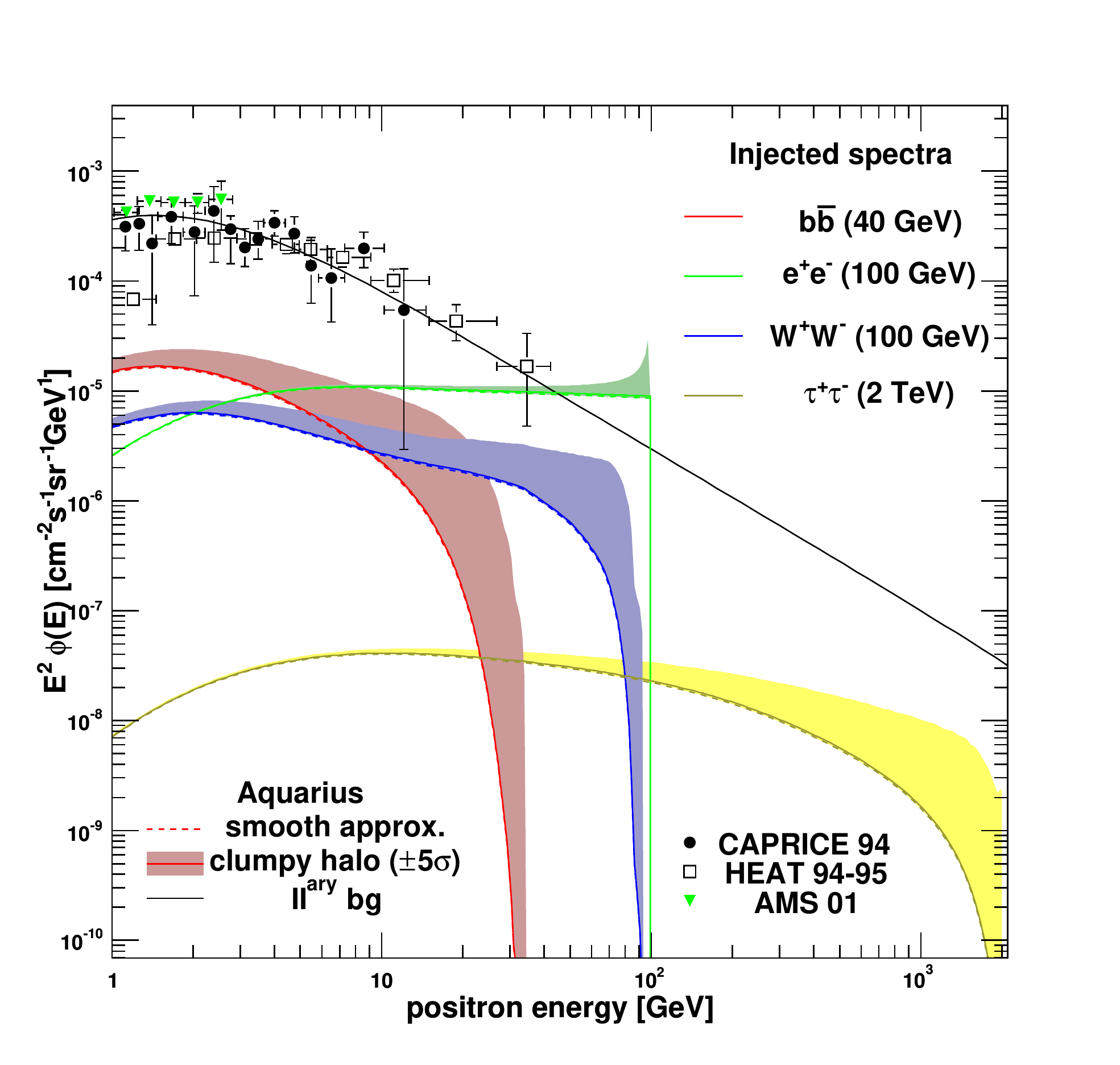}
\includegraphics[width=0.67\columnwidth, clip]{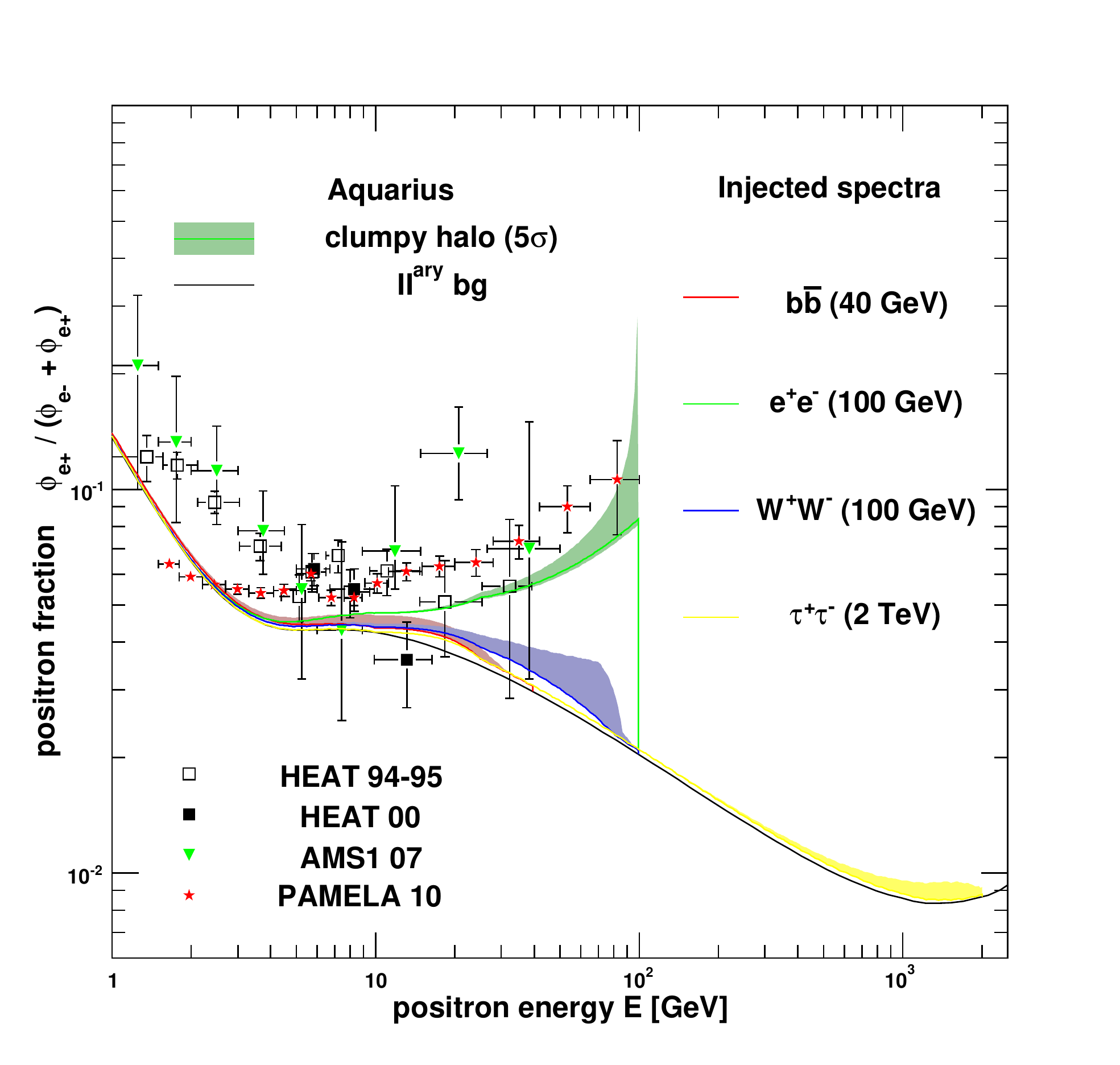}
\includegraphics[width=0.67\columnwidth, clip]{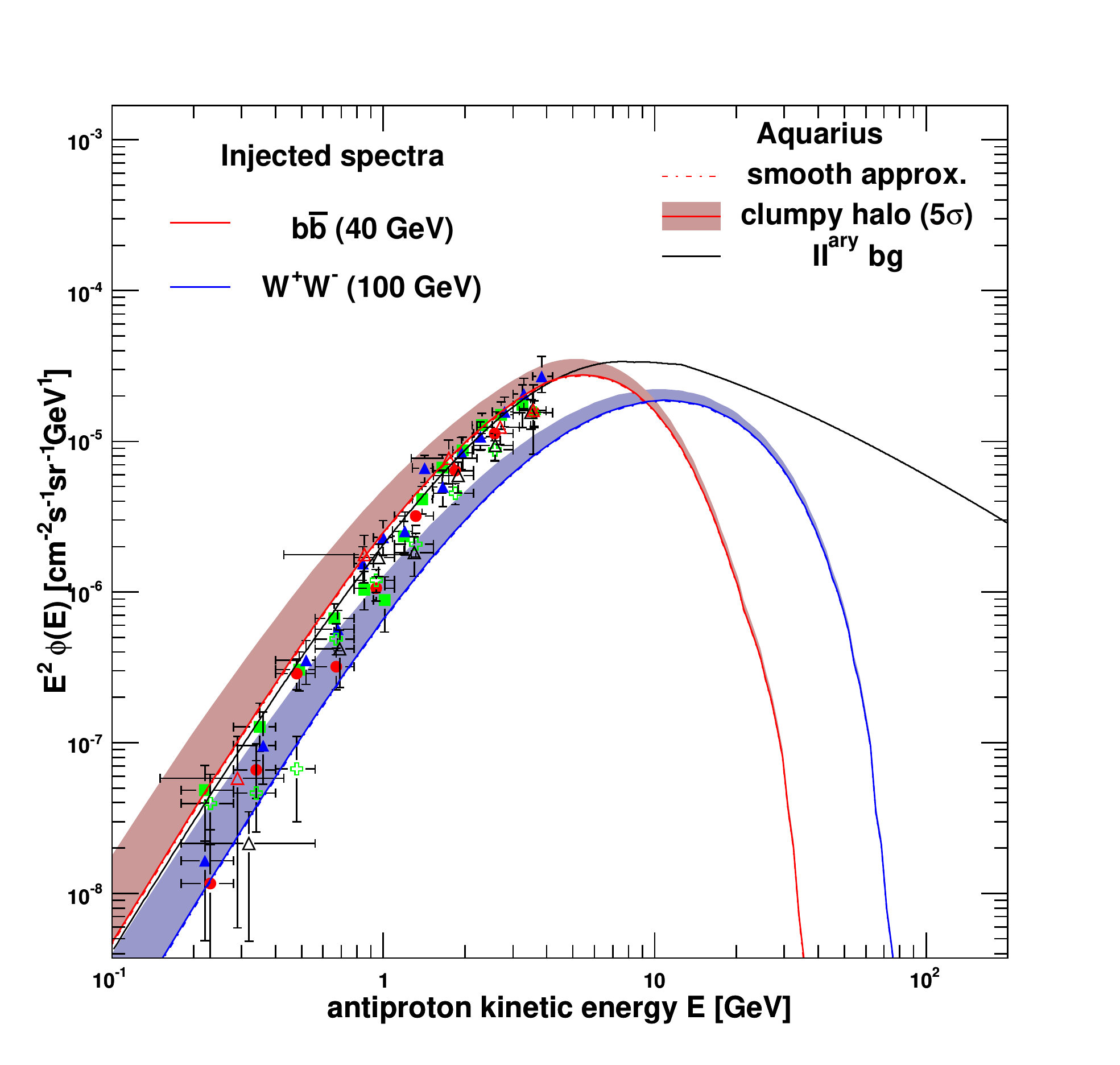}
\caption{\small Same as \citefig{fig:cr_flux} but when plugging the Roche 
  tidal disruption effects.}
\label{fig:cr_flux_tide}
\end{center}
\end{figure*}

\begin{figure*}[t]
\begin{center}
\includegraphics[width=0.67\columnwidth, clip]{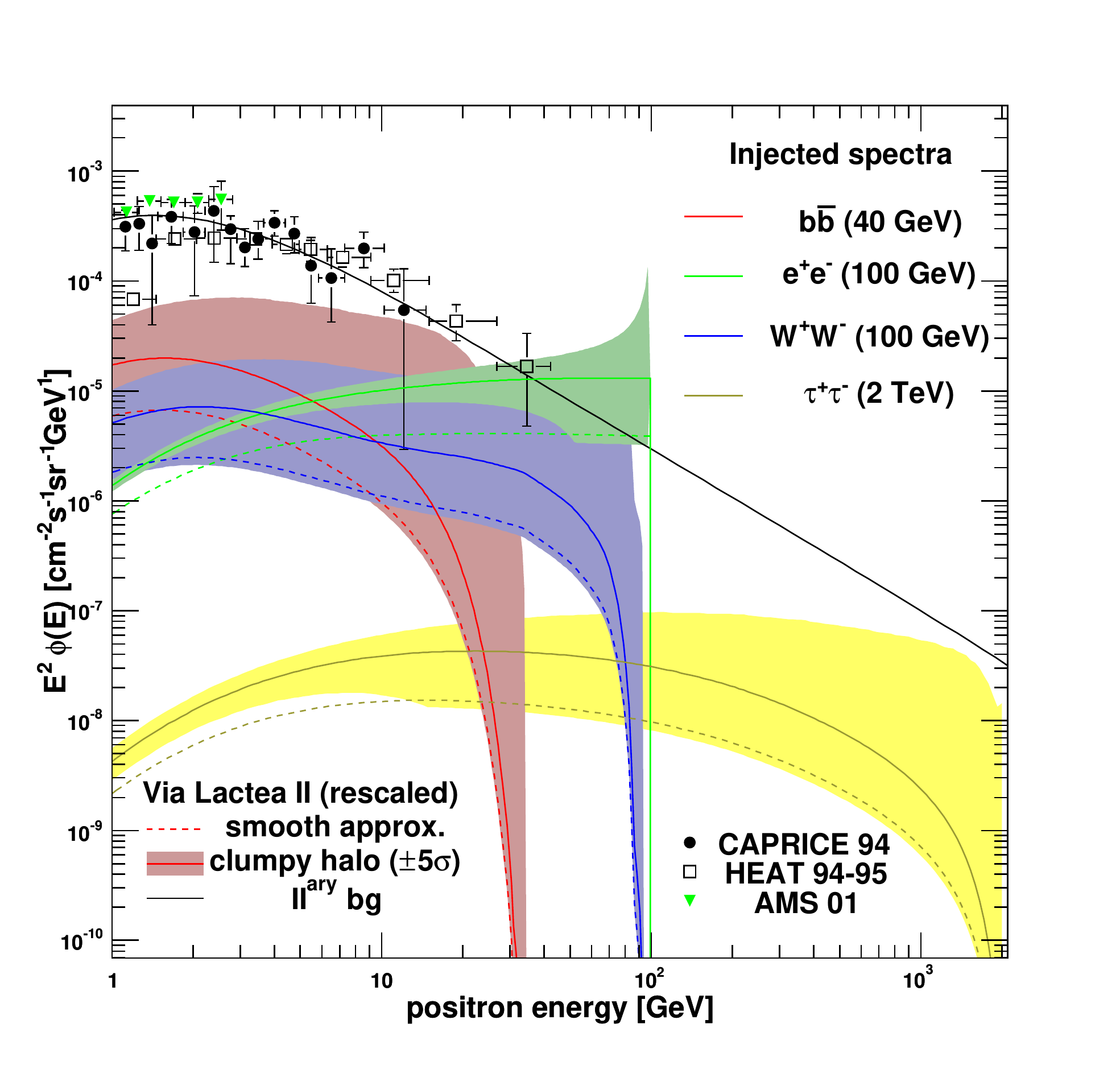}
\includegraphics[width=0.67\columnwidth, clip]{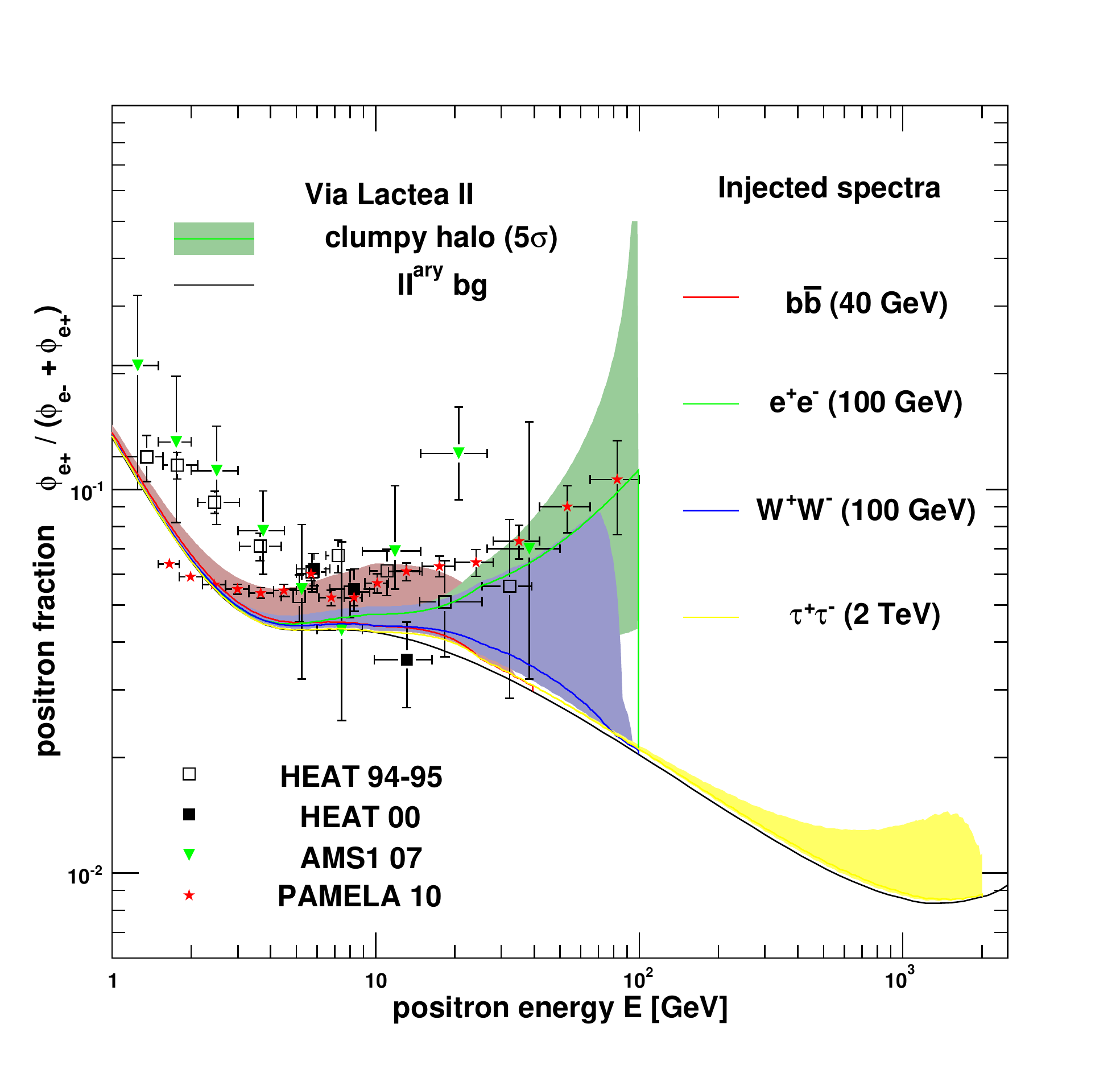}
\includegraphics[width=0.67\columnwidth, clip]{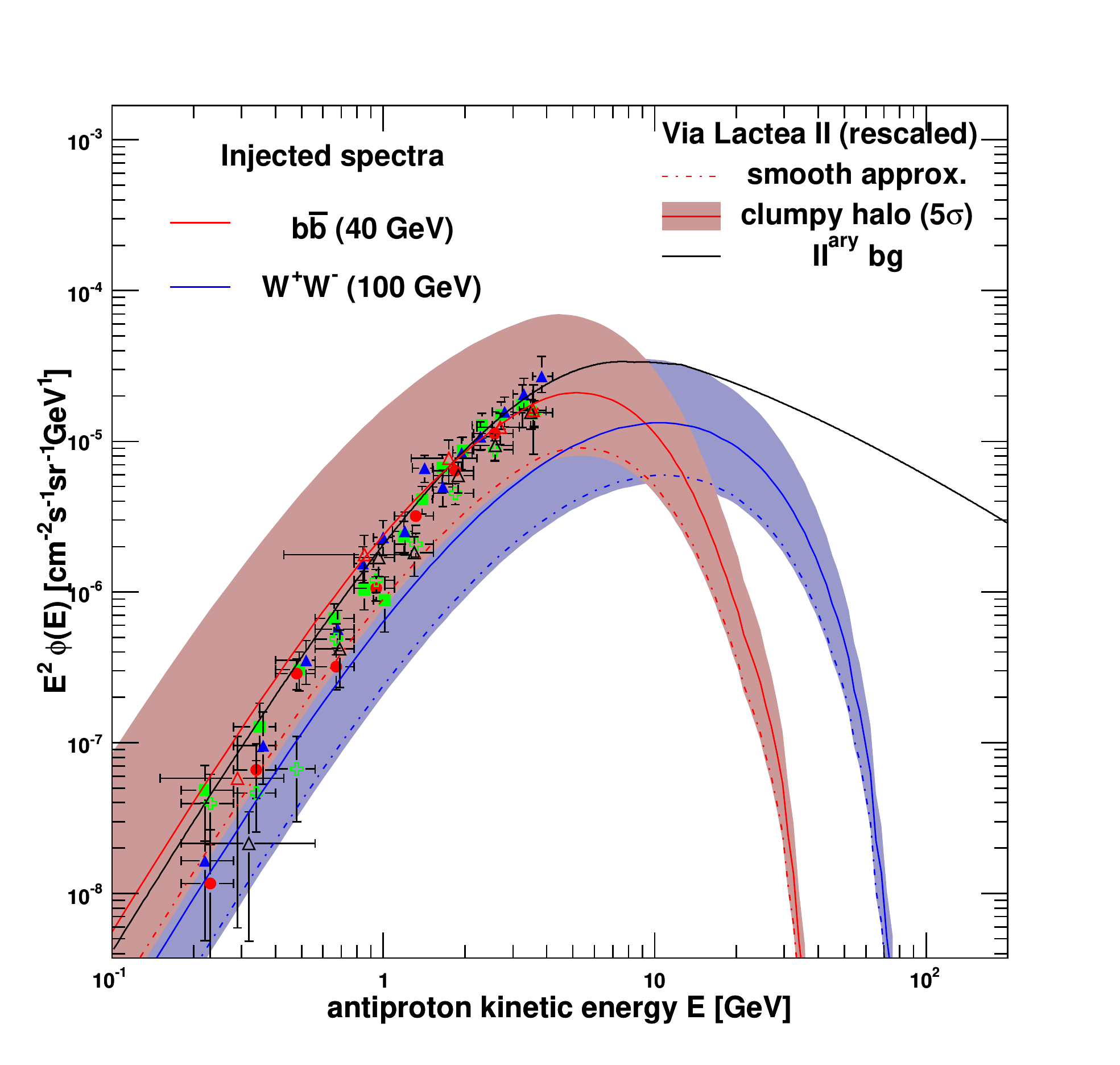}
\includegraphics[width=0.67\columnwidth, clip]{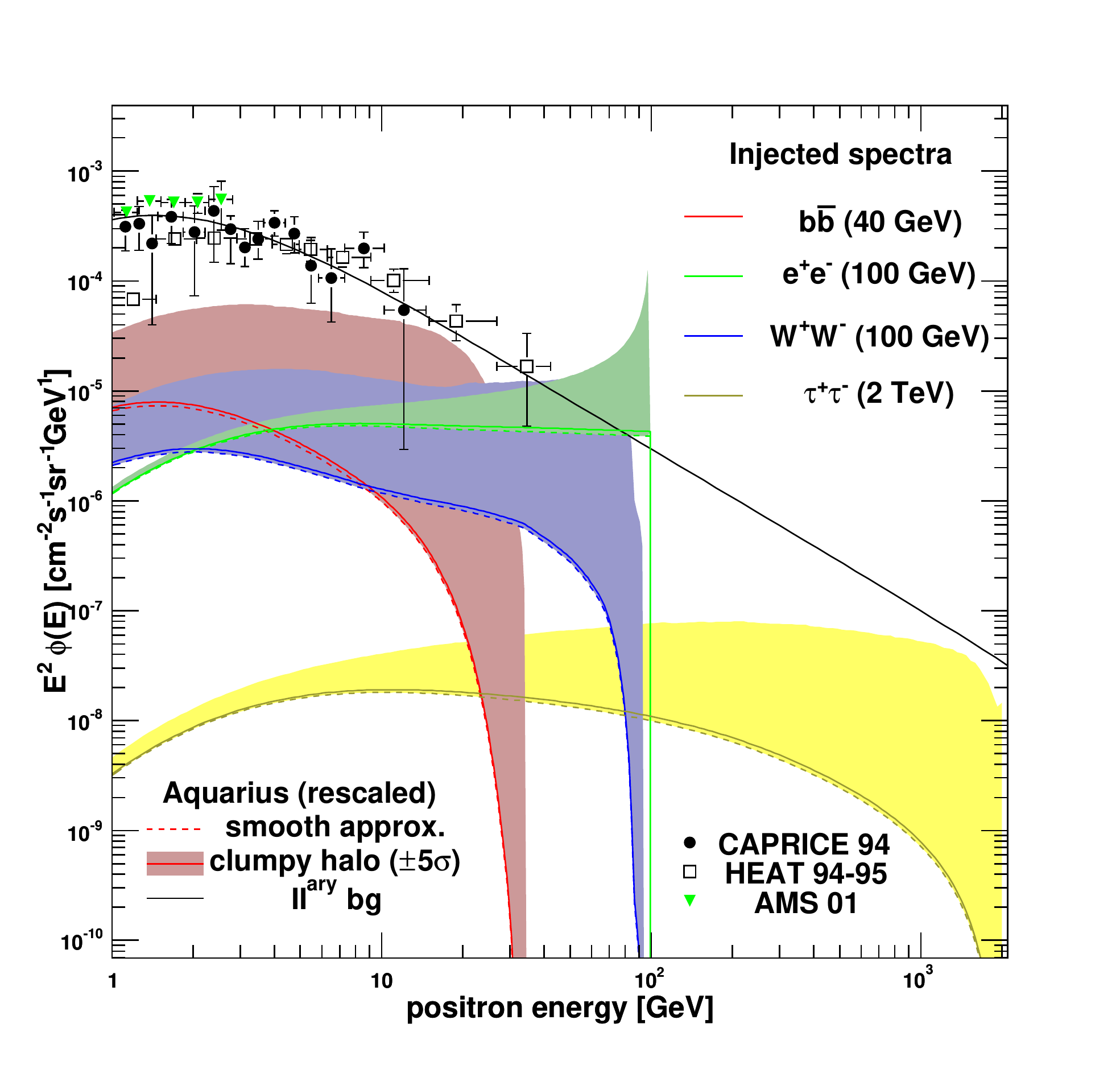}
\includegraphics[width=0.67\columnwidth, clip]{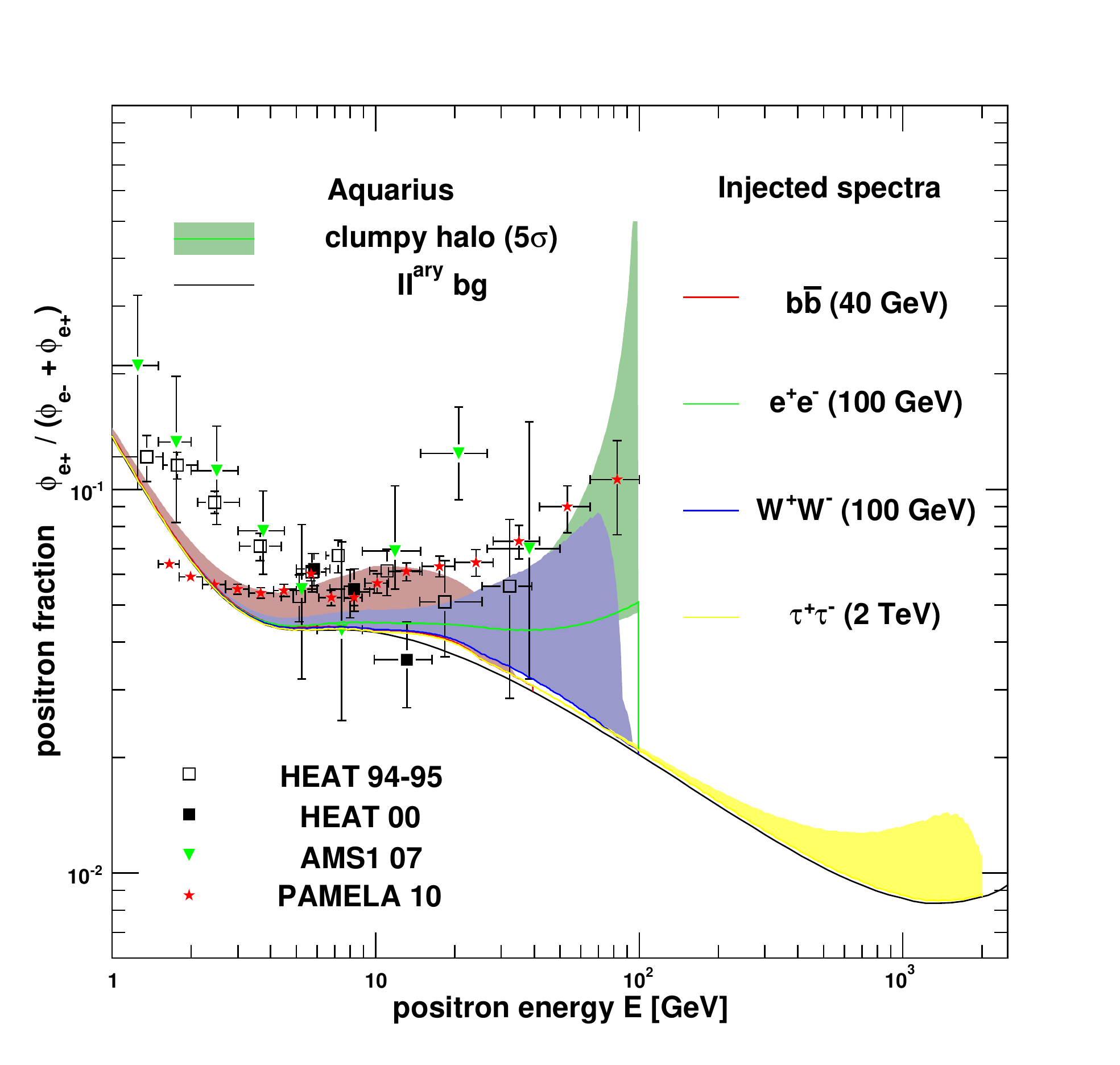}
\includegraphics[width=0.67\columnwidth, clip]{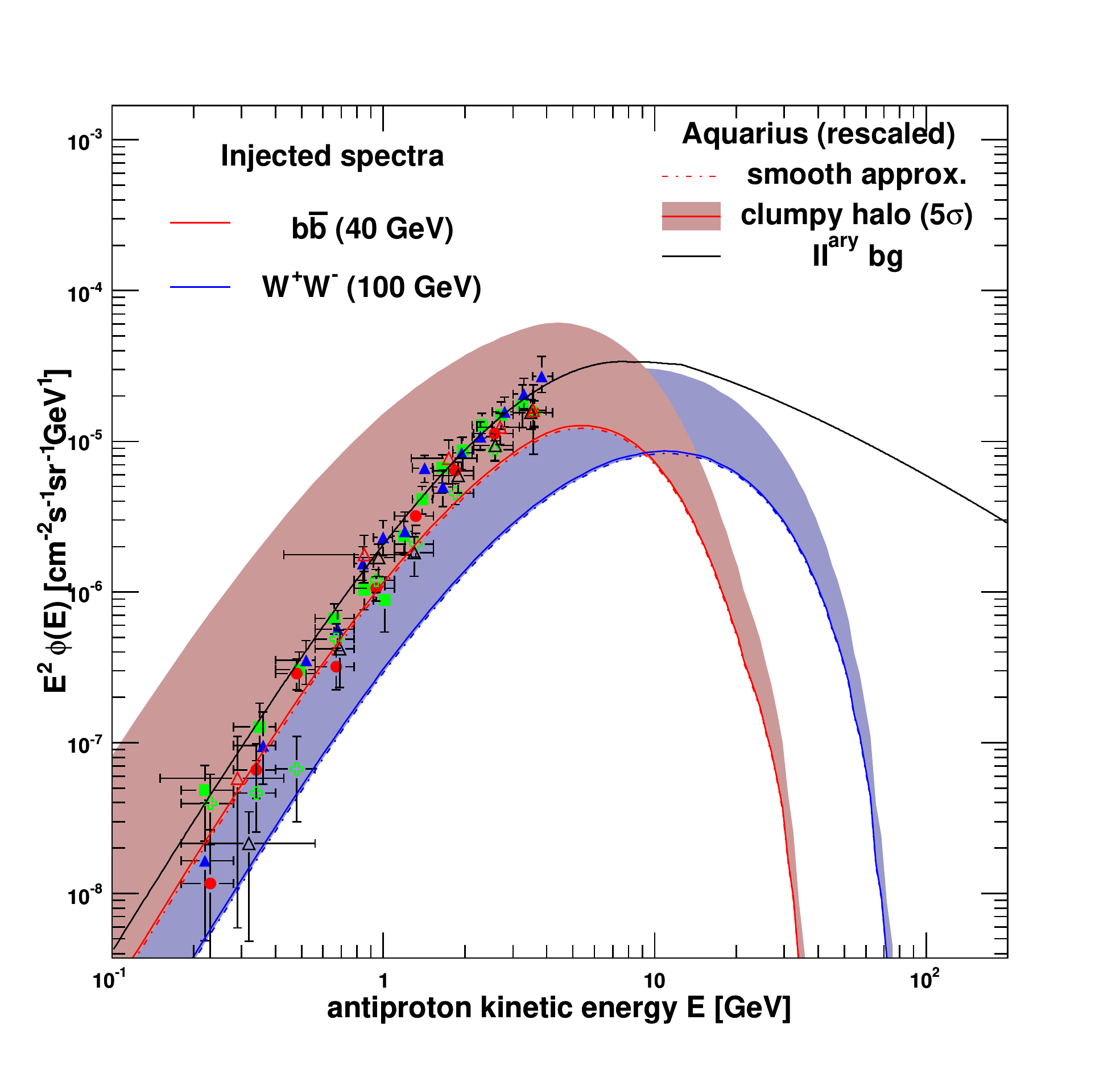}
\caption{\small Same as \citefig{fig:cr_flux} but after having rescaled the 
\vlii~and \aquarius~configurations to the same local dark matter density 
$\rhosun = 0.385\,{\rm GeV.cm^{-3}}$.}
\label{fig:cr_flux_rescaled}
\end{center}
\end{figure*}

We have computed the expected antimatter flux for all the benchmark WIMP 
candidates discussed in \citesec{sec:bench}. The results are shown in 
\citefigs{\ref{fig:cr_flux},~\ref{fig:cr_flux_tide}~and 
\ref{fig:cr_flux_rescaled}}, where we have applied a force field of 600 MV to
account for solar modulation. \citefig{fig:cr_flux} displays
our full results for both the \vlii~and~\aquarius~configurations ---
\citefig{fig:cr_flux_tide} shows the effect of plugging the Roche disruption 
and \citefig{fig:cr_flux_rescaled} is the same as \citefig{fig:cr_flux} 
when the local dark matter density of both configurations is rescaled to
$\rhosun = 0.385\,{\rm GeV cm^{-3}}$. In each figure, the left, middle and right
panels show the positron flux, the positron fraction and the antiproton flux, 
respectively, with the associated $5\sigma$ fluctuations caused by the presence 
of subhalos; the plots in the upper (lower) row are those associated with 
the \vlii~(\aquarius) setup. The solid colored curves correspond to the mean
values predicted for our benchmark WIMP models when subhalos are included 
(the colored bands encompass the $5\sigma$ contours), while the dashed colored 
curves are the expectations in the absence of subhalos. Our theoretical 
predictions are compared with the observational data taken 
from~\cite{2000ApJ...532..653B,2001ApJ...559..296D,2002PhR...366..331A} for 
positron flux, from \cite{1997ApJ...482L.191B,2004PhRvL..93x1102B,2007PhLB..646..145A,2009Natur.458..607A} for the positron fraction and from~\cite{2000PhRvL..84.1078O,2001APh....16..121M,2002PhRvL..88e1101A,2005ICRC....3...13H,2002PhR...366..331A} for the antiproton flux. We also report the secondary background flux
predictions from \cite{2009A&A...501..821D} for positrons and from 
\cite{2001ApJ...563..172D} for antiprotons.

Interestingly enough, we remark that the antiproton predictions provide rather
tight constraints on our dark matter modeling (see right panels). Indeed, the 
averaged predictions obtained for the 40 GeV $b\bar{b}$ model are in  
tension with the data in both \vlii~and \aquarius~configurations (top right and 
top left panels, respectively). Additional contributions from subhalo 
fluctuations ($5\sigma$) are excluded since clearly overshooting the data, even 
when tidal disruption towards the Galactic center is implemented 
(see \citefig{fig:cr_flux_tide}). The observational constraints are less 
severe for the 100 GeV $W^+W^-$ model, the averaged predictions of which lie
at a factor of $\sim 2$ below the secondary prediction. Note however that
large fluctuations above the mean predicted fluxes would again be in tension 
with the data, reaching the secondary background prediction when tidal 
effects are plugged (even exceeding it when they are neglected).

We have verified that the subhalos observable in $\gamma$-rays, the coordinates
and properties of which were extracted from our MC runs, are not dominating
the overall antiproton flux, as expected. Indeed, since the antiproton range
is large at the GeV-TeV energy scale, the bulk of the antiproton flux does in 
fact originate mostly from the large population of unresolved subhalos at 
low and intermediate energies, and from the Galactic central regions at higher 
energies (see the boost factor predictions in the right panel of 
\citefig{fig:boost_crs}). This means that as soon as subhalos dominate the 
overall contribution, which is the case in the \vlii~configuration, 
rescaling the smooth dark matter density down would not be sufficient to 
decrease the predictions. The comparison of \citefig{fig:cr_flux} with 
\citefig{fig:cr_flux_rescaled} is a rather striking illustration: predictions 
associated with the \vlii~setup are poorly affected by a rescaling of the 
smooth dark matter density, whereas those with the \aquarius~setup are clearly 
decreased. Therefore, a \vlii-like setup for subhalos (steep mass profile) 
associated with low WIMP mass models coupling to quarks seems clearly 
disfavored by the current data. Nevertheless, we should also keep in mind that
cosmic ray propagation is affected by large theoretical uncertainties, the 
impacts of which are much stronger on the dark matter yields than on the 
background predictions. For instance, lowering the vertical extent of the 
diffusion zone (and lowering the diffusion coefficient accordingly to fulfill
the B/C constraints) or increasing the convection velocity would result in 
lower flux predictions \cite{2004PhRvD..69f3501D}, which could rehabilitate 
such scenarios.

As regards the positron flux, it is interesting to note that the predicted
mean fluxes are much lower than the secondary background expectation by
$\sim 1$ order of magnitude for all benchmark models but the one annihilating 
into electron-positron pairs. Disregarding this latter case for the moment, it
turns out that $5\sigma$ statistical fluctuations could still lead to 
observational spectral features for the $W^+W^-$ model, as made clearer in the 
positron fraction plots. Nevertheless, this statistical effect is actually 
cancelled as soon as tidal effects are implemented, as shown in 
\citefig{fig:cr_flux_tide}. Indeed, the probability that a nearby single and 
very luminous subhalo (with typical mass $\gtrsim 10^{6-7}\Msun$) dominates 
the overall flux is much smaller in that case (see the allowed maximal 
masses in \citefig{fig:roche}). Note that at variance with the antiproton
signal, a single nearby object can dominate the high energy flux in the case 
of positrons, due to their short propagation range.

Focusing on the 100 GeV positron line model, we can further compare our two 
dark matter distribution configurations, for which the predictions generically 
exceed the secondary background. This can be understood easily by 
deriving the general analytical expression of the flux for an injected 
positron line in the limit $E\rightarrow \mchi$, which reads for standard 
quantities:
\ben
\phi^{\chi}_{e^+}(E\rightarrow\mchi) &=& \frac{\delta \beta c}{4\pi}
\frac{\tau_{\rm loss} E_0}{E^2} \frac{\sigv}{2} 
\left( \frac{\rhosun}{\mchi} \right)^2\nn\\
&\approx& 3\times 10^{-10} {\rm cm^{-2}s^{-1}GeV^{-1}sr^{-1}}\times \nn\\
 && \frac{\tau_{\rm loss}}{10^{16}{\rm s}}
\left\lbrack \frac{ \rhosun}{0.3\;{\rm GeV/cm^3} } \right\rbrack^2 \times \nn\\
&& \left\lbrack \frac{\mchi }{ 100\;{\rm GeV}} \right\rbrack^{-4}
\frac{\sigv }{3\times 10^{-26} {\rm cm^3/s} } \;.
\label{eq:poslineflux}
\een
Surprisingly enough, it is exactly the value of the predicted background flux 
$\phi^{\rm bg}_{e^+}(E=100\,{\rm GeV}) \approx 3\times 10^{-10}{\rm cm^{-2}s^{-1}GeV^{-1}sr^{-1}}$ at 100 GeV in the median model of~\cite{2009A&A...501..821D}. 
This formula is readily applied to the \vlii~setup by using 
$\rhosun = 0.42 \,{\rm GeV cm^{-3}}$ and multiplying by the local boost factor, 
which can be taken from \citefig{fig:boost_crs}. We find
$\phi^{\chi}_{e^+}(\mchi = 100 \, {\rm GeV}) 
\approx 1.7 \times 10^{-9}{\rm cm^{-2}s^{-1}GeV^{-1}sr^{-1}}$, which is larger 
than the secondary positron flux by a factor of $\sim 6$. We note that this 
asymptotic flux prediction is only valid for $E\rightarrow\mchi$ and falls 
thereby very quickly with $\mchi$ like $m_{\chi}^{-4}$. We will further comment 
on this when discussing the positron fraction in \citesec{subsubsec:pf}.

Finally, we stress again that the theoretical uncertainties on the propagation 
parameters are still large, and the resulting uncertainty in terms of primary 
positron flux can easily reach one order of magnitude 
\cite{2004PhRvD..69f3501D,2008PhRvD..77f3527D}

\subsubsection{Comments on the positron fraction}
\label{subsubsec:pf}

The excess in the positron fraction above a few GeV, recently made clearer 
with the release of PAMELA data~\cite{2009Natur.458..607A}, but previously 
hinted by the HEAT~\cite{1997ApJ...482L.191B} and AMS 
data~\cite{2007PhLB..646..145A}, has triggered an impressive number of 
studies, most of them dedicated to a possible DM 
interpretation (\eg~\cite{2008PhRvD..78j3520B}). Most of the predictions 
rely on the assumption that the DM annihilation rate is boosted, 
essentially from the non-relativistic Sommerfeld effect, with large branching 
ratios to leptons. Some others invoke instead DM decay with a tuned 
lifetime. All of these assumptions are somewhat fine-tuned, and 
most of them do not treat the background consistently with the 
primary component. Here, we provide self-consistent predictions for some 
benchmarks mostly motivated by particle physics  (except for one leptophilic 
model) and, more important, do not demand a good fit to the PAMELA data. 
Instead, we aim at testing the potential imprints and the detectability of such 
scenarios in the antimatter spectrum.

In the central panel of \citefig{fig:cr_flux} (see also 
\citefigs{\ref{fig:cr_flux_tide}~and \ref{fig:cr_flux_rescaled}}), we have 
plotted the results for the positron flux in terms of the corresponding 
positron fraction, \ie~$\phi_{e^+}/(\phi_{e^+}+\phi_{e^-})$. 
The actual denominator of the positron fraction can be obtained by fitting the 
Fermi data on the sum of cosmic electrons plus positrons above 20 GeV 
\cite{2009PhRvL.102r1101A}, avoiding thereby to invoke any model of primary 
or/and secondary component. At lower energy, we have constrained the electron 
spectrum from a fit on the AMS data~\cite{2002PhR...366..331A}, and for 
positrons, we have taken the sum of our primaries plus secondaries, assuming 
that there are no other sources of positrons. This assumption is conservative 
in the sense that below 10 GeV, the secondaries dominate over all our 
primaries. We have linked the two domains by interpolating over a range of a 
few GeV.

As already discussed in \citesec{subsubsec:cr_fluxes}, only the WIMP model
annihilating in $e^+e^-$, with $\mchi = 100$ GeV, provides a sizable 
contribution to the positron fraction, significantly overtopping the secondary 
background above 10 GeV. The other benchmark models contribute at most at the 
percent level, except when considering $5\sigma$ subhalo fluctuations and 
neglecting tidal disruption. Of course, our leptophilic model could afford for 
a large part of the PAMELA data without artificial boost factor, only by
considering theoretically constrained dark matter distributions. This is 
a very appealing result. Nevertheless, we remind the reader that viable 
astrophysical explanations exist, like the contribution of local pulsars
\cite{1989ApJ...342..807B,2010arXiv1002.1910D}, and 
it might be difficult to distinguish between those different solutions to 
the positron excess, given the limited sensitivities and energy 
resolutions of the current experiments (even a $\chi^2$ analysis is hardly 
relevant due to the large theoretical uncertainties). Our leptophilic 
model would give a sharp cut-off in the positron fraction as well as in the 
positron flux above 100 GeV, which would nevertheless be smeared by energy 
resolution effects. Positron data at higher energy would really help 
to clarify this issue of the possible contribution of DM annihilation to 
the positron spectrum. Indeed, if the excess is still prominent above say 200 
GeV, a leptophilic model could hardly fuel the dominant contribution with 
conventional parameters, due to the $m_\chi^{-4}$ scaling of the primary flux, 
when $E\rightarrow\mchi$, compared to the $E^{-3.5}$ scaling of the secondary 
background (see \citeeqp{eq:poslineflux}).

\section{Inverse Compton Scattering of electrons and positrons from DM 
  annihilation on galactic photons}
\label{sec:ics}

We finally consider the Inverse Compton Scattering (ICS) process which happens when 
the high energy electrons and positrons produced in the DM annihilation scatter 
on the low energy photons of our Galaxy. The resulting energies of the 
upscattered photons will be increased by a factor $\propto \gamma^2$ where 
$\gamma$ is the Lorentz factor $\gamma$ =  $E_e \over m_e$.

Galactic target photons include starlight photon in the optical wavelength from 
the stars of the galactic disk (hereafter SL), infrared radiation produced by 
the interaction of such starlight photons with the galactic dust (IR) and the 
homogeneous bath of cosmic microwave background photons (CMB).

Now consider that SL/IR/CMB photons have typical mean energy of $0.3 $/ 
$3.5 \times 10^{-3}$/$2.5\times 10^{-4}$ eV respectively. The maximum Lorentz 
factor for electrons and positrons produced by the annihilation of our benchmark
candidates (got when $E_e = m_\chi$) is of the order of $\gamma \sim  8 \times 
10^{4}$, $2 \times 10^{5}$ and $4 \times 10^{6}$ for benchmarks A ($m_\chi= 40 
\GeV$), B and C ($m_\chi= 100 \GeV$) and D ($m_\chi = 2 \TeV$) respectively.
Thus, the photons can be scattered up to energies of $10^{-3}$/$9\times10^{-3}$/
3.6 GeV (CMB), $2 \times 10^{-2}$/$10^{-1}$/53 GeV (IR) and $2$/$12$/462 GeV 
(SL) for benchmark A, B\&C and D respectively. These are rough estimates of 
the relevant energy ranges to discuss, though we have to keep in mind that 
blackbody distributions spread beyond their mean values. Since we are interested
in the $\gamma$-ray flux at the GeV scale, this implies that benchmark A will 
not give any contribution at all. As far as benchmark B is concerned, we expect 
its ICS contribution to $\gamma$-rays to be negligible at the energies of 
interest, since the bulk of electrons and positrons will have energies much 
lower than $m_\chi$ (we recall that the annihilation channel is $W^+ W^-$).

We will then restrict ourselves to the computation of ICS processes for 
benchmarks C and D, which on the other hand do not produce any significant 
$\gamma$-ray flux from the prompt emission.

\subsection{The Inverse Compton Scattering computation}

The ICS $\gamma$-ray spectrum is given by
\ben
\frac{d\Phi}{d\epsilon_1} = \frac{1}{\epsilon_1} \int_{\Delta \Omega} d\Omega 
\int_{\rm los} ds\, \frac{j(\epsilon_1,r)}{4\pi}
\een
Here $\epsilon$ is the energy of the original photon and $\epsilon_1$  the 
energy of the scattered photon. $j(\epsilon_1,r)$ is the emissivity given by
\ben
j(\epsilon_1,r)=2\int_{m_e}^{M_{\rm DM}}dE\ \mathcal{P}(\epsilon_1,E,r)\ n_e(r,E)
\;,
\een
which carries units of inverse volume and inverse time.
Here $\mathcal{P}(\epsilon_1,E, r)$ is the differential power emitted into 
photons of energy $\epsilon_1$ by an electron with energy $E$. We refer to 
\cite{1970RvMP...42..237B,2009NuPhB.821..399C} for its explicit form. We 
just recall that the emitted power is given by the rate of scattering 
$dN_{E,\epsilon}/dtd\epsilon_1$ of high energy electrons on photons of energy 
$\epsilon$ into photons of energy $\epsilon_1$, times the energy lost in a 
scattering $(\epsilon_1 - \epsilon)$, integrated over all initial photon 
energies. The rate of scattering is proportional to the energy density 
of the photon bath $n(\epsilon,r)$. Finally, $n_e(r,E)$ is the electron number 
density.

At variance with \cite{2009NuPhB.821..399C}, we compute $n_e(r,E)$ by solving 
the diffusion-loss equation in the galactic disk, as explained in 
\citesec{subsec:cr_prop}. Indeed, we have computed the electron-positron flux 
at 3 different distances from the GC both in the diffusionless and in the 
complete case. We find that the difference between the two cases gets larger 
when getting closer to the GC, as expected, and as illustrated in the left
panel of \citefig{fig:ics}. This confirms that the 
diffusionless approximation is not valid at all in the steady state regime 
when the injection rate varies significantly over a typical diffusion length, 
like for example in the Galactic center.

Following \cite{2009NuPhB.821..399C}, we model the total radiation density as a 
superposition of three blackbody-like spectra of the form
\ben
n_{a}(\epsilon,r) = {\mathcal N}_a(r)\ 
\frac{\epsilon^2}{\pi^2}\frac{1}{(e^{\epsilon/T}-1)}
\een
with different temperatures: $T_{\rm CMB}=2.5\times 10^{-4}$ eV,  
$T_{\rm IR}= 3.5 \times 10^{-3}$ eV and $T_{\rm SL}=0.3$ eV. Yet, here we are not 
interested on averaging our signal on large parts of the sky, while we want to 
study the angular dependence of the ICS flux for solid angles $\Delta \Omega =
10^{-5}$ sr, to be comparable with the prompt $\gamma$-ray flux and to be 
predictive for Fermi all-sky observations. Hence, we cannot set 
${\mathcal N}_a$ as a constant as in  \cite{2009NuPhB.821..399C}.

We fit the results of \cite{2005ICRC....4...77P,2008ApJ...682..400P} to model 
the $\rho$ and $z$ dependence of  ${\mathcal N}_a$, where $\rho$ and $z$ are 
cylindrical coordinates along and perpendicular to the galactic plane. 
We find that ${\mathcal N}_a (\rho,z)$ can be written as a function of a 
constant ${\mathcal N}_a^{GC}$ computed at $\rho = z = 0$.
Values of ${\mathcal N}_a^{GC}$ can be found in \cite{2009NuPhB.821..399C}: 
${\mathcal N}_{CMB}^{GC} = 1$, ${\mathcal N}_{IR}^{GC} = 7 \times 10^{-5}$ 
and ${\mathcal N}_{SL}^{GC} = 1.7 \times 10^{-11}$. \\
For the CMB, obviously ${\mathcal N}_{CMB}(\rho,z)  = 
{\mathcal N}_{CMB}^{GC} = 1$.

In the SL case, we obtain the following fits:
\ben
 {\mathcal N}_{SL} (\rho, z)= 
\begin{cases}
{\mathcal N}_{SL}^{GC} e^{- \alpha_{SL} \rho} \ \ \ \ & z \leq 0.2 \kpc \\
A_{SL} \times {\mathcal N}_{SL}^{GC} e^{- \alpha_{SL} \rho} e^{- \beta_{SL} z} 
& z > 0.2 \kpc
\end{cases}\nn\\
\een
For the IR case we get
\ben
 {\mathcal N}_{IR} (\rho, z)= 
\begin{cases}
({\mathcal N}_{IR}^{GC} - B_{IR} \rho) e^{- \beta_{IR} z} \ \ \ \ & 
\rho \leq 4 \kpc\\
A_{IR} \times {\mathcal N}_{IR}^{GC} e^{- \alpha_{IR} \rho} e^{- \beta_{IR} z} 
& \rho > 4 \kpc
\end{cases}\nn\\
\een
with $\alpha_{SL} \sim 0.47$, $\beta_{SL} \sim 0.57$, $A_{SL} \sim 1.12$, 
$\alpha_{IR} \sim 0.33$, $\beta_{IR} \sim 0.43$, $A_{IR} \sim 2.29$ and  
$B_{IR} \sim 6.6 \times 10^{-6}$.

\subsection{Results}

We have computed the $\gamma$-ray flux from ICS for our particle physics 
benchmarks C and D. The results are shown in \citefig{fig:ics,fig:ics2}. The 
right panel of \citefig{fig:ics} displays the flux predicted in the direction of
the GC (for benchmark D only), while  \citefig{fig:ics2}  shows the longitude 
(left) and latitude (right) dependence on the integrated flux at null latitude 
and longitude, respectively.

We note that the ICS $\gamma$-ray flux produced by electrons and 
positrons deriving from the annihilation of particles described by our 
benchmarks C and D are not in conflict with the available data on 
$\gamma$-rays. In fact, \cite{2009NuPhB.821..399C} have shown that benchmark C 
(D) would need a  cross section 1 (3) order(s) of magnitude larger than the 
thermal one in order to overproduce photons with respect to the EGRET and Fermi 
diffuse background. We note also that the predictions performed in the 
\aquarius~configuration overtop the ones calculated in the \vlii~configuration
towards the central regions of the Galaxy, where the dark matter annihilation 
rate is set by the smooth halo, whereas the hierarchy reverses at larger 
longitudes where subhalos come into play. This is due to the larger mass 
fraction and the relative domination of the lightest subhalos occurring in the 
\vlii~setup.

We have also checked our predictions against the preliminary Fermi observation 
of the 1$^\circ$ region around the Galactic Center 
\cite{cohentanugi_fermisymp09}. We find 
that benchmark D is orders of magnitude below such a signal. In the case of 
benchmark C, the predicted flux does not exceed the observed one, although it 
reaches one third of its value at energies around 20 GeV. We therefore conclude
that the ICS contribution to the $\gamma$-ray flux is likely not observable for
the leptophilic benchmarks that we selected in this paper.

\begin{figure*}[t]
\begin{center}
\includegraphics[width=\columnwidth, clip]{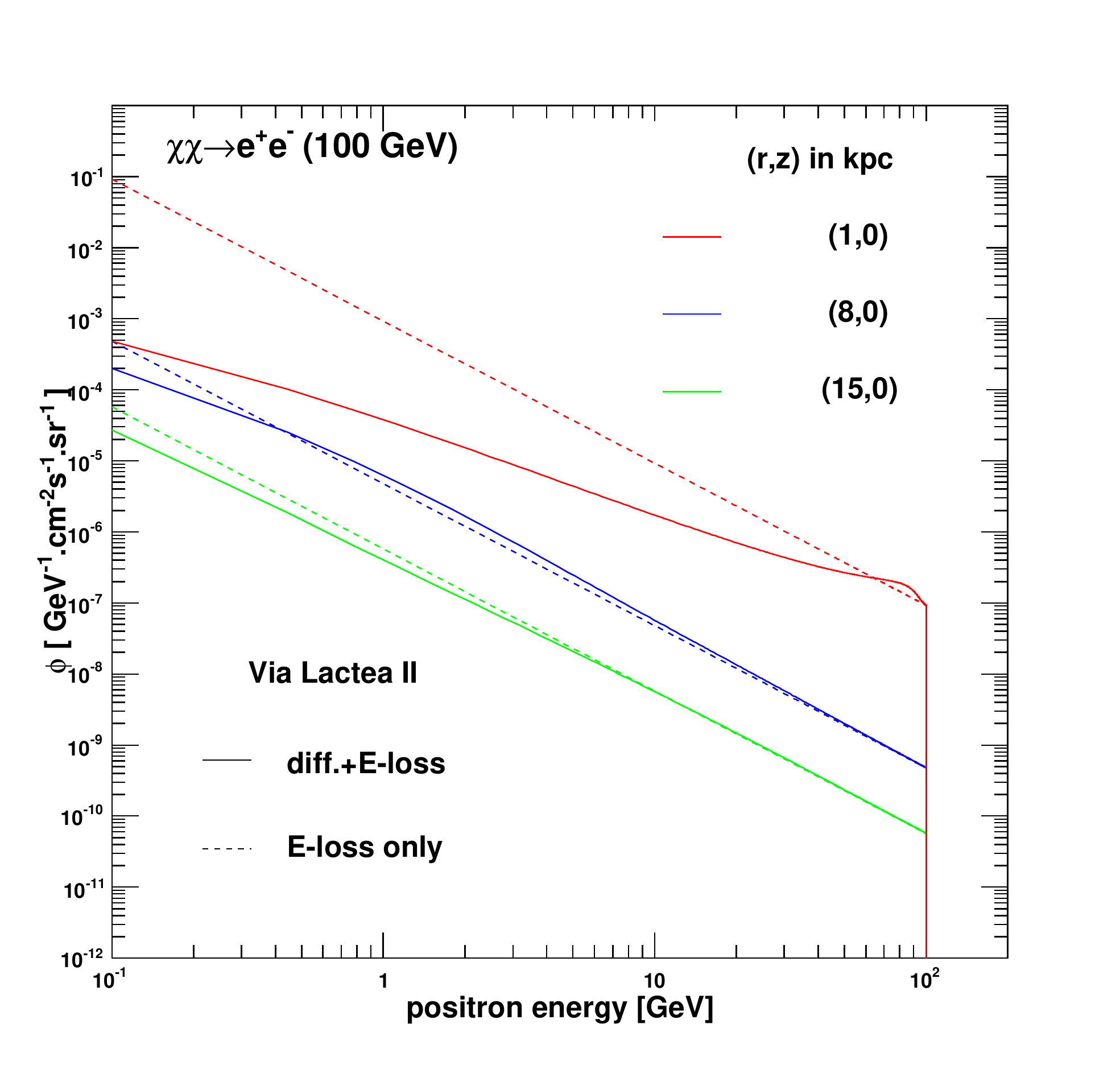}
\includegraphics[width=\columnwidth, clip]{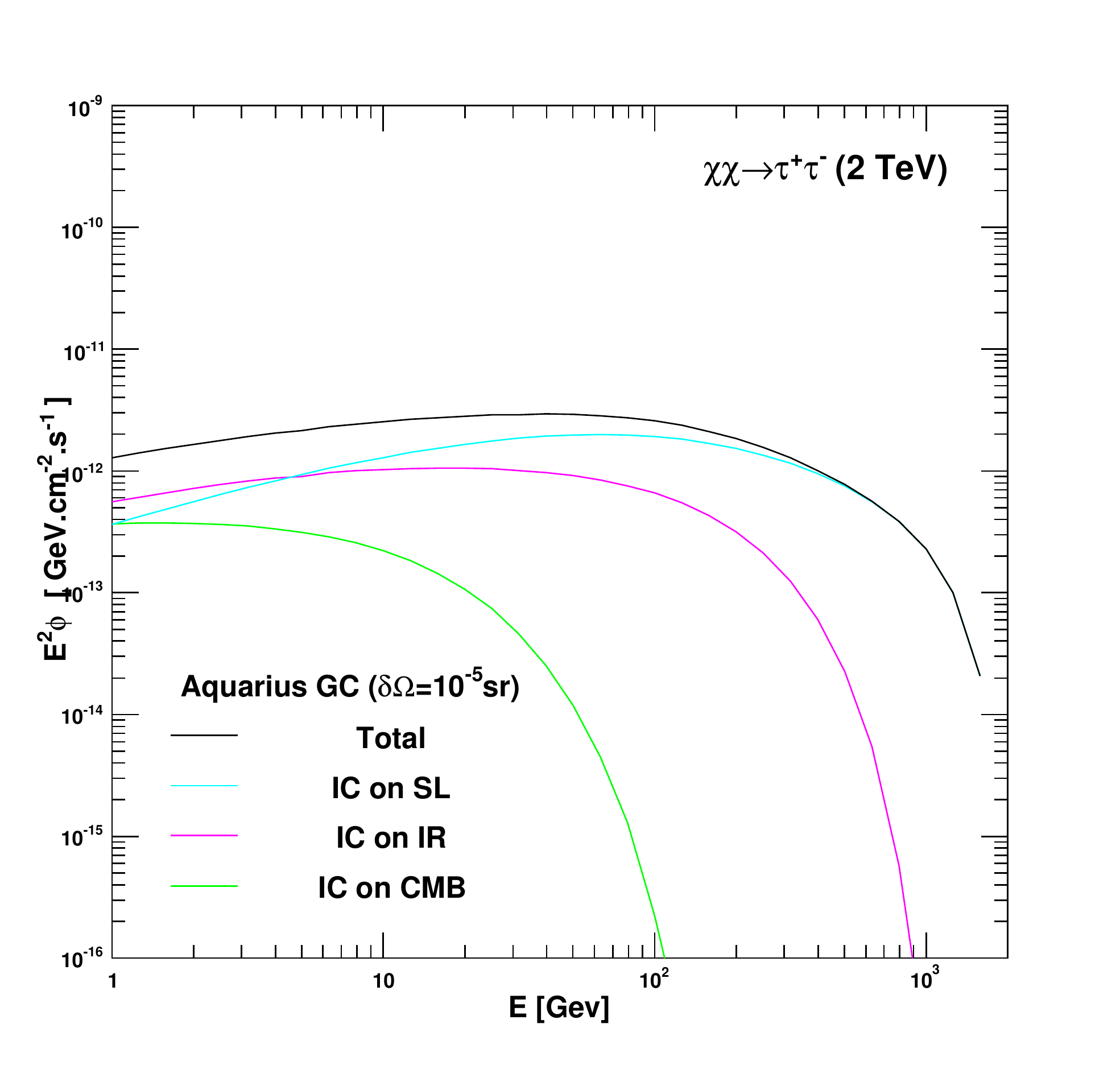}
\caption{\small Left: Template comparison between the positron fluxes predicted 
  at different radii in the \vlii~setup (in the smooth approximation) in
  a full propagation model (solid curves) and in a model considering the energy 
  losses only. Right: Predictions of the IC flux generated by benchmark D 
  in the direction of the GC for both the \aquarius~and \vlii~configurations.}
\label{fig:ics}
\end{center}
\end{figure*}

\begin{figure*}[t]
\begin{center}
\includegraphics[width=\columnwidth, clip]{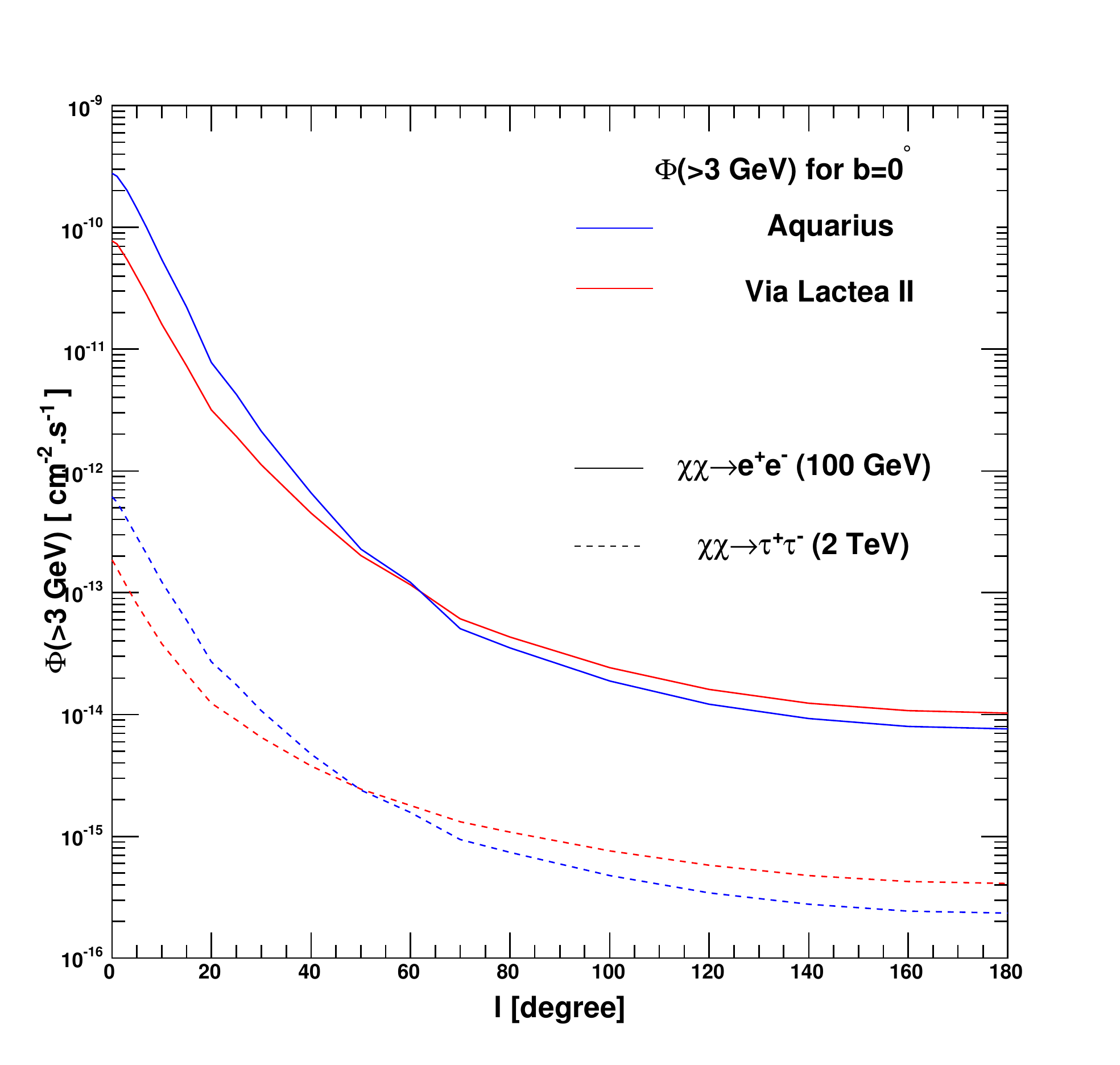}
\includegraphics[width=\columnwidth, clip]{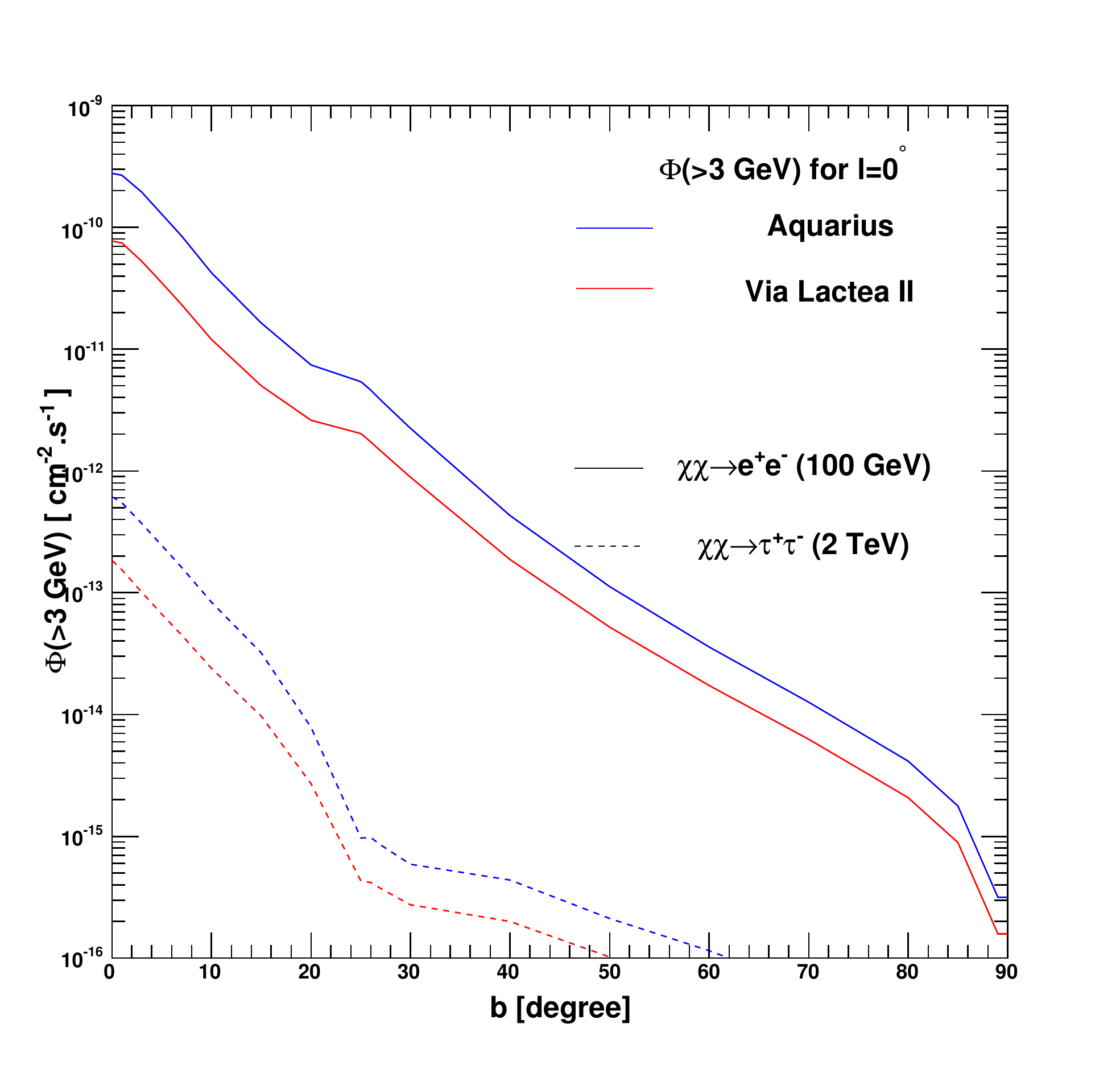}
\caption{\small Left: Longitudinal dependence of the IC integrated flux ($b=0^\circ$)
  obtained for \aquarius~and~\vlii.
  Right: Latitude dependence ($l=0^\circ$).}
\label{fig:ics2}
\end{center}
\end{figure*}

\section{Conclusions}
\label{sec:concl}

We have studied the $\gamma$-ray and antimatter fluxes arising in two
astrophysical setups built on the \aquarius~and \vlii~
high-resolution N-body simulations. Our aim was to quantify the differences
between the two simulations in terms of prospects for detection, 
and to assess the impact of extrapolating the mass function and concentration 
of subhalos down to their minimum mass, that for common DM candidates can 
be 10 order of magnitude smaller than the mass resolution of simulations.
We stress that the two simulations are in remarkable agreement with each 
other, and that the biggest differences in the prospects for detection
arise from the different extrapolations of physical quantities suggested 
by different groups of authors (as in the case of the mass function of 
subhalos), and from the application of the results of the simulations to the 
specific MW halo. We present our results for the \aquarius~and \vlii~setups, 
meaning that they correspond to models that we have built on simulations, and 
not to the simulations themselves.

We have studied the different contributions to the $\gamma$-ray flux arising 
from the smooth DM halo of the MW, resolved subhalos (that we have generated
with a Monte Carlo procedure), unresolved subhalos and extragalactic halos
and subhalos. The smooth component dominates the annihilation flux in
the inner regions of the galaxy. In the \vlii~setup, the resolved 
and unresolved subhalos dominate the annihilation flux at angles larger than
$\sim 20$ degrees from the GC, all the way to the anti-center, where the
extragalactic flux becomes comparable, though never dominant. In the
\aquarius~setup, the substructures component is suppressed, and 
never exceeds the smooth component. The extragalactic flux becomes instead 
dominant at angles larger than $\sim 60$ degrees from the GC.

We have provided full-sky maps that can be used as templates for DM
searches with current experiments such as Fermi. If the search is concentrated
towards the GC, there is little difference between the two simulations, 
in the sense that the profile of the annihilation flux is very similar in 
the two setups, while the normalization must be kept free given the
uncertainties on the particle physics parameters. The optimal strategy
to search for an annihilation signal is to take the sky maps provided 
by Fermi, that will constitute the 'background', and take our DM templates
as signal. One can then easily estimate the size and shape of the 
region around the GC that maximize the S/N ratio.

For a fixed particle physics model, the annihilation flux from the central 
regions of the Galaxy in the \vlii~setup is slightly smaller with respect to 
\aquarius. This is due to the smaller local density, \ie\ DM density in the 
solar neighborhood, in the \vlii~setup and the smaller total mass. If one 
rescales the \vlii~and \aquarius~setup to match the most recent determinations 
of the local density, and if the same subhalo mass fraction $f_{cl}=0.18$ is 
adopted for both simulations, then the annihilation maps look almost identical.

Should the search for the diffuse emission from the GC fail because
of the complicated astrophysical backgrounds in what is probably the
most crowded region of the sky, the possibility remains to search for
unidentified $\gamma$-ray sources, that would appear as non-variable bright spots 
with no astrophysical counterpart, possibly correlated with dwarf galaxies,
and with identical spectra. The number of detectable sources in both 
simulation setups is very similar, and for an optimistic DM scenario is between 
1 and 10 for the Fermi-LAT in 5 years of operation.

Finally, we have calculated the antimatter fluxes in both simulation 
setups, and we found that the boost factor often invoked to provide
a viable DM interpretation of the cosmic leptons puzzle, are completely
unrealistic. The only annihilation channel that provides a sizeable 
enhancement of the positron ratio is direct annihilation to $e^+ e^-$ around
100 GeV, which provides a flux significantly higher than the secondary 
background for the set of propagation parameters used here, even without the 
help of any subhalo contribution. We have also verified that the associated ICS 
contribution to the $\gamma$-ray flux was not violating the current 
observational constraints. Although this model seems an appealing possibility, 
we still stress that (i) it has been tuned to provide an exception case to 
the usual need of large boost factor to interpret the PAMELA data (see 
\citeeqp{eq:poslineflux} and comments below) without any particle physics 
motivation, (ii) slightly increasing the DM particle mass above 100 GeV would 
completely erase such a peak with respect to the background because of the 
dependency of the peak amplitude in $\mchi^{-4}$ and the steep decrease of the 
flux at lower energies and (iii) we did not include the contributions of other 
astrophysical primary sources, like pulsars, which are likely sizable.

\section*{Acknowledgments}
We would like to thank J. Diemand and M. Fornasa for useful discussions.
EB thanks the Institut d'Astrophysique de Paris for the kind hospitality. LP also thanks the Department of Physics of the University of Trento for the kind hospitality. EB and LP 
acknowledge financial contribution from contract ASI-INAF/TH-018. 
This work was supported in part by the French ANR project ToolsDMColl,
BLAN07-2-194882.

\appendix
\section{Smooth versus subhalo mass density profiles for antibiased relations}
\label{app:antibiased}
The spatial distribution of subhalos has long been modeled with a cored
isothermal profile, \eg\ as suggested in Ref.~\cite{2004MNRAS.352..535D}.
Nevertheless, it turns out that such a spatial distribution is hardly 
consistent with a global NFW fit on the overall DM distribution, which scales 
like $r^{-3}$ at large radii, at variance with the $r^{-2}$ isothermal 
behavior. Indeed, subhalos are usually found to dominate the mass profile at 
radii larger than the scale radius ($\sim$ 20 kpc), so one could expect their 
mass density profile to track the $r^{-3}$ shape of the overall fit.
In Ref.~\cite{2007ApJ...667..859D}, the authors quoted the same previous 
reference and proposed the following empirical spatial distribution for the 
subhalo number density $n(r)$, the so-called {\em antibiased} distribution:
\ben
n(r) \propto r\,\rho_{\rm host}(r)\;.
\een
Considering that $\rho_{\rm host}$ is the overall fit, this would lead to the same
issue as above: for a global NFW profile, the subhalo distribution would 
decrease like $r^{-2}$ beyond the scale radius, which is inconsistent with the 
fact that they are found to dominate the mass profile on large radii. For 
consistency, $\rho_{\rm host}$ should thereby be the {\em smooth} DM component 
instead.

In this appendix, we sketch an analytical method to model any antibiased 
subhalo distribution, given a defined overall density profile.

\subsection{General case}
\label{subapp:gen}

Let us consider that a global fit on an N-body galaxy made of pure dark matter 
provides an analytical shape for the overall mass density profile, 
$\rhotot(r)$. This density profile must therefore obey
\ben
4\,\pi\int^{\Rvir}_0 dr\,r^2 \,\rhotot(r) =  \MMW \;.
\label{eq:mmw}
\een
In the following, we will consider that \MMW, \Rvir~and \rhotot~are known.

Now, let us assume that this overall profile is in fact made of two 
sub-components, one describing the smooth distribution of dark matter, 
$\rhosm(r)$, and another accounting for the mass density carried by subhalos, 
$\rhosub(r)$. If we know the mass fraction of resolved subhalos in any N-body
Galaxy and if we further know the mass distribution of these objects, then 
assuming a scale invariant mass profile allows us to determine the total mass 
fraction $\fsub$ for any arbitrary minimal mass for subhalos. Therefore,
the total mass density profile may be rewritten as
\ben
\label{eq:def_gs}
\rhotot(r) &=& \rhosm(r) + \rhosub(r)\\ 
&=& (1-\fsub) \, \MMW \, \gsm(r) + \fsub \, \MMW \, \gsub(r)\;,\nn
\een
where $\gsm(r)$ and $\gsub(r)$ are normalized to unity inside a spherical 
volume delineated by \Rvir. We have merely used the fact that the total mass 
in the form of subhalos is $\msubtot = \fsub \, \MMW $.

If we further assume that the subhalo spatial distribution $\gsub(r)$ is 
antibiased with respect to the smooth distribution $\gsm(r)$, then we have
the following relation:
\ben
\gsub(r) = K\,r\,\gsm(r)\;,
\label{eq:def_gsub}
\een  
where $K$ is a constant which ensures the normalization to unity inside a 
sphere of radius \Rvir.

From this, we can readily express $\gsm(r)$ or $\gsub(r)$ in terms of the 
known quantities. For the former one, we find:
\ben
\gsm(r) 
= \frac{\rhotot(r)}{(1-\fsub)\, \MMW\left[ 1 + \frac{r}{r_b}\right]}
\;,
\label{eq:gsm}
\een
where we have defined the {\em bias radius} $r_b$ as follows:
\ben
r_b \equiv \frac{(1-\fsub)}{\fsub\,K} \;.
\label{eq:def_rb}
\een
All parameters are known, except the constant $K$. However, it can easily be 
derived from this equation by demanding that $\gsm(r)$ is normalized to unity 
inside a sphere of radius \Rvir. We can further verify that this value obtained 
for $K$ automatically ensures the normalization of $\gsub(r)$ as expressed in 
\citeeq{eq:def_gsub}.

It is interesting to derive the expression for the smooth density profile
\ben
\rhosm(r) &=& (1-\fsub)\,\MMW\,\gsm(r) \nn\\
&=& \frac{\rhotot(r)}{(1 + r/r_b)}\;.
\label{eq:rhosm}
\een
Similarly, we can determine the averaged subhalo mass density profile:
\ben
\rhosub(r) &=& \fsub \, \MMW \, K \, r \,\gsm(r) \nn\\
&=& \frac{\rhotot(r)\,(r/r_b)}{(1 + r/r_b)}\;.
\label{eq:rhosub}
\een
One can readily check that $\rhosm(r) + \rhosub(r) = \rhotot(r)$, as required.

We can now interpret the physical meaning of the bias radius. This radius 
actually provides the scale beyond which the smooth profile departs from the
total density profile, more precisely beyond which the smooth density decreases
one power of $r$ faster than the overall density, since the mass density in 
subhalos starts to dominate. In \vlii, where the overall density is 
well fitted with an NFW profile, the bias radius is of order of the NFW scale 
radius. Thus, the smooth profile is found to scale like $r^{-1}$ in the central 
regions of the Galaxy, while it falls like $r^{-4}$ in the outer skirts, so 
faster than the $r^{-3}$ behavior of the NFW shape.

\subsection{Application to \vlii}
\label{subapp:appl}

As a useful application, we will use the \vlii~setup, for
which the parameters are recalled in \citetab{tab:vl2}. We remind that
the NFW profile is given by
\ben
\rhotot(r) = \frac{\rho_s\,(r_s/r)}{\left( 1 + r/r_s\right)^2} \;.
\een
If we assume a mass profile for subhalos of the form $\propto m^{-\alpha}$, 
and impose that a certain fraction $\fref$ of the Milky Way mass is carried
by subhalos in the mass range $[m_{\rm ref},m_{\rm max}]$, then the total 
mass fraction in subhalos \fsub~is entirely fixed by the minimal subhalo 
mass $m_{\rm min}$.

In the antibiased hypothesis, the smooth density profile is defined by 
\citeeqs{\ref{eq:gsm},\ref{eq:rhosm}}. In particular, we have defined function 
$\gsm$ such that it is normalized to unity over the Galaxy volume. It 
turns out that this volume integral has an analytical form in the \vlii~setup:
\ben
1 &=& 4\,\pi\int_0^{R_{\rm vir}}dr\,r^2 \,g_{\rm sm}(r) \\
&=&
4\,\pi\,r_b\,r_s^3 \,\rho_s \times \nn\\
&& \frac{\left[ \Rvir ( r_s - r_b ) + 
  r_b (\Rvir + r_s) 
  \ln\left\{ \frac{ r_b (\Rvir + r_s)}{r_s(r_b + \Rvir)}\right\} \right]}
{(1-\fsub)\MMW (r_b-r_s)^2(\Rvir + r_s)}\;,\nn
\een
such that we can easily compute the value of $r_b$ that ensures the 
normalization to unity. The full results are summarized in \citetab{tab:vl2}.

\begin{table*}
\centering
\begin{tabular}{ccccccccccc}
\hline
$R_{\rm MW}$ & $M_{\rm MW}$ & $r_s$ & $r_b$ & $\rho_s$ & slope $\alpha$ & 
\msubtot  & \fsub  & $\rho_{\odot}$ & $\rho_{\odot,{\rm sm}}$ & 
$\rho_{\odot,{\rm sub}}$\\
(kpc) & ( $ M_\odot$ ) & (kpc) & (kpc) & (GeV/cm$^3$) & & ($M_\odot$ ) & (\%) & 
(GeV/cm$^3$) & (GeV/cm$^3$) & (GeV/cm$^3$) \\
 402 & $1.9\times 10^{12}$  & 21  & 85.5 & 0.31  & 2.0 & $10^{12}$ & 53.3 & 
 0.42 & 0.39 & 0.03 \\
\hline
\end{tabular}
\caption{Parameters relevant for the \vlii~setup. The Sun position is taken 
  at 8 kpc.}
\label{tab:vl2}
\end{table*}

\bibliography{lavalle_bib}


\end{document}